\def\mearth{{\rm\,M_\oplus}}

\def\deg{^\circ}
\def\micron{\mu {\rm m}}

\documentclass[traditabstract]{aa}

\usepackage{graphicx}
\usepackage{txfonts}

\usepackage{natbib}
\usepackage{aas_macros}

\begin{document}

\titlerunning{Debris disks as signposts of terrestrial planet formation II}
\authorrunning{Raymond et al.}

\title{Debris disks as signposts of terrestrial planet formation}

\subtitle{II. Dependence of exoplanet architectures on giant planet and disk properties}

\author{Sean N. Raymond\inst{1,2}, 
Philip J. Armitage\inst{3,4},
Amaya Moro-Mart{\' i}n\inst{5,6},
Mark Booth\inst{7,8},
Mark C. Wyatt\inst{7},
John. C. Armstrong\inst{9},
Avi M. Mandell\inst{10},
Franck Selsis\inst{1,2}, \&
Andrew A. West\inst{11,12}
}

\institute{
Universit{\'e} de Bordeaux, Observatoire Aquitain des Sciences de l'Univers, 2 rue de l'Observatoire, BP 89, F-33271 Floirac Cedex, France; \email{raymond@obs.u-bordeaux1.fr}
\and
CNRS, UMR 5804, Laboratoire d'Astrophysique de Bordeaux, 2 rue de l'Observatoire, BP 89, F-33271 Floirac Cedex, France
\and
JILA, University of Colorado \& NIST, Boulder CO 80309, USA
\and
Department of Astrophysical and Planetary Sciences, University of Colorado, Boulder CO 80309, USA
\and
Department of Astrophysics, Center for Astrobiology, Ctra. de Ajalvir, km 4, Torrej{\' o}n de Ardoz, 28850, Madrid, Spain
\and
Department of Astrophysical Sciences, Princeton University, Peyton
Hall, Ivy Lane, Princeton, NJ 08544, USA
\and
Institute of Astronomy, Cambridge University, Madingley Road, Cambridge, UK
\and
University of Victoria, 3800 Finnerty Road, Victoria, BC, V8P 1A1 Canada
\and
Department of Physics, Weber State University, Ogden, UT, USA
\and
NASA Goddard Space Flight Center, Code 693, Greenbelt, MD 20771, USA
\and
Department of Astronomy, Boston University, 725 Commonwealth Ave, Boston, MA, 02215 USA
\and
Visiting Investigator, Department of Terrestrial Magnetism, Carnegie Institute of Washington, 5241 Broad Branch Road, NW, Washington, DC 20015, USA
}

\date{}

\abstract{
We present models for the formation of terrestrial planets, and the collisional evolution of 
debris disks, in planetary systems that contain multiple marginally unstable gas giants. We 
previously showed that in such systems, the dynamics of the giant planets introduces a correlation 
between the presence of terrestrial planets and cold dust, i.e., debris disks, which is particularly 
pronounced at $\lambda \sim 70 \micron$. Here we present new simulations that show that this connection
is qualitatively robust to a range of parameters: the mass distribution of the giant planets, the width 
and mass distribution of the outer planetesimal disk, and the presence of gas in the disk when the giant 
planets become unstable. We discuss how variations in these parameters affect the evolution.
We find that systems with equal-mass giant planets undergo the most violent
instabilities, and that these destroy both terrestrial planets and the outer planetesimal disks that produce 
debris disks. 
In contrast, systems with low-mass giant planets efficiently produce both terrestrial planets and debris disks. A
large fraction of systems with low-mass ($M \lesssim 30 \mearth$) outermost giant planets have final 
planetary separations that, scaled to the planets' masses, are as large or larger than Uranus and 
Neptune in the Solar System. We find that the gaps between these planets are not only dynamically stable to test particles, 
but are frequently populated by planetesimals. The possibility of 
planetesimal belts between outer giant planets should be taken into account when interpreting debris disk
SEDs. In addition, the presence of $\sim$ Earth-mass ``seeds'' in outer planetesimal disks causes the disks to
radially spread to colder temperatures,  and leads to a slow depletion of the outer planetesimal disk 
from the inside out. We argue that this may 
explain the very low frequency of $>1$ Gyr-old solar-type stars with observed $24 \micron$ excesses.  
Our simulations do not sample the full range of plausible initial conditions for planetary 
systems. However, among the configurations explored, 
the best candidates for hosting terrestrial planets at $\sim$ 1 AU are stars older than 0.1-1 Gyr with bright debris
disks at $70 \micron$ but with no currently-known giant planets. These systems combine evidence for the 
presence of ample rocky building blocks, with giant planet properties that are least likely to 
undergo destructive dynamical evolution.  Thus, we predict two correlations that should be detected by upcoming surveys: an anti-correlation between debris disks and eccentric giant planets and a positive correlation between debris disks and terrestrial planets.

}{}{}{}{}

\keywords{planetary systems: formation --- methods: n-body simulations --- circumstellar matter --- infrared stars --- Kuiper belt --- Solar System --- astrobiology}

\maketitle

\section{Introduction}

The Solar System's distinctive architecture, in which rocky terrestrial planets lie interior to gas and ice giants,
with the Kuiper Belt of smaller bodies beyond, is not unexpected. The mass in protoplanetary feeding zones increases
with orbital distance, but the resulting tendency toward the formation of larger planets further out is eventually
frustrated both by the lengthening time scale for accretion~\citep[e.g.,][]{lissauer93,kokubo02}, 
and by the increased ability of planetary cores to scatter planetesimals inward~\citep{levison01}. The competition between
these effects plausibly leads to the relatively slow ($\sim$100~Myr) assembly of a handful of terrestrial planets
inside a few~AU, the faster growth of several planetary cores with $M \gtrsim 5 \ M_\oplus$ inside $\sim$10~AU, and
the persistence of a belt of unconsolidated debris further out. Simple arguments of this kind fail to establish how
often planetary cores grow fast enough to admit the formation of fully fledged gas giants, but empirical estimates
based on extrapolations of radial velocity and microlensing surveys suggest that gas giant formation is
common~\citep{cumming08,gould10}. Debris disks \citep{wyatt08} are also observed around a significant fraction of 
young stars -- despite the existence of both dynamical and collisional processes that can destroy them 
on a time scale short compared to the main sequence lifetime of Solar-type stars -- but the abundance 
of Earth-mass planets remains to be measured.

For the Solar System, we have access to a unique array of observational constraints. Even with 
these advantages the exact nature of the interactions between the giant planets, the 
terrestrial planets, and the Kuiper belt remain under debate. In the inner Solar System, 
at a {\em minimum} secular resonances with the giant planets   
would have influenced terrestrial planet formation~\citep{nagasawa05,raymond09c}. There could, 
however, have been stronger effects. Gas driven migration of
the giant planets could have brought them (temporarily) closer to the Sun \citep{walsh11}, directly reducing the 
supply of raw material in the Mars region and preventing the growth of a larger planet \citep{hansen09}. 
In the outer Solar System, early studies focused on the dynamics of Neptune, which is of  
low enough mass that inward scattering of planetesimals can drive substantial outward orbital 
migration \citep{fernandez84}. The migration can deplete the mass in the
Kuiper Belt and result in the resonant capture of Pluto and other bodies by
Neptune~\citep{malhotra93}. Subsequent work introduced the idea of larger-scale dynamical 
instability among the Solar System's giant planets \citep{thommes99}. In the most-developed 
models, early outer Solar System evolution is characterized by a combination of planetesimal 
migration, close encounters between planets, and resonant interactions \citep{tsiganis05,levison11}.

Theoretically, attempts to construct equally detailed models for extrasolar planetary system evolution 
are hampered by uncertainties in the distribution of the initial disk conditions, and by our 
poor knowledge of the evolution of gas disks~\citep{armitage11} and formation mechanism for  
planetesimals~\citep{chiang10}.
Several observed properties of extrasolar planetary systems, however, including the existence of hot Jupiters (whose  
orbits are sometimes misaligned with respect to the stellar spin axis) and 
the prevalance of eccentric orbits, 
favor scenarios in which large-scale orbital evolution of giant planets is the
norm~\citep{winn10,triaud10,schlaufman10}.  It is therefore of interest to determine 
the diversity of outcomes that could arise given initial conditions and dynamical processes 
similar to those of the 
Solar System, and to examine how those outcomes depend upon parameters such as the masses of the giant 
planets, and the properties of primordial planetesimal belts. Doing so is the goal of the 
present paper. We are particularly interested in studying how terrestrial planets form, and 
debris disks evolve, in the presence of dynamically active giant planet systems. In an 
earlier paper~\citep[``Paper 1"][]{raymond11} we showed that if giant planets
form in or near dynamically unstable configurations, there are striking correlations between the nature of
the terrestrial planets that form, and the properties of outer debris disks whose ongoing collisional evolution can
be observed out to ages of several Gyr~\citep[e.g.][]{wyatt08,krivov10}. The dynamically calm conditions that favor
the formation of massive terrestrial planet systems also result in long-lived outer debris disks, that remain bright
in cold dust emission (e.g. at $\lambda \sim 70 \micron$) to late times. In systems that suffer more dramatic
dynamical evolution, we identified a channel for the formation of unusual terrestrial planet systems in which a
single planet exhibits large oscillations in eccentricity and inclination due to secular coupling to a scattered
giant.  Here, we consider a broader range of models within the same qualitative class, and study 
how robust our earlier conclusions are to changes in the poorly-constrained model parameters.

Our paper is structured as follows.  In section 2 an outline of our methods is presented (the reader is referred to
Paper~1 for more details and numerical tests).  In subsequent sections we present results of different sets of
simulations to test the effect of: the giant planet mass and mass distribution (section 3),  the width and mass
distribution of the outer planetesimal disk (section 4), and the presence of gas during giant planet instabilities
(section 5). In section 6, we discuss the implications of our models for debris disks and terrestrial 
planet systems. We conclude in section 7.

\section{Methods}
The initial conditions for our simulations assume that the location and mass of the rocky material in the 
terrestrial planet zone, and the mass of planetesimals in the outer disk, are fixed at values similar 
to those employed in Solar System models. We assume that the masses of the giant planets, on the other 
hand, have a broad dispersion (extending up to masses above those realized in the Solar System) that is 
uncorrelated with the mass in either the terrestrial planet region or in the outer planetesimal disk. 
All of our simulations contain three radially-segregated components orbiting a solar-mass star: 
\begin{enumerate}
\item The building blocks of terrestrial planets: $9 \mearth$ in 50 planetary embryos and 500 planetesimals from 0.5 to 4 AU, with equal mass in each component and a radial surface density profile $\Sigma \sim r^{-1}$.  The initial eccentricities were chosen at random from 0-0.02 and the initial inclinations from $0-0.5^\circ$.
\item Three giant planets at Jupiter-Saturn distances: the innermost planet is placed at 5.2 AU and the two others
are spaced outward by 4-5 mutual Hill radii.  We adopt three-planet initial conditions because this is the simplest
plausible configuration that evolves dynamically to match the measured eccentricity distribution of massive
extrasolar planets~\citep{chatterjee08}.  The planets were placed on initially circular orbits with randomly-chosen inclinations of $0-1^\circ$.
\item An outer disk of planetesimals thought to be analogous to the primitive Kuiper belt.  This belt consists of 1000 planetesimals with a total mass of 50 $\mearth$. The belt starts 4 Hill radii beyond the outermost giant planet and extends radially for 10 AU, and also follows an $r^{-1}$ radial surface density profile.  f The initial eccentricities were chosen at random from 0-0.01 and the initial inclinations from $0-0.5^\circ$.
\end{enumerate}

Adopting these initial conditions amounts to making implicit assumptions about the typical outcome 
of planet formation. First, our terrestrial, giant planet, and outer disk zones are located such that 
they are in immediate dynamical contact with each other. This is reasonable only if planetesimal 
formation results in a smooth, gap-less distribution of bodies in $0.5 \ {\rm AU} < a \lesssim 20 \ {\rm AU}$, 
{\em and} if giant planet migration is limited. Substantial giant planet migration, of the kind envisaged in 
models by \citet{masset01}, \citet{walsh11} and \citet{pierens11}, could create dynamical separation between the giant and terrestrial planets prior to the gas-less phase of evolution that we simulate. Second, we assume 
non-resonant initial conditions for the giant planets. Mean-motion resonances can be established 
readily if there is significant migration, due to either gas disk torques or planetesimal scattering, 
and a plausible alternate class of models could be constructed in which fully resonant initial 
conditions were the norm \citep{morbidelli07}. We do not consider this possibility further here.

Embryo and giant planet particles feel the gravitational attraction of all other bodies in the simulation.  Planetesimal particles, both in the inner and outer disk, feel the gravity of embryos and giant planets but do not self-gravitate.  This commonly-used approximation allows for an adequate treatment of collective particle effects and dramatically reduces the required computation time.
Our methods are outlined in detail in Paper~1.  Here we summarize the key points.  Each simulation was integrated
for 100-200 million years using the hybrid version of the Mercury code~\citep{chambers99} with a 6 day timestep. Collisions between particles were treated as inelastic mergers.  
Particles were removed from a simulation when they either
came within 0.2 AU of the central (solar-type) star at which point they are assumed to have collided with the star,
or ventured farther than 100 AU from the central star (1000 AU for the {\tt widedisk} runs discussed in Section 4.2), at which point the
particle is assumed to have been ejected from the system.

\begin{table*}
\caption{Summary of the simulations}
\begin{tabular}{|l|l|c|c|c|}
\hline
Set & N(sims)$^\star$ & Giant planet mass distribution & Giant planet spacing & Gas?\\
\hline
{\tt mixed} & 156 & $dN/dM \sim M^{-1.1}$ from $1 M_{\rm Sat}$ to $3 M_{\rm Jup}$ & $4-5 R_{H,m}$ & no \\
{\tt equal} & 68 & $M = 30 M_\oplus, 1 M_{\rm Sat}, 1 M_{\rm Jup}\, {\rm or} \, 3 M_{\rm Jup}$& $2.5-3 R_{H,m}$& no\\
{\tt lowmass} & 86 &  $dN/dM \sim M^{-1.1}$ from $10 M_\oplus$ to $1 M_{\rm Jup}$ & Same as {\tt mixed} & no \\
{\tt widedisk} & 91 &  Same as {\tt mixed} & Same as {\tt mixed} & no \\
{\tt smallseed} & 44 & Same as {\tt mixed} & Same as {\tt mixed} & no \\
{\tt bigseed} & 47 & Same as {\tt mixed} & Same as {\tt mixed} & no \\
{\tt gas} & 45 & Same as {\tt mixed} & crossing orbits & yes \\
\hline
\end{tabular}
\\ $^\star$Note that N(sims) represents the number of simulations that met our criteria for run time of $>$ 100 Myr and energy conservation of $dE/E < 10^{-2}$ (see Appendix A of Paper~1 for numerical tests).  Our initial batches of simulations were somewhat larger: 200 simulations in {\tt mixed}, 20 for each {\tt equal} set, 100 for {\tt lowmass} and {\tt widedisk}, and 50 for {\tt gas}, {\tt smallseed} and {\tt bigseed}.
\label{tab:init}
\end{table*}

In Paper~1 we presented the results from our fiducial {\tt mixed} set of simulations.  In these simulations the giant planet masses are drawn randomly from the observed exoplanet mass distribution~\citep{butler06,udry07b}:
\begin{equation}
\frac{{\rm d}N}{{\rm d}M} \propto M^{-1.1},
\end{equation}
where masses were chosen between one Saturn mass and 3~Jupiter masses.  In these runs, the masses of individual planets were chosen independently.  

We also computed a number of alternate models in which either the range of the mass function, the assumption of independent masses, or the properties of the outer disk were altered (see Table~\ref{tab:init}):
\begin{itemize}

\item The {\tt lowmass} simulations represent systems with low-mass giant planets.  For these cases the giant planet
masses also follow the observed exoplanet distribution, but with masses between 10~$M_\oplus$ and 1~$M_{\rm
Jup}$. The initial conditions for the giant planets in these {\tt lowmass} simulations are the same as the
``mixed2'' simulations in~\cite{raymond08b,raymond09a,raymond09b,raymond10}.  Abnormally low giant 
planet masses, relative to the mass in terrestrial planet-forming material, might occur physically in 
disks around stars where stronger than average photoevaporation limits the disk lifetime.
\item The {\tt equal} simulations comprise four sets of simulations, each containing three giant planets with fixed
masses of $30 \mearth$, $1 M_{Sat}$, $1 M_J$, or $3 M_J$.  For the {\tt equal} simulations the planets were placed
in a slightly more compact configuration (separated by 3.5-4 mutual Hill radii rather than 4-5) to ensure that they
would become unstable.  This variation mimics the reality that the conditions that favor the growth of
one gas giant to high masses -- for example early core formation or a long disk lifetime -- probably apply also to
other planets in the same system.  Past work suggests that these simulations should produce the most violent instabilities~\citep{ford03,raymond10}.
\item The {\tt widedisk} simulations test the effect of planetesimal disks that are 20 AU wide rather than 10, and twice as massive (so that there is the same mass in the first 10 AU of the annulus as the fiducial case).  In these simulations, the radius beyond which an object is considered ejected was 1000 AU rather than 100 AU.  The giant planets' initial orbits are identical to the {\tt mixed} set.
\item The {\tt seeds} simulations test the effect of the mass distribution within the planetesimal disk by including five or ten equally-spaced equal-mass fully self-gravitating seeds of either $2 \mearth$ or $0.5 \mearth$ respectively.  The total mass and width of the planetesimal disk was held fixed at 10 AU.  
\item The {\tt gas} simulations test the effect of the presence of a gas disk during and after the giant planet instabilities.  (The methodology is discussed in \S 5.)
\end{itemize}  

Each simulation was post-processed to calculate the spectral energy distribution of dust in the system following the
method of~\cite{booth09} with a few small changes (see Section 2.3 in Paper~1).  To do this each planetesimal particle was assumed to represent a population of objects with sizes between $2.2 \micron$ and 2000 km.  This population was assumed to be in collisional
equilibrium such that the differential size distribution can be written as $n(D) \propto D^{-3.5
}$~\citep{dohnanyi69}.  The radial distribution of dust was calculated by a simple combination of the
planetesimal orbital distribution (and by sampling eccentric orbits at multiple intervals along their orbit equally spaced in mean anomaly).  The spectral energy distribution was calculated by assuming that the dust grains in each radial bin emit as blackbodies based on their effective temperature.  At each simulation timestep the collisional timescale $t_c$ was calculated for the largest objects ($D = $ 2000 km) for the population of both asteroidal and cometary planetesimals.  In practice, $t_c$ represents the mean timescale between collisions energetic enough to disrupt an object of a given size at a given orbital radius.  $t_c$ is a function of the mass, width, and orbital distribution of the planetesimal belt as well as the physical properties of planetesimals themselves, in particular their $Q_D^\star$, the impact energy needed to catastrophically disrupt a planetesimal of size $D$~\citep[for details, see][]{wyatt99,wyatt07b,booth09,raymond11,kains11}.  Once $t_c$ was calculated for a given timestep, the effective dust mass of each population was decreased by a factor of $\left[1 + t/t_c(D_c)\right]^{-1}$.  This decrease in the dust mass is not self-consistent because the planetesimal mass in the simulations is constant.  This effect can be important for the asteroidal planetesimals because their collisional timescale, $t_c \sim 10^4$ years, is short compared to the interesting timescales for dynamical evolution. The opposite ordering typically applies for the outer, cometary planetesimals, whose collisional timescale is $t_c \gtrsim 10^8$ years. The dust fluxes used in the analysis later in the paper are dominated by the cometary component so this inconsistency has little to no effect on our results.  In addition, for the case of low-mass giant planets that migrate due to planetesimal scattering (i.e., the {\tt lowmass} simulations), the timescale for dynamical mass loss from the outer planetesimal disk is roughly an order of magnitude shorter than the timescale for the calculated collisional mass loss in that same region.  Thus, our assumption that the planet-planetesimal disk dynamics is not affected by the collisional cascade appears reasonable for the outer disk. 

This simple model is based on previous studies that fit the statistics of debris disks using models for the collisional evolution of planetesimals~\citep{dominik03,krivov05,krivov06,wyatt07b,wyatt08,lohne08,kains11}.  Our model agrees to within a factor of 2-3 at $24 \micron$ and $70 \micron$ with more detailed calculation of dust production during the collisional evolution~\citep{kenyon08,kenyon10} and also with dust fluxes observed around solar-type
stars~\citep{habing01,beichman06,moor06,trilling08,hillenbrand08,gaspar09,carpenter09}.  However, due to our
incomplete knowledge of the physical properties of planetesimals, there remains uncertainty in the dust fluxes of up
to an order of magnitude for a given system~\citep[see ][]{booth09}.  Our model does not include outgassing from comets (i.e., outer disk planetesimals) when they enter the inner Solar System.  Thus, the dust fluxes that we calculate during planetesimal bombardments are significantly underestimated.  Indeed, the $24 \micron$ dust flux during the late heavy bombardment calculated by~\cite{booth09} with our method reaches a peak that is roughly an order of magnitude lower than that calculated by~\cite{nesvorny10}, who accounted for cometary dust production.  Cometary outgassing is of importance at mid-infrared wavelengths ($10 \lesssim \lambda \lesssim 50 \micron$) during and shortly after bombardments.  As our results focus on the steady-state production of cold dust in outer planetesimal disks rather than on bombardments, we are not strongly affected by this effect.  

An important point is the fact that our sets of simulations systematically over-predict the frequency of debris disks by a factor of roughly 2-4.  Given {\it Spitzer}'s detection limits, the observed frequency of debris disks at $70 \micron$ around Solar-type stars older than 1 Gyr is 16.4\%~\citep{trilling08,carpenter09}.  At $24 \micron$, the observed frequency is much lower, only 2-3\% for stars older than 300 Myr~\citep{carpenter09,gaspar09}.  Various aspects of and potential solutions to this issue will be discussed in more detail throughout the paper.  

\subsection{An example {\tt mixed} simulation}
Here we briefly present a simulation from the {\tt mixed} set to facilitate comparison against example simulations from other sets that are presented in later sections.  The chosen system started with giant planets of 0.96, 0.46 and 0.64 $M_J$ in order of increasing orbital distance, and an outer planetesimal disk that extended out to 22.8 AU.  

\begin{figure*}
\includegraphics[width=0.63\textwidth]{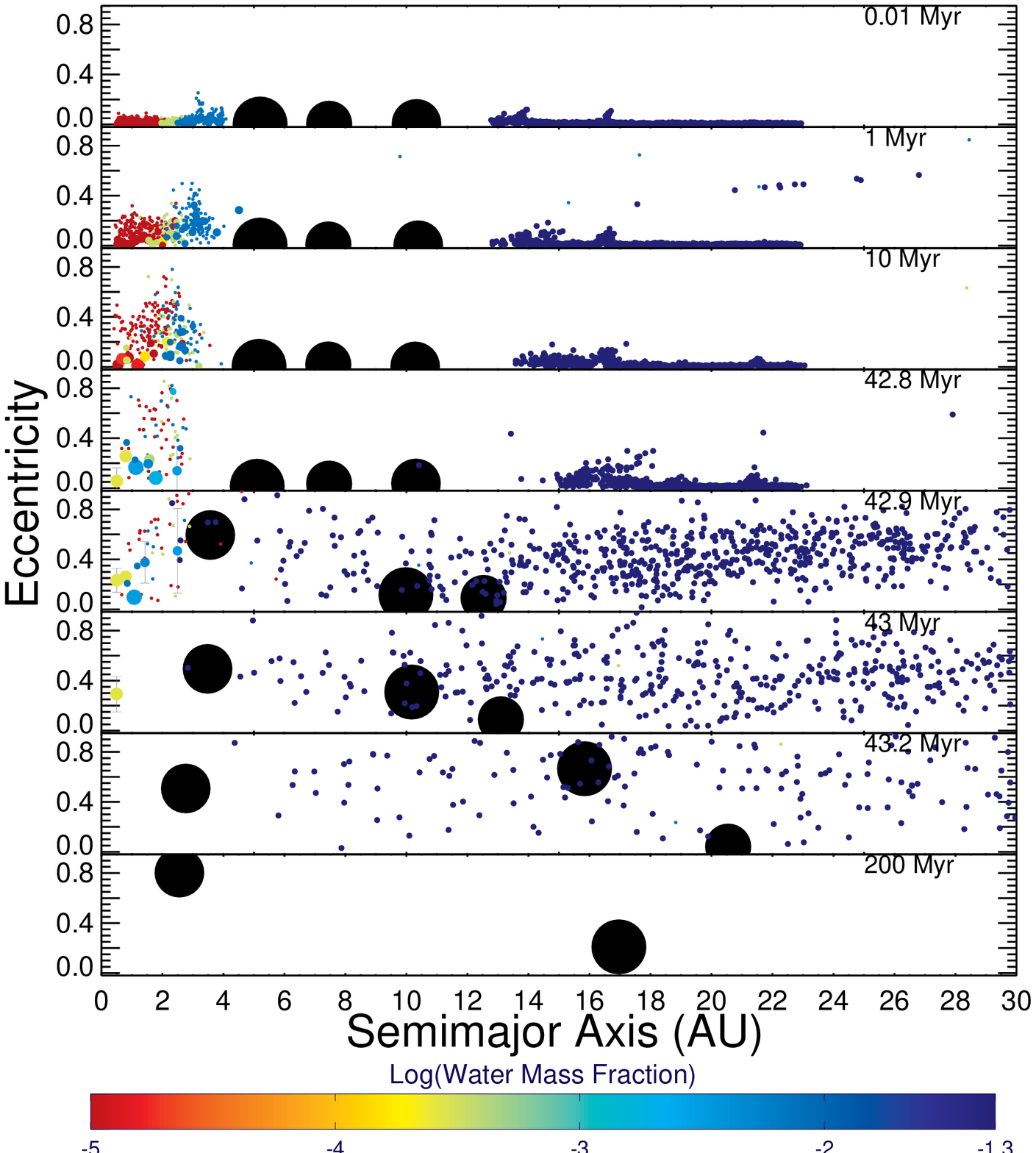}
\vskip -5in
\hspace {0.63\textwidth}
\includegraphics[width=0.37\textwidth]{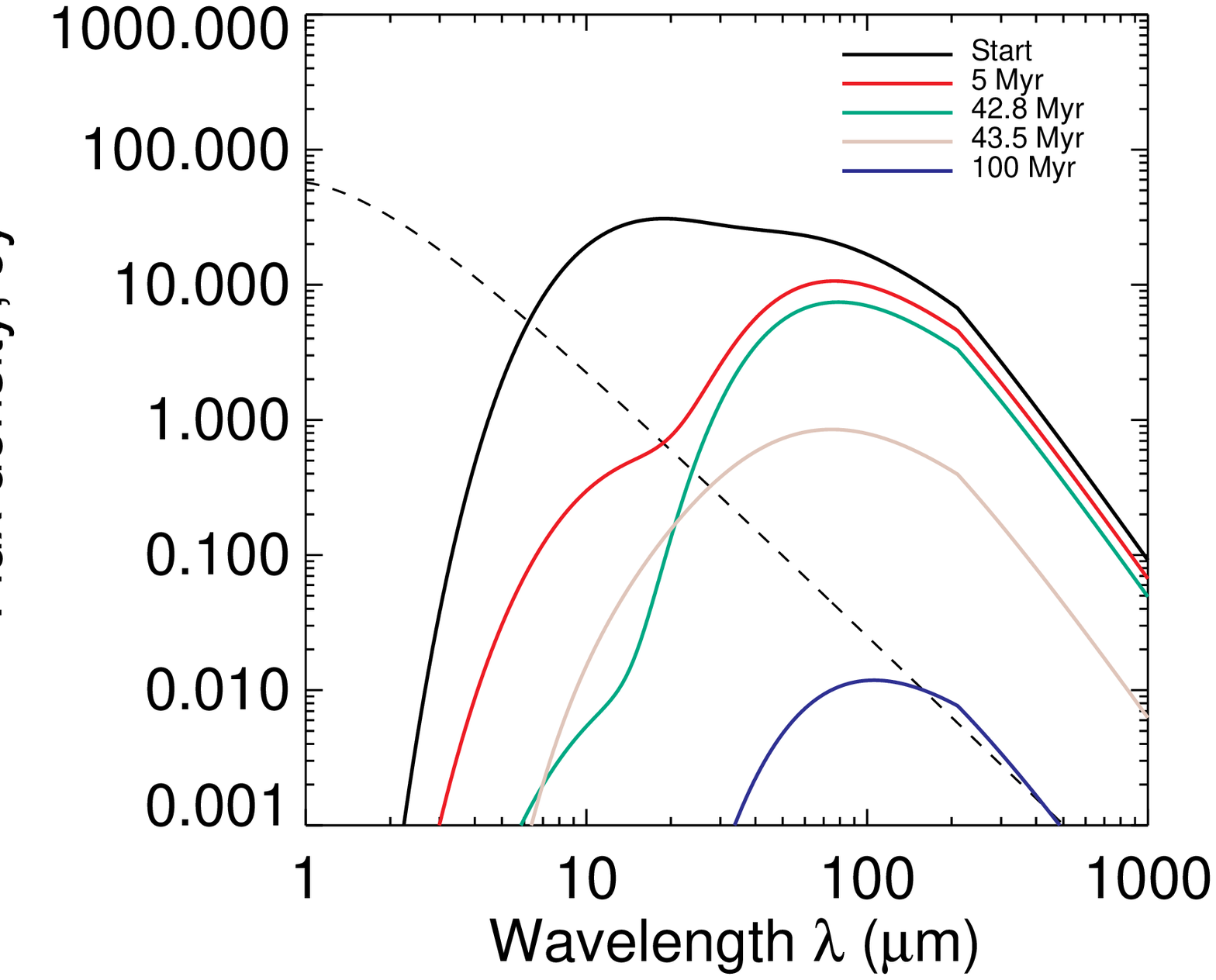}
\vskip 0.01in 
\hspace {0.63\textwidth} 
\includegraphics[width=0.37\textwidth]{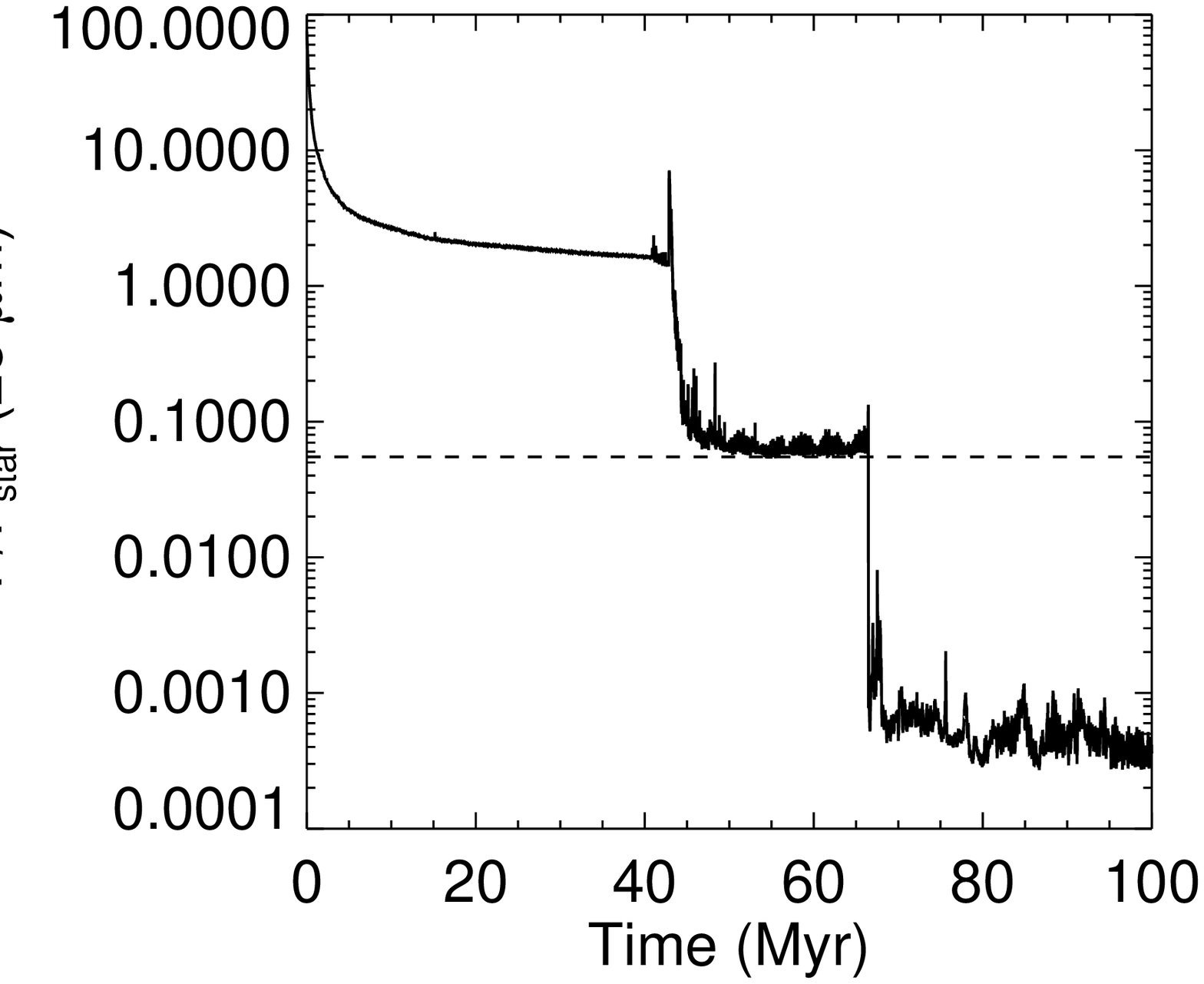}
\vskip 0.3in 
\caption{Evolution of a reference {\tt mixed} simulation in which the giant planets underwent a violent instability after 42.8 Myr of evolution.  The initial giant planet masses were, in order of increasing orbital distance, 0.96, 0.46 and $0.64 M_J$. {\bf Left:} Orbital eccentricity vs. semimajor axis of each body in the simulation.  The size scales with the mass$^{1/3}$ and the color corresponds to the water content, using initial values taken to Solar System data~\citep{raymond04} and re-calculated during impacts by mass balance.  The giant planets are shown as the large black bodies and are not on the same size scale.  {\bf Top right:} The spectral energy distribution of the dust during five simulation snapshots.  The dashed line represents the stellar photosphere.  {\bf Bottom right:} The ratio of the dust-to-stellar flux $F/F_{star}$ at $25 \micron$ as a function of time.  The rough {\it Spitzer} observational limit is shown with the dashed line~\citep{trilling08}. A movie of this simulation is available at http://www.obs.u-bordeaux1.fr/e3arths/raymond/scatterSED\_199.mpg.}
\label{fig:aeit_mixed199}
\end{figure*}

Figure~\ref{fig:aeit_mixed199} shows the evolution of the simulation's dynamics and calculated dust flux.  The system remained stable for 42.8 Myr at which time it underwent a strong dynamical instability that started with a close encounter between the middle and outer giant planets that triggered a series of planet-planet scattering events over the next 400,000 years.  The instability culminated in the ejection of the middle giant planet, and the surviving two giant planets swapped orbits (i.e., the innermost planet became the outermost and vice versa).  At the end of the simulation both planets' eccentricities are large: $e_{inner}$ oscillates between 0.61 and 0.83 and $e_{outer}$ between 0.15 and 0.37.  Despite their large eccentricities, both of the planets' inclinations with respect to the initial orbital plane remain modest (at least in comparison with the planets' eccentricities): $i_{inner}$ oscillates between 3.5$^\circ$ and 14$^\circ$ and $i_{outer}$ between zero and $3.1^\circ$.  With a semimajor axis of 2.55 AU, the orbit of the inner giant planet is well-represented by the more eccentric of the known exoplanets, while the outer planet would probably not be currently detectable.

Before the instability the inner planetary system was undergoing standard terrestrial accretion.  Embryos' eccentricities remained damped by dynamical friction from planetesimals and embryos grew by frequent planetesimal impacts and occasional giant embryo impacts.  At the time of the instability the accretion process was relatively mature, as only eight embryos remained inside 2 AU with masses between 0.08 and $0.91 \mearth$, as well as three smaller ($0.07-0.2 \mearth$) embryos in the asteroid belt.  In the immediate aftermath of the instability, ten of the eleven embryos collided with the central star.  The mechanism that drove the embryos into the star was strong eccentricity pumping by a combination of close encounters and secular pumping by the (initially outermost) inward-scattered giant planet.  The one embryo that was not driven into the star was the outermost one, which had a pre-instability semimajor axis of 2.6 AU and was ejected after a series of close encounters with the scattered giant planets.  

The pre-instability outer planetesimal disk was for the most part dynamically calm.  The inner edge of the disk was slowly eroded during this period as the giant planets cleared out planetesimals that were unstable on long timescales.  A few other resonances (e.g., the 3:1 resonance with the outer giant planet at 21.4 AU) also acted to increase the eccentricities of long-term stable planetesimals.  A slow trickle of planetesimals was also destabilized by certain strong mean motion resonances, notably the 2:1 resonance.  When the instability started, the (previously middle) giant planet was scattered out into the planetesimal disk on an eccentric orbit.  Its eccentricity was further pumped by a series of close encounters with the (initially innermost) planet such that for a period of several hundred thousand years (several thousand cometary orbits) the giant planet's orbit completely crossed the initial planetesimal disk.  The outer planetesimal disk was entirely destabilized by secular interactions and close encounters.  These planetesimals' eccentricities increased drastically until they either hit the sun (this occurred about 25\% of the time), were scattered out beyond 100 AU and removed from the simulation by presumed hyperbolic ejection ($\sim75\%$), or collided with a giant planet ($\sim 0.5\%$).  More than 80\% of the outer disk planetesimals were destroyed within 500,000 years and 97\% within 5 Myr.  Only 23 planetesimal particles survived more than 5 Myr after the instability on orbits that were unstable on 10 Myr timescales, typically with high inclinations and eccentricities.  At the end of the simulation a single planetesimal survived, although it is almost certainly unstable on longer timescales as its orbit crosses the outer giant planet's.  

The system's spectral energy distribution (SED) -- shown at right in Figure~\ref{fig:aeit_mixed199} -- reflects its dynamical evolution.  The asteroidal planetesimals are quickly ground to dust, as their collisional timescales are only $10^4-10^5$ years.  This causes a rapid decline in flux at short wavelengths ($\lambda \lesssim 20 \micron$).  The erosion of the inner edge of the outer planetesimal disk causes a continued decrease in flux at shorter wavelengths.  The flux at $\lambda \gtrsim 50 \micron$ is dominated by the mass in outer disk planetesimals and is only very weakly affected by the grinding of asteroids or the slow erosion of the inner edge of the planetesimal disk.  When the instability occurs and the outer planetesimal disk is destabilized, a large number of planetesimals are temporarily placed on high-eccentricity (and therefore small-periastron) orbits which produce hot dust.  This burst in hot dust changes the shape of the SED by increasing the flux at short wavelengths.  This is the cause of the spike in flux seen at $25 \micron$ in the lower right panel of Fig.~\ref{fig:aeit_mixed199}. As noted already, the magnitude of the spike is underestimated because we do not account for the outgassing that occurs when an icy body is heated~\citep[see][]{nesvorny10}.  However, as the outer planetesimal disk is removed the flux drops dramatically.  The dust flux in the 30 Myr after the instability is maintained at a relatively high level by a single planetesimal that survived from 43 to 66.5 Myr between the giant planets with a perihelion that dropped periodically below 3 AU (on a retrograde orbit).  This single close-in planetesimal produced enough dust to keep the system above the $25 \micron$ detection threshold during this period.  We note that our dust production scheme does not allow for the collisional evolution of a single planetesimal so its dust flux is certainly overestimated.  

This example is relatively extreme in terms of the planets' final orbital eccentricities and in that all the terrestrial and cometary particles were destroyed.  However, as we will see below, this simulation allows for a convenient comparison with upcoming examples because the instability is delayed and so the evolution of the dust flux from the quiescent outer disk is unperturbed at early times.  

\begin{figure*}
\includegraphics[width=0.48\textwidth]{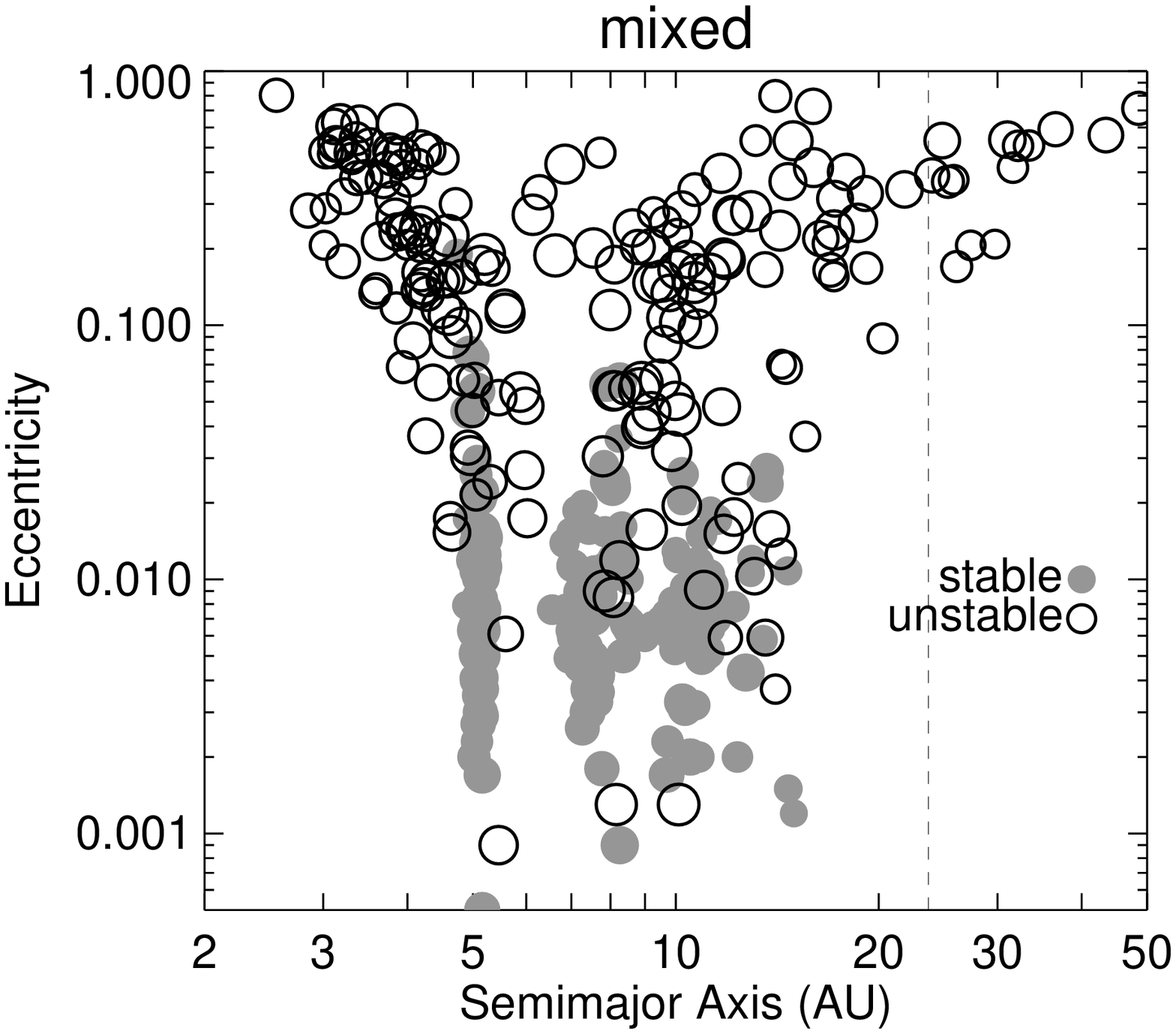}
\hfill
\includegraphics[width=0.48\textwidth]{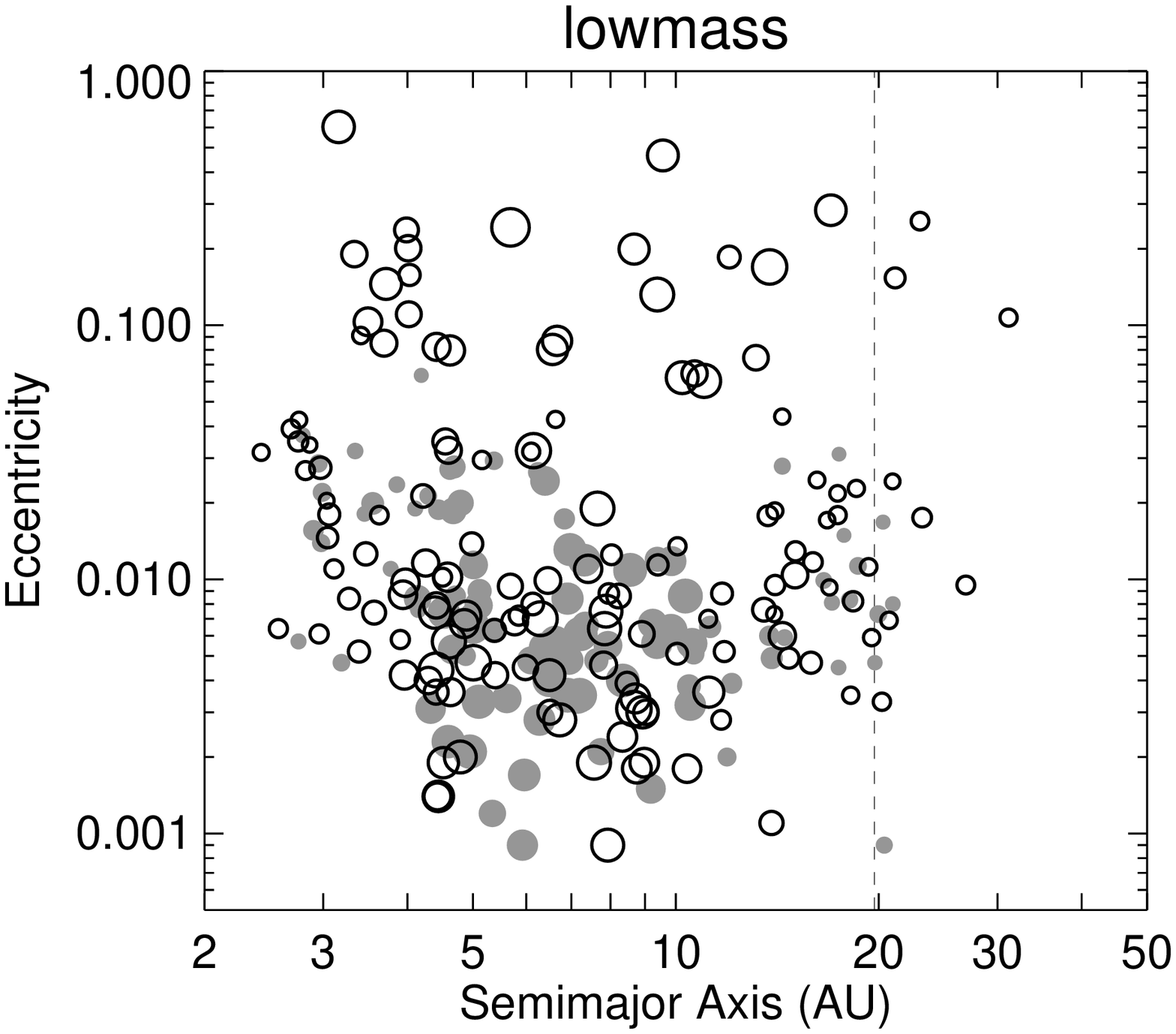}
\caption{Eccentricity vs. semimajor axis for the surviving giant planets in the {\tt mixed} (left panel) and {\tt lowmass} (right) simulations.  Black circles represent unstable simulations -- defined as simulations in which at least one giant planet-planet scattering event occurred -- underwent and grey dots are stable simulations.  The size of each circle is proportional to the logarithm of the planet mass.  The dashed vertical lines represent the median outer edge of the planetesimal disk for each set of simulations.  Note that the outer edge varied from simulation to simulation depending on the giant planet masses, from 20.6 to 27 AU (with a median of 23.7 AU) for the {\tt mixed} simulations and 17.3 to 23.3 AU for the {\tt lowmass} simulations (median of 19.7 AU).}
\label{fig:aegp}
\end{figure*}

\section{Effect of the giant planet masses and mass distribution}
We now analyse two sets of simulations that explore alternate giant planet mass distributions than the {\tt mixed} set analysed in Paper~1. The {\tt mixed} set included three planet systems with the masses of the planets being chosen randomly and independently in the range between a Saturn mass and 3 Jupiter masses. In the {\tt equal} simulations, planet masses within a given system are the same. We consider masses between $30 \mearth$ and $3 M_J$ for different systems.  Assuming that the masses within individual systems are perfectly correlated has the effect of maximizing the strength of dynamical instabilities, compared to systems where there is a range in masses. In the {\tt lowmass} simulations, planet masses are drawn randomly from the observed distribution but only in the range of $10 \mearth$ to $1 M_J$ (in contrast to the range of $M_{Sat}$ to $3 M_J$ for the {\tt mixed} set). The inclusion of the lower mass planets increases the fraction of systems for which dynamical interactions between planets and the planetesimal disk are important.  These interactions can take two forms.  First, ``planetesimal-driven migration'' changes the orbital radius of a planet due to the back-reaction of planetesimals that are gravitationally scattered by the planet, and thus changes a planet's orbital radius while maintaining a small eccentricity~\citep{fernandez84,malhotra93,hahn99,gomes04,kirsh09,levison10}.  Second, the orbit of an eccentric planet can be re-circularized by ``secular friction'', a process by which an eccentric planet excites the eccentricities of the outer disk planetesimals and causes a corresponding decrease in the planet's eccentricity~\citep{thommes99,levison08}. A low-mass planet can therefore be gravitationally scattered by another planet in the inner part of a planetary system and have its eccentricity decreased on a much wider orbit by secular friction with the outer planetesimal disk~\citep[e.g.,][]{thommes99,raymond10}.

The {\tt low mass} and {\tt equal} sets of simulations are of particular interest because a combination of the two can produce an alternate sample that matches the observed exoplanet distribution.  We explore the two sets of simulations independently (sections 3.1 and 3.2) and later combine them into a sample to compare with exoplanet statistics (called case B in section 6).  

\subsection{Systems with low-mass giant planets (the {\tt lowmass} simulations)}
Figure~\ref{fig:aegp} shows that, because of planetesimal-driven migration and secular friction, surviving low-mass giant planets populate different regions of parameter space than more massive giant planets.  Given that all of our simulated giant planets started in the 5-15 AU region, massive giant planets are only able to alter their semimajor axes by energy exchange during close encounters to essentially follow curves with perihelia or aphelia at the encounter distance.  In other words, high-mass planets at large $a$ necessarily have large $e$.  However, low-mass planets can have large $a$ and small $e$ by either 1) planetesimal-driven migration, which maintains planets' low $e$ out to large $a$ (to the outer edge of the planetesimal disk), or 2) being scattered outward in a dynamical instability but having their eccentricities damped by secular friction with the outer planetesimal disk.  Similarly, massive planets that are scattered interior to 5.2 AU necessarily have large $e$ but low-mass inner giant planets can undergo inward changes in $a$ and survive on low-$e$ orbits. Indeed many low-mass planets do just that, ending up at $a = 2.5-4$ AU.  Note that secular friction is only relevant in unstable systems in which a planet acquires a large orbital eccentricity.  On the other hand, planetesimal-driven migration is mainly relevant for stable systems although periods of migration may occur in some cases {\em after} secular friction has already re-circularized the orbit of a scattered low-mass giant planet.

Thus, the surviving high-mass giant planets retain a memory of their initial conditions: the stable planets and many unstable planets are clustered at their original locations.  However, given the ease and inevitability of planetesimal-driven migration and secular friction for low-mass giant planets, the initial conditions are erased.  

\begin{figure*}
\includegraphics[width=0.63\textwidth]{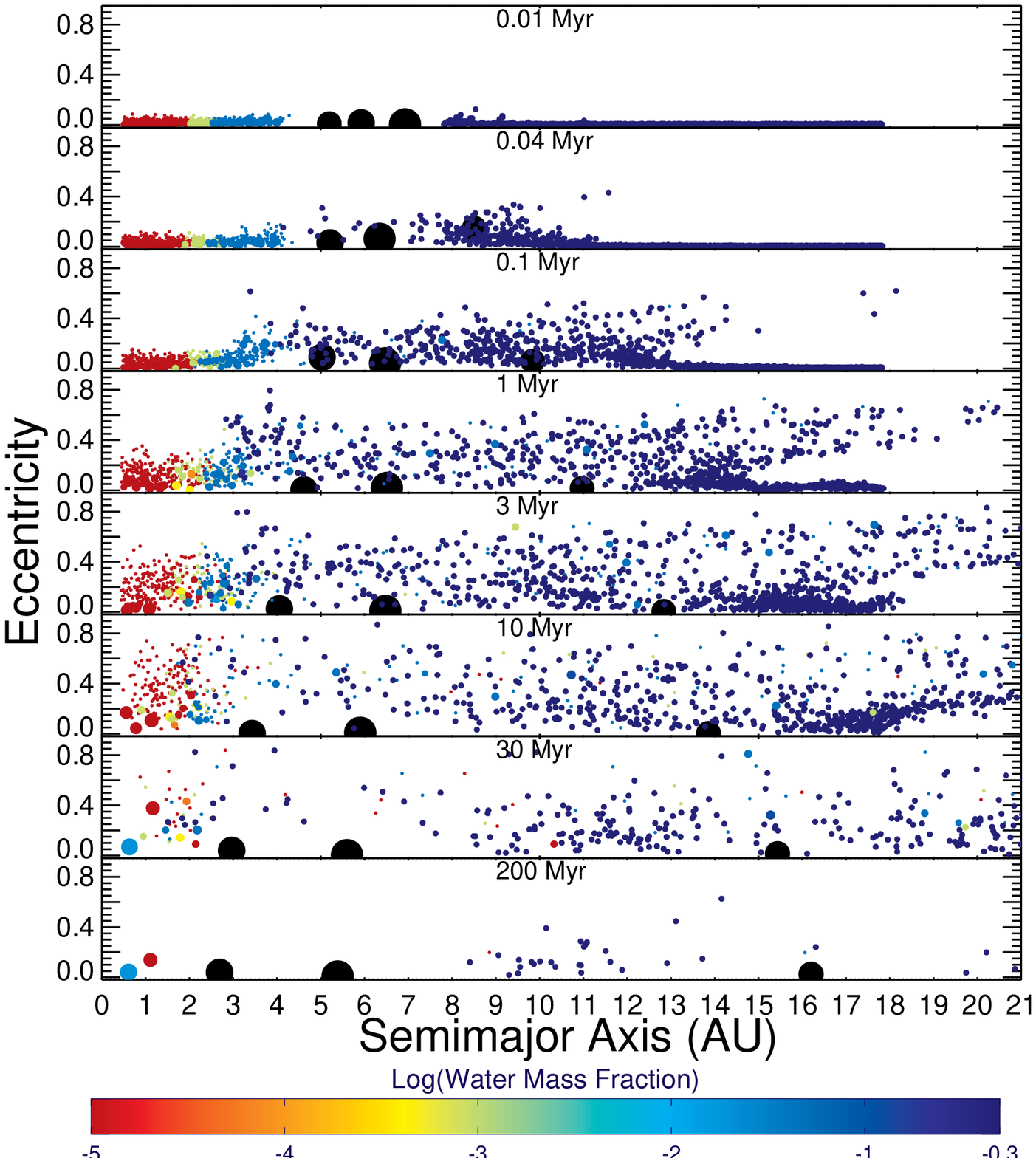}
\vskip -5in
\hspace {0.63\textwidth}
\includegraphics[width=0.37\textwidth]{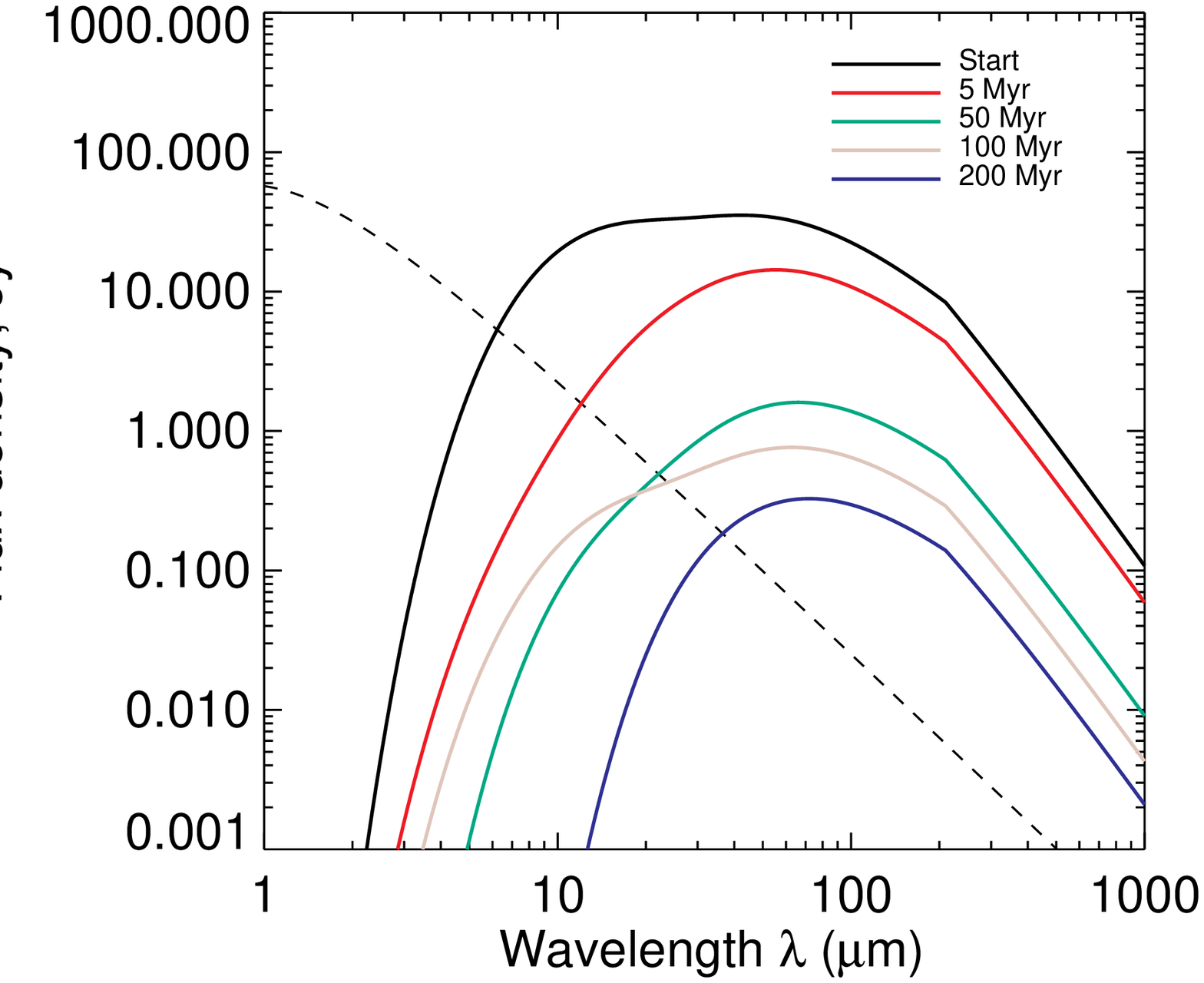}
\vskip 0.01in 
\hspace {0.63\textwidth} 
\includegraphics[width=0.37\textwidth]{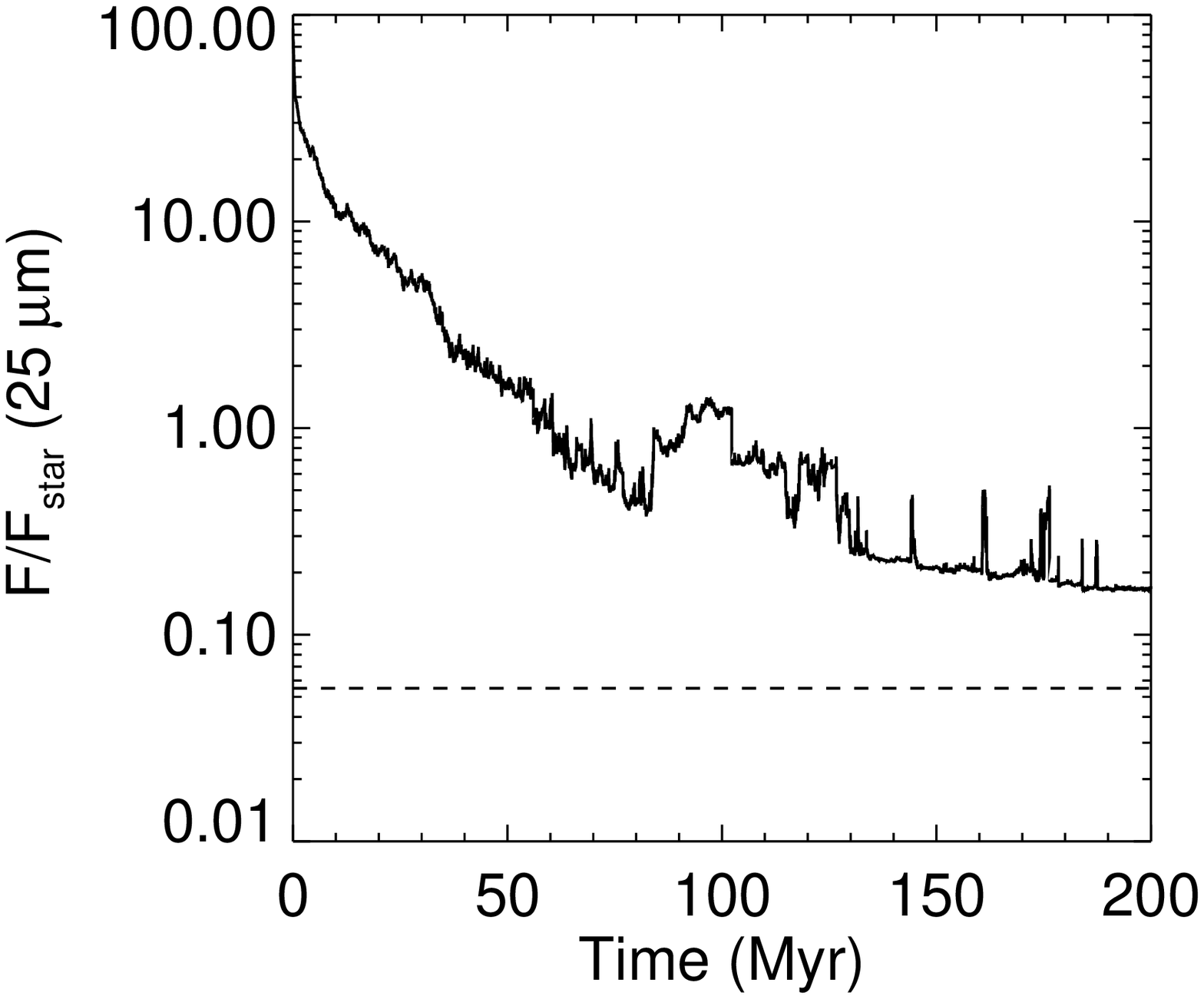}
\vskip 0.3in 
\caption{Evolution of a simulation with low-mass giant planets.  {\bf Left:} Orbital eccentricity vs. semimajor axis of each body in the simulation.  The size scales with the mass$^{1/3}$ and the color corresponds to the water content, using initial values taken to Solar System data~\citep{raymond04} and re-calculated during impacts by mass balance.  The giant planets are shown as the large black bodies and are not on the same size scale.  {\bf Top right:} The spectral energy distribution of the dust during five simulation snapshots.  The dashed line represents the stellar photosphere.  {\bf Bottom right:} The ratio of the dust-to-stellar flux at 25 microns as a function of time.  The rough observational limit of the {\it MIPS} instrument on NASA's {\it Spitzer Space Telescope} is shown with the dashed line~\citep{trilling08}. A movie of this simulation is available at http://www.obs.u-bordeaux1.fr/e3arths/raymond/scatter\_lowmass\_54.mpg.}
\label{fig:aeit_lowmass}
\end{figure*}

Figure~\ref{fig:aeit_lowmass} shows the evolution of a {\tt lowmass} simulation with initial giant planet masses of
12.4 $\mearth$ (inner), 18.6 $\mearth$ (middle), and 35.9 $\mearth$ (outer).  In this simulation the giant planets
underwent an instability after 33,000 years, which threw the inner planet into the outer planetesimal disk.  Once
interacting with the planetesimal disk, the planet scattered planetesimals inward and migrated outward for roughly
20 Myr, then slowed when it came within $\sim$3.5 Hill radii of the outer edge of the planetesimal disk.  As this
represents the approximate boundary for dynamical stability~\citep{marchal82,gladman93,chambers96}, the number of planetesimals
available to be scattered by the planet decreased and the planet's migration slowed drastically.  The planetesimals
that were scattered inward from the outer disk were for the most part subsequently scattered outward by the inner
giant planets, causing the two inner planets to migrate inward.  However, some of the inward-scattered
planetesimals were trapped on low-eccentricity orbits between the two inner giant planets and the outer giant as the
outer planet migrated outward by continued planetesimal scattering. This is similar to the mechanism that
may have been responsible for populating the Solar System's asteroid belt during the {\em outward} migration of
Jupiter and Saturn~\citep{walsh11}. The inner giant planets' inward migration was fueled by the planetesimals that
ended up with high-eccentricity, low-perihelion orbits after being scattered inward, although the inner giants also
scattered some embryos and planetesimals from the inner disk.  The inner giant planet migrated in to 2.69 AU but
maintained an eccentricity lower than 0.1 throughout and less than 0.05 during the last phases.

Two terrestrial planets formed in the simulation shown in Figure~\ref{fig:aeit_lowmass}: a 1.25 $\mearth$ planet at 0.61 AU and a 0.69 $\mearth$ planet at 1.11 AU.  The eccentricities of the inner and outer planet are 0.06 and 0.11, respectively, with peak to peak oscillation amplitudes of 0.11 and 0.20.  The inner planet underwent its last giant (embryo) impact after 40.6 Myr but the outer planet did not undergo any giant impacts after 3.8 Myr.  The inner planet is wet: it accreted a small amount of water from material that originated in the inner asteroid belt (an embryo and two planetesimals from $\sim $ 2 AU) but the bulk of its water came from a single cometary impact.  The outer planet did not accrete any material from beyond 2 AU and so is considered to be dry~\citep[see ][]{raymond04,raymond07a}.  The closest giant planet's semimajor axis is only 1.6 AU larger than the outer terrestrial planet's but given the giant planet's small mass ($19.2 \mearth$) and modest eccentricity (0.035 with oscillations of 0.02 in full amplitude) the system is stable.  

In the simulation from Fig.~\ref{fig:aeit_lowmass}, only a relatively small fraction of the initial terrestrial mass was incorporated into the two surviving terrestrial planets.  The majority of the initial terrestrial mass (57\%) was ejected from the system and an additional 10\% collided with the central star.  This contrasts with the unstable systems with higher-mass giant planets, in which terrestrial material preferentially collides with the star (as in the simulation from Fig.~\ref{fig:aeit_mixed199}).  This difference is due to the fact that more massive giant planets pump the eccentricities of terrestrial bodies efficiently and can thus drive down their perihelion distances on short timescales.  Low-mass giant planets require longer to excite high eccentricities and also migrate in reaction to scattering such bodies, inward in this scenario.  Thus, in systems with unstable low-mass giant planets terrestrial material is more easily transported outward than inward, to be ejected after many encounters with the outer giant planets.  

In addition, in the simulation from Fig.~\ref{fig:aeit_lowmass} two Mars-sized embryos were scattered out and survived on distant orbits, at 29.2 and 34.8 AU, and one embryo collided with the middle giant planet. 

At the end of the simulation there are two surviving planetesimal belts: a low-eccentricity belt between the two inner and the outer giant planets and an outer disk of higher-eccentricity objects exterior to the outer giant planet.  The outer belt is analogous to the Solar System's scattered disk~\citep{luu97,duncan97}, having been scattered by the outward-migrating giant planet.  The scattered belt contains 3.7 $\mearth$ in 72 particles with a median eccentricity of 0.27 and a median inclination of 11.5$\deg$.  This scattered disk also contains two embryos from the inner disk, including one that originated inside 2 AU.  The inner belt of planetesimals -- located between roughly 8 and 14 AU -- contains 1.3 $\mearth$ in 27 planetesimals with a median eccentricity of 0.12 and a median inclination of 16.8$\deg$.  The orbital distributions and surface densities of these two populations are quite different, and we suspect that a wide diversity of planetesimal belt structures must exist around other stars. 


The evolution of the system's dust brightness is shown in Figure~\ref{fig:aeit_lowmass}.  The SED of the system decreases systematically as the system loses mass, but changes shape after roughly 80 Myr when four separate icy planetesimals entered the very inner planetary system and remained on orbits interior to the innermost giant planet (with perihelion distances as small as 0.3 AU) for several tens of Myr before being ejected.\footnote{Note that the plateau in brightness seen in the $25 \micron$ plot of Fig.~\ref{fig:aeit_lowmass} may be slightly overestimated because our method does not adequately account for collisional grinding of bodies on extremely close-in orbits when they are isolated particles.  In this case, these four isolated particles were at close enough distances to dominate the flux at wavelengths shorter than $\sim 50 \micron$ from about 80-120 Myr.}  At wavelengths longer than $\sim 50 \micron$, the dust brightness decreased monotonically in time.  However, shorter wavelengths (such as $25 \micron$; Fig.~\ref{fig:aeit_lowmass}) show the additional structure caused by the icy planetesimals entering the inner planetary system because they are sensitive to hot dust.

\begin{figure}
\center\includegraphics[width=0.45\textwidth]{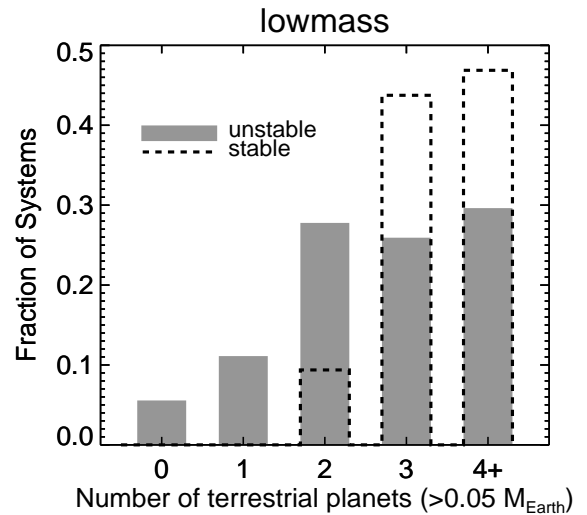}
\caption{Distribution of the number of surviving terrestrial planets for the unstable (grey) and stable (dashed) simulations in the {\tt lowmass} set of simulations. 
}
\label{fig:hist_nterr_lowmass}
\end{figure}

As a whole, the {\tt lowmass} simulations were extremely efficient at forming terrestrial planets and also at creating long-lasting debris disks.  Out of the 86 total simulations, 82 (95.3\%) formed terrestrial planet systems containing a total of at least 0.5 $\mearth$.  Of the four remaining systems, three destroyed their terrestrial planets entirely and the fourth formed a single planet of $0.25 \mearth$.  Of the 86 {\tt lowmass} simulations, 76 (88.4\%) were above the Spitzer detection threshold at $70 \micron$ after 1 Gyr (73 remained bright after 3 Gyr; recall that this is far higher than the observed frequency of 16.4\%).  Of the 54 (62.7\%) unstable simulations, only 4 did not yield at least 0.5 $\mearth$ in terrestrial planets and only 10 were not detectable at $70 \micron$ after 1 Gyr.  The few systems with destructive instabilities were those that by chance contained several massive planets, and so essentially overlapped with the {\tt mixed} distribution.  Figure~\ref{fig:hist_nterr_lowmass} shows that all 32 stable simulations finished with two or more terrestrial planets and also with bright debris disks.  Note that in this figure we use a low mass cutoff of just $0.05 \mearth$ in our definition of a terrestrial planet.  Any surviving planetary embryo is therefore considered a planet.  This allows for a consistent comparison with other sets of simulations including more violent instabilities (like the {\tt equal} simulations) in which surviving embryos are common.  In the Solar System, Mars is thought to be a surviving embryo~\citep{dauphas11}.

Among systems with planets less massive than 50-100 $\mearth$ ($\approx 0.5-1 M_{Sat}$), there was little difference in the final outcome between systems that underwent planet-planet scattering and those that did not.  This is because the planetesimal disk provides strong enough damping to quickly decrease the planets' eccentricities back to near zero.  Systems containing a single relatively massive giant planet ($M \gtrsim M_{Sat}$) also ended in a dynamically calm state because instabilities caused the lower-mass giant planets to be scattered and, again, their eccentricities and inclinations are quickly damped.  The only situation that preserved large eccentricities was the relatively infrequent combination of multiple massive giant planets in the same system.  In those systems the large eccentricity caused by strong scattering between giant planets could not be damped (the low-mass giant planets in such systems are usually scattered, sometimes to be ejected or sometimes re-circularized in the outer planetesimal disk).  Thus, unlike massive giant planets, the eccentricities of low-mass planets do not retain a memory of the system's dynamical history.  In addition, multiple giant planets must exist in the same system to yield eccentric giant planets.  The abundance of observed eccentric planets~\citep[e.g.][]{wright09} thus points to the frequency of strong instabilities~\citep[although alternate models exist; see][for a thorough review]{ford08}.  

\begin{figure*}
\includegraphics[width=0.48\textwidth]{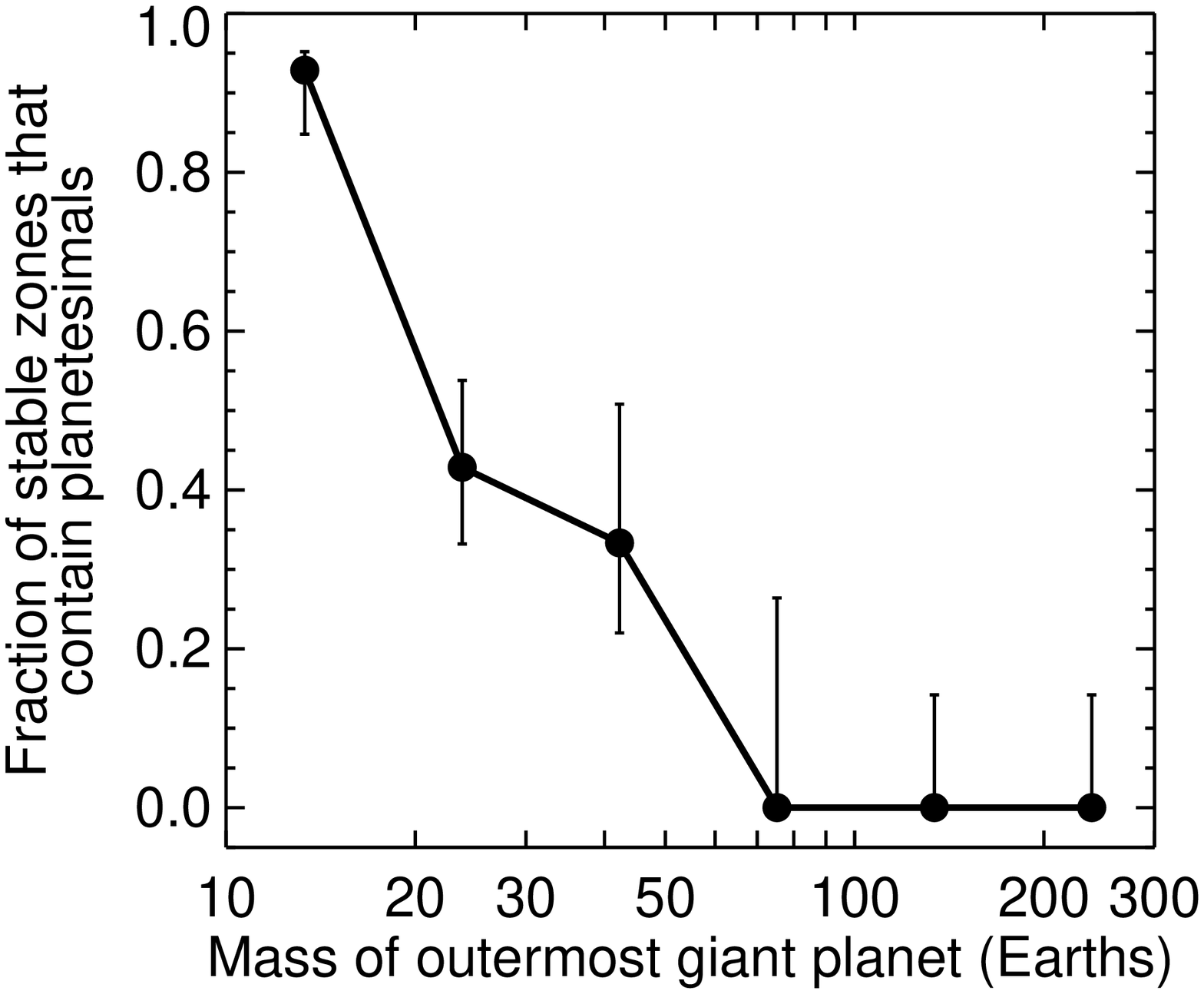}
\hfill
\includegraphics[width=0.48\textwidth]{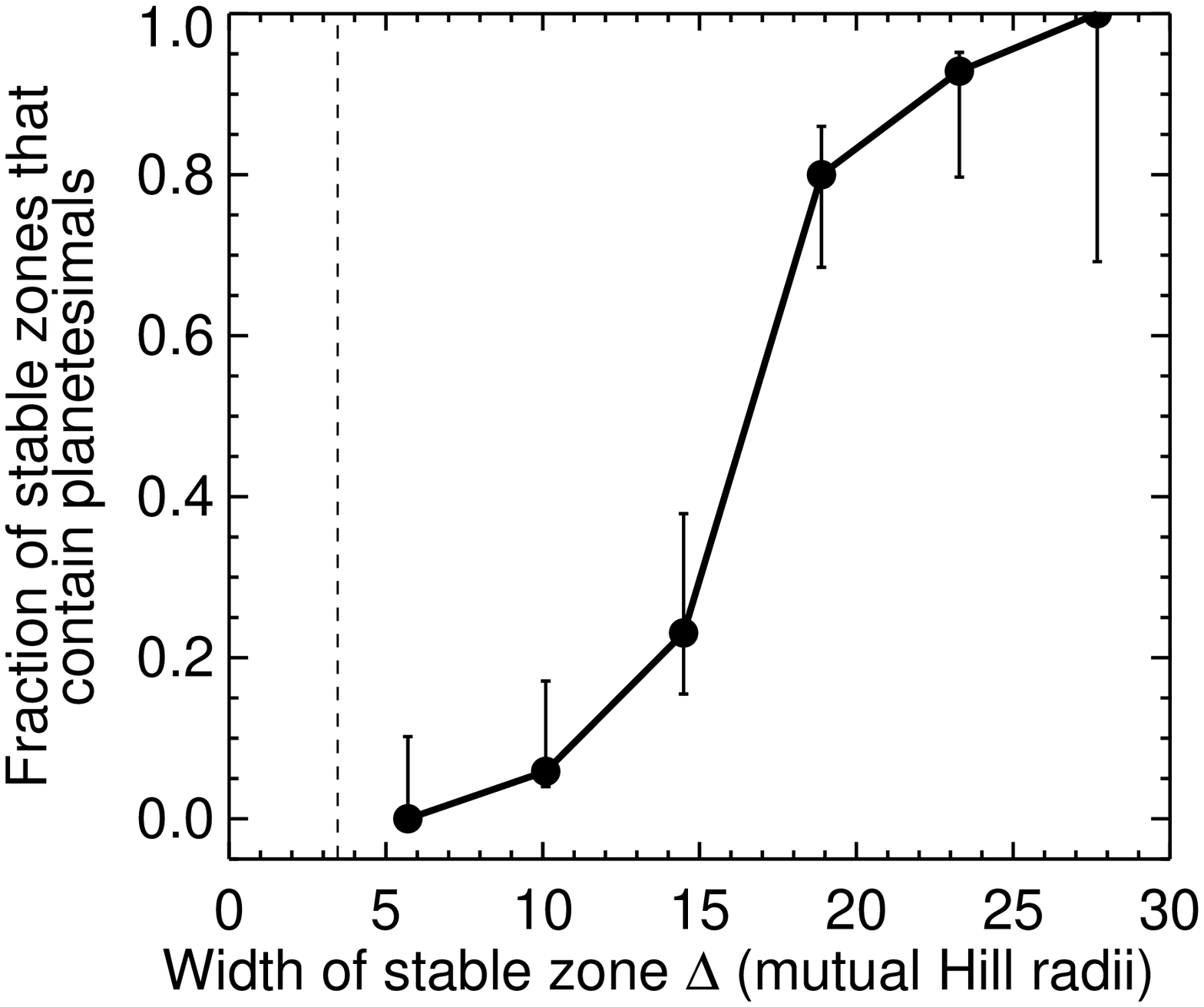}
\caption{In the {\tt lowmass} planetary systems, the fraction of stable zones between pairs of outer giant planets that contain at least one planetesimal on a stable orbit as a function of the mass of the outermost giant planet (left panel) and the width of the stable zone $\Delta$ in units of mutual Hill radii (right panel).  For the left panel, bins were evenly logarithmically spaced from $10 \mearth$ to $1 M_J$.  For the right panel, bins were evenly spaced from $\Delta = 3.5$ to 30.  The error bars were calculated using binomial statistics following~\cite{burgasser03}.  The dashed line marks the two planet stability boundary~\citep{marchal82,gladman93}.}
\label{fig:plgaps}
\end{figure*}

Most of the {\tt lowmass} systems have large gaps between giant planets, and many of these gaps contain planetesimals on stable orbits, such as the belt between 8-14 AU in the system from Fig.~\ref{fig:aeit_lowmass}.  The existence of isolated belts alters the radial distribution of dust and can be inferred from the spectral energy distribution.  There are several systems known to contain belts of dust that appear to be radially confined, presumably by known or as-yet undetected planets~\citep[e.g.,][]{beichman05,lisse08,su09}.   

The majority of the {\tt lowmass} systems (60/85 = 71\%) have gaps between the outer two giant planets with
separations $\Delta$ of 10 or more mutual Hill radii (in one system just a single giant planet survived and so is
not counted).  Given that two planets must be separated by at least $\Delta > 2 \sqrt{3} \approx 3.5$ mutual Hill
radii for long-term dynamical stability~\citep{marchal82,gladman93},  the existence of a zone between two planets' orbits that is stable for planetesimals requires at a minimum an 
interplanetary separation $\Delta \gtrsim 10$.  In practice, somewhat wider gaps are needed in realistic systems. The separation 
between Saturn and Uranus, and between Uranus and Neptune, amounts to 14 mutual Hill radii, and there is 
only a small region that is stable over long timescales between the latter two planets \citep{holman93}. 
In our runs we frequently find gaps that are not only wider than their Solar System counterparts, 
but which are also populated with primordial material despite the relatively coarse sampling of the 
outer planetesimal disk population.
The {\tt lowmass} simulations yield a range of separations from $\Delta = 3.9$ to 30 and outer
giant planet masses of $11 \mearth$ to $0.97 M_J$.  From the entire sample, the stable zone between the outer two
giant planets contained at least one particle in half of all {\tt lowmass} simulations (38/85 = 45\%, although 5 stable
zones contained just a single planetesimal).  The closest separation between two planets for which planetesimals
existed on stable orbits between the two was $\Delta = 11.9$, and the widest separation for which no planetesimals
existed between two planets was $\Delta = 21.5$.  The total mass in these planetesimal belts ranged from the mass of
one planetesimal particle ($0.05 \mearth$) to $5.9 \mearth$, and in one case an embryo from the inner disk survived
in such a belt.  

The two most important factors that determine whether a stable zone between outer giant planets contains a planetesimal belt are first, the width of the stable zone and second, the mass of the outermost giant planet.  Figure~\ref{fig:plgaps} shows that the probability that a stable zone contains at least one planetesimal particle on an orbit that is stable for long timescales ( i.e., whose orbit does not come within 4 Hill radii of any giant planet's orbit) as a function of the mass of the outermost giant planet and the width of the stable zones (as quantified by $\Delta$).  Stable zones are preferentially filled for larger $\Delta$ values simply because there is a larger region of parameter space into which planetesimals can be scattered and survive, and the fraction of stable zones that is filled increases dramatically for $\Delta \gtrsim 15$.  Stable zones are also preferentially filled in systems with outermost giant planets less massive than $\sim 50 \mearth$, and almost 100\% of stable zones are filled when the outermost giant planet is less than about $20 \mearth$ (FIg.~\ref{fig:plgaps}).  As they interact with the outer planetesimal disk, lower-mass planets scatter planetesimals onto lower eccentricities than do higher-mass planets, and these planetesimals are more likely to avoid encountering giant planets and to remain on stable orbits than if their obits are more eccentric.  

Do other system parameters influence the probability that a stable zone between giant planets will contain planetesimals?  We tested the importance of several other parameters using a suite of Kolmogorov-Smirnov (K-S) tests that compared different characteristics of stable zones with and without planetesimals.  Using a cutoff of $p < 0.01$ for a statistically significant difference between the two populations, we found that five parameters influence whether a stable zone will contain planetesimals.  In order of most important (lowest $p$ value) to least important (highest $p$ value), these are 1) the width of the stable zone ($\Delta$), 2) the mass of the outermost giant planet, 3) the semimajor axis of the outermost giant planet, 4) the mass ratio of the outer two giant planets (systems with a large inner/outer mass ratio preferentially contain planetesimal belts), and 5) the total mass in the surviving giant planets (the probability of stable zones being empty increases for higher total mass).  The effect of these five parameters is statistically significant, although several are correlated (e.g., $\Delta$ correlates with the semimajor axis of the outermost giant planet, although $\Delta$ is about three orders of magnitude more important in determining whether a stable zone will be filled).  We tested five additional parameters whose effect turned out to be unimportant ($p > 0.1$ in each case): the mass and semimajor axis of the giant planet marking the inner boundary of the stable zone, the eccentricity of the outermost giant planet, and the mass-weighted eccentricities of all surviving giant planets.  

\subsection{Systems with equal-mass giant planets (the {\tt equal} simulations)}
While the {\tt lowmass} simulations represent a calm environment conducive to the production of both terrestrial planets and bright debris disks, the {\tt equal} simulations were destructive on both counts.  This is simply because scattering among equal-mass giant planets is the most violent planet-planet instability~\citep{raymond10}, and strong instabilities destroy small bodies in both the inner and outer disks, typically by driving a large fraction of inner bodies into the central star and ejecting the majority of outer bodies.  Indeed, the eccentricity distribution of the surviving {\tt equal} giant planets with masses of $M_{Sat}$ or larger was skewed toward higher values than the {\tt mixed} sample, reaching values as high as 0.89 and with a median of 0.35 (compared with 0.21 for the {\tt mixed} giant planets).  An exception to this rule are systems with equal-mass giant planets that are themselves low-mass.  In systems containing three giant planets of 30 $\mearth$, the outcome was similar to the {\tt lowmass} systems because planet-disk interactions trumped planet-planet scattering.  Indeed, the low-mass {\tt equal} systems were dominated by planetesimal-driven migration of the outer giant planet and behaved very similarly to the {\tt lowmass} simulations.  Given this strong dichotomy, we now consider just the high-mass {\tt equal} simulations, as the lower-mass cases are more appropriately included with the {\tt lowmass} set.

\begin{figure}
\center\includegraphics[width=0.45\textwidth]{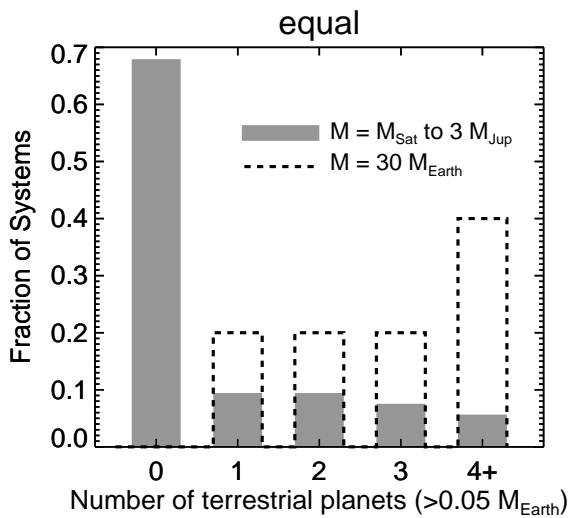}
\caption{Distribution of the number of surviving terrestrial planets for the {\tt equal} set of simulations.  Systems with giant planets of $M_{Sat}$ or larger are shown in grey, and systems with giant planets of $30 \mearth$ are shown by the black dashed line.
}
\label{fig:hist-nterr_equal}
\end{figure}

Figure~\ref{fig:hist-nterr_equal} shows that, of the 53 simulations with giant planet masses of $M_{Sat}$, $M_J$, or $3 M_J$, 36 (68\%) destroyed all terrestrial material.  Five simulations (9\%) formed a single terrestrial planet, although these were all small, roughly Mars-mass $\sim 0.1 \mearth$ planets, and 4/5 of these were lone surviving embryos.  The remaining 12 simulations with high-mass giant planets (21\%) formed two or more terrestrial planets.  In contrast, not a single simulations with a $30 \mearth$ giant planet destroyed all of its terrestrial material.  Rather, these systems usually formed several terrestrial planets -- only 3/15 systems formed just one.

Of the 53 {\tt equal} systems, only 17 (32\%) contained detectable amounts of cold dust at $70 \micron$ after 1 Gyr (only 14/53 = 26\% after 3 Gyr, although again note that this is higher than the observed frequency of 16.4\%).  Of the 17 systems with detectable dust at 1 Gyr, 11 (65\%) formed terrestrial planets.  Of the 36 systems with no detectable dust, only 6 (17\%) formed terrestrial planets.  Thus,  there exists a natural connection between debris disks and terrestrial planets that spans the domain of giant planet mass.  This same debris disk-terrestrial planet correlation was found in Paper~1 for the {\tt mixed} simulations.  The details of this correlation depend on the system parameters and are of course confined to the context of our initial conditions, in particular to systems with relatively massive outer planetesimal disks (see discussion in Section 6). 

The {\tt equal} systems are violently unstable by construction and so the outcomes tend to be extreme.  This is precisely the type of behavior that is required by exoplanet observations, in particular the trend for more massive planets to have more eccentric orbits than lower-mass planets~\citep{jones06,ribas07,wright09,raymond10}.  If planet masses within individual planetary systems were random (as in the {\tt mixed} simulations), then lower-mass giant planets should have higher eccentricities than higher-mass giants~\citep{raymond10}; this is the opposite of what is observed.  Thus, the {\tt equal} simulations provide a key ingredient in constructing a sample of simulations that matches the observed giant exoplanets, as discussed in Section 6.

\section{Effect of the properties of outer planetesimal disks}
We now turn our attention to the effect of the properties of outer planetesimal disks.  We first examine the {\tt seeds} simulations -- subdivided into the {\tt bigseed} and {\tt smallseed} sets -- that contained a population of $\sim$Earth-mass embryos in their outer planetesimal disks.   We then test the effects of doubling the width (and total mass) of the planetesimal disk in the {\tt widedisk} simulations.  

\subsection{The mass distribution of the planetesimal disk (the {\tt seeds} simulations)}
In the {\tt seeds} simulations a small number of fully-interacting massive bodies were included in the outer planetesimal disk (in contrast to planetesimal particles, which interact gravitationally with massive bodies but not with each other).  The disk maintained the same total mass and numerical resolution (the masses of individual planetesimals were decreased to maintain a constant total mass).  In 50 simulations comprising the {\tt bigseed} set, five icy embryos of $2 \mearth$ each were included at 11, 13, 15, 17 and 19 AU.  In 50 additional {\tt smallseed} simulations 10 seeds of $0.5 \mearth$ each were spaced with 1 AU of separation from 10.5 to 19.5 AU.  These seed masses are broadly consistent with calculations of accretion in outer planetesimal disks~\citep{kenyon08,kenyon10}.  

\begin{figure*}
\includegraphics[width=0.63\textwidth]{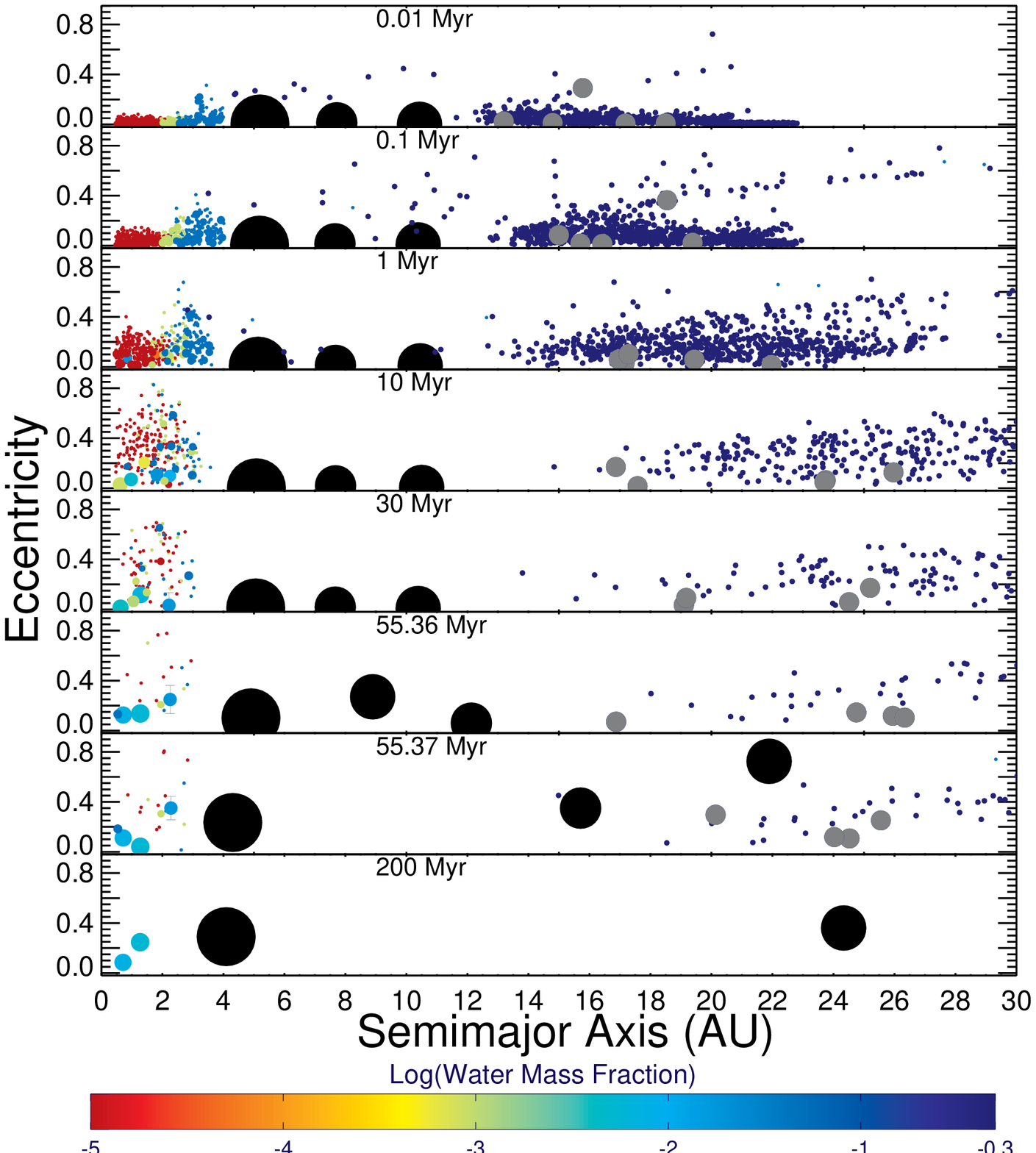}
\vskip -5in
\hspace {0.63\textwidth}
\includegraphics[width=0.37\textwidth]{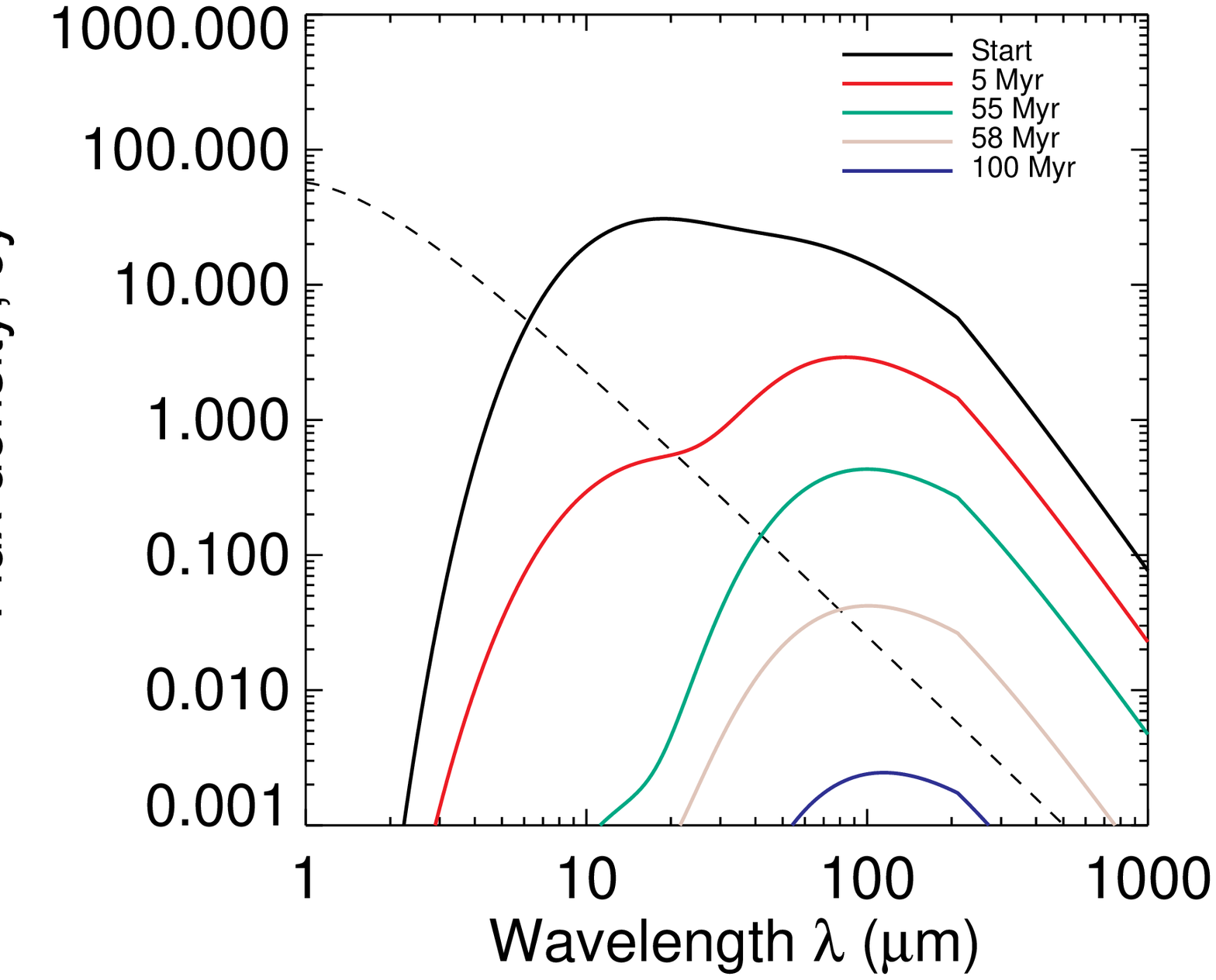}
\vskip 0.01in 
\hspace {0.63\textwidth} 
\includegraphics[width=0.37\textwidth]{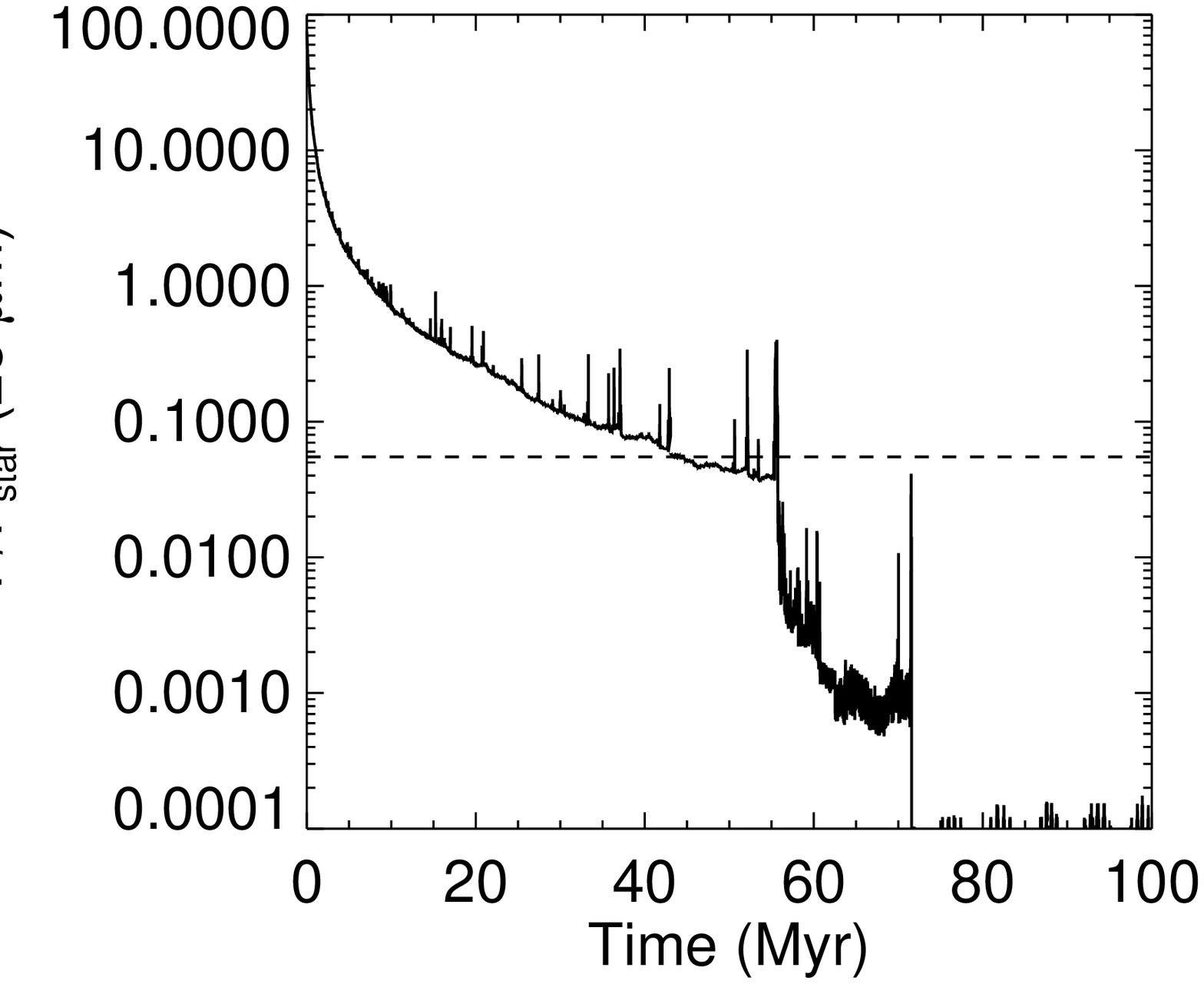}
\vskip 0.3in 
\caption{Evolution of a simulation that includes 5 icy embryos (shown in grey) in the outer planetesimal disk.  {\bf Left:} Orbital eccentricity vs. semimajor axis of each body in the simulation.  The size scales with the mass$^{1/3}$ and the color corresponds to the water content, using initial values taken to Solar System data~\citep{raymond04} and re-calculated during impacts by mass balance.  The giant planets are shown as the large black bodies and are not on the same size scale. The two surviving terrestrial planets have $a$ = 0.71 and 1.27 AU, $m$ = 1.3 and 1.58 $\mearth$, and Myr-averaged $e$ = 0.09 and 0.17, respectively.  Both terrestrials have substantial water contents accreted from hydrated asteroidal material. The two surviving giant planets have $a =$4.1 and 24.4 AU, $m =$1.27 and 0.45 $M_J$, and Myr-averaged $e =$0.32 and 0.33, respectively.  {\bf Top right:} The spectral energy distribution of the dust during five simulation snapshots.  The dashed line represents the stellar photosphere.  {\bf Bottom right:} The ratio of the dust-to-stellar flux $F/F_{star}$ at $25 \micron$ as a function of time.  The rough {\it Spitzer} observational limit is shown with the dashed line~\citep{trilling08}. An animation of this simulation is available at http://www.obs.u-bordeaux1.fr/e3arths/raymond/scatterSED\_seed13.mpg.}
\label{fig:aeit_seed}
\end{figure*}

Figure~\ref{fig:aeit_seed} shows the evolution of one {\tt bigseed} simulation that became unstable after a long delay and therefore allows for a comparison with the evolution of both stable (prior to the instability) and unstable systems.  The evolution is qualitatively similar to the {\tt mixed} simulations in the inner disk such as the system from Fig.~\ref{fig:aeit_mixed199}, as embryos maintain smaller eccentricities than planetesimals via dynamical friction and accrete as their feeding zones begin to overlap.  The instability started after 55.35 Myr with a close encounter between the two outer giant planets.  After a series of close encounters lasting 40,000 years, the middle giant planet was ejected.  At the end of the simulation, the system contains two giant planets each with $e \approx 0.3$ and two roughly Earth-sized terrestrial terrestrial planets (details in caption of Fig.~\ref{fig:aeit_seed}).

The evolution of the outer disk differs significantly from simulations without seeds.  The presence of the seeds introduces an effective viscosity into the outer planetesimal disk that is driven by scattering of planetesimals by close encounters with the icy embryos.  This causes the disk to spread out radially.  Within 1 Myr the outer edge of the low-eccentricity portion of the disk has expanded from 22 to 26 AU, and beyond 30 AU in the next few Myr (Fig.~\ref{fig:aeit_seed}).  This outward expansion is balanced by the loss of a large portion of the total disk mass that is scattered inward to encounter the giant planets and be ejected from the system.  The planetesimal disk is further depleted on a 20-50 Myr timescale by instabilities in the system of icy embryos that increases their eccentricities -- these instabilities are analogous to but far weaker than instabilities between the giant planets (and are weaker still in the {\tt smallseeds} systems).  When the giant planets go unstable the bulk of the planetesimal disk is rapidly ejected and the last icy planetesimal is removed just after 100 Myr.  

The dust production in the {\tt seeds} simulations is also different than in simulations with calmer planetesimal disks.  The planetesimal population closest to the giant planets -- the inner several AU of the outer planetesimal disk -- is depleted in a few Myr as the disk ``viscously'' spreads.  Thus, warm dust has a far shorter lifetime than in simulations with no seeds.  This is reflected in the evolution of the $25 \micron$ flux in Fig.~\ref{fig:aeit_seed}, that drops below the detection threshold after $\sim$ 45 Myr while the system is still stable.  In contrast, in the example {\tt mixed} (Fig.~\ref{fig:aeit_mixed199}) and {\tt lowmass} (Fig.~\ref{fig:aeit_lowmass}) simulations, the $25 \micron$ flux was more than an order of magnitude higher after 40-50 Myr of evolution.  

\begin{figure*}
\includegraphics[width=0.48\textwidth]{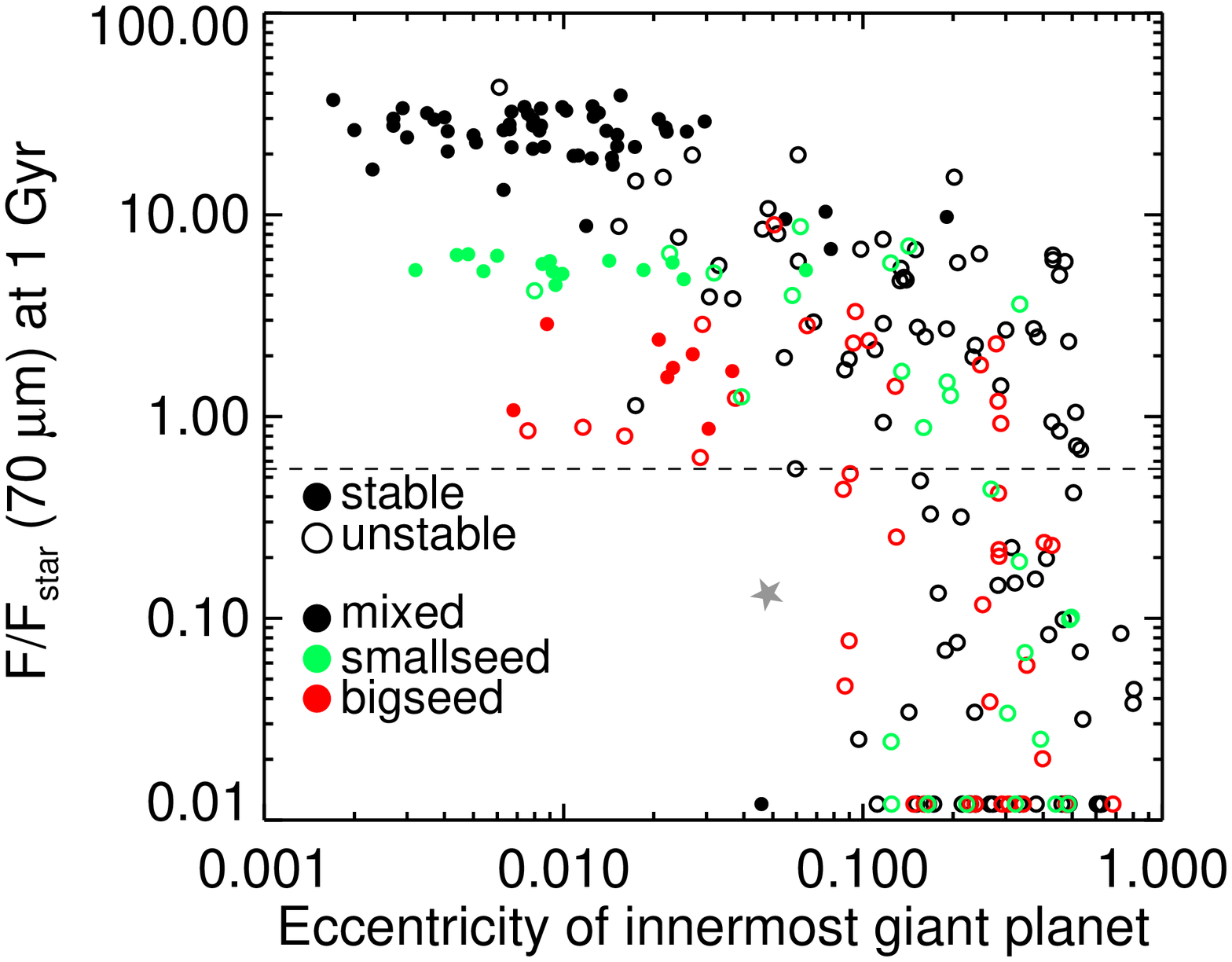}
\hfill
\includegraphics[width=0.48\textwidth]{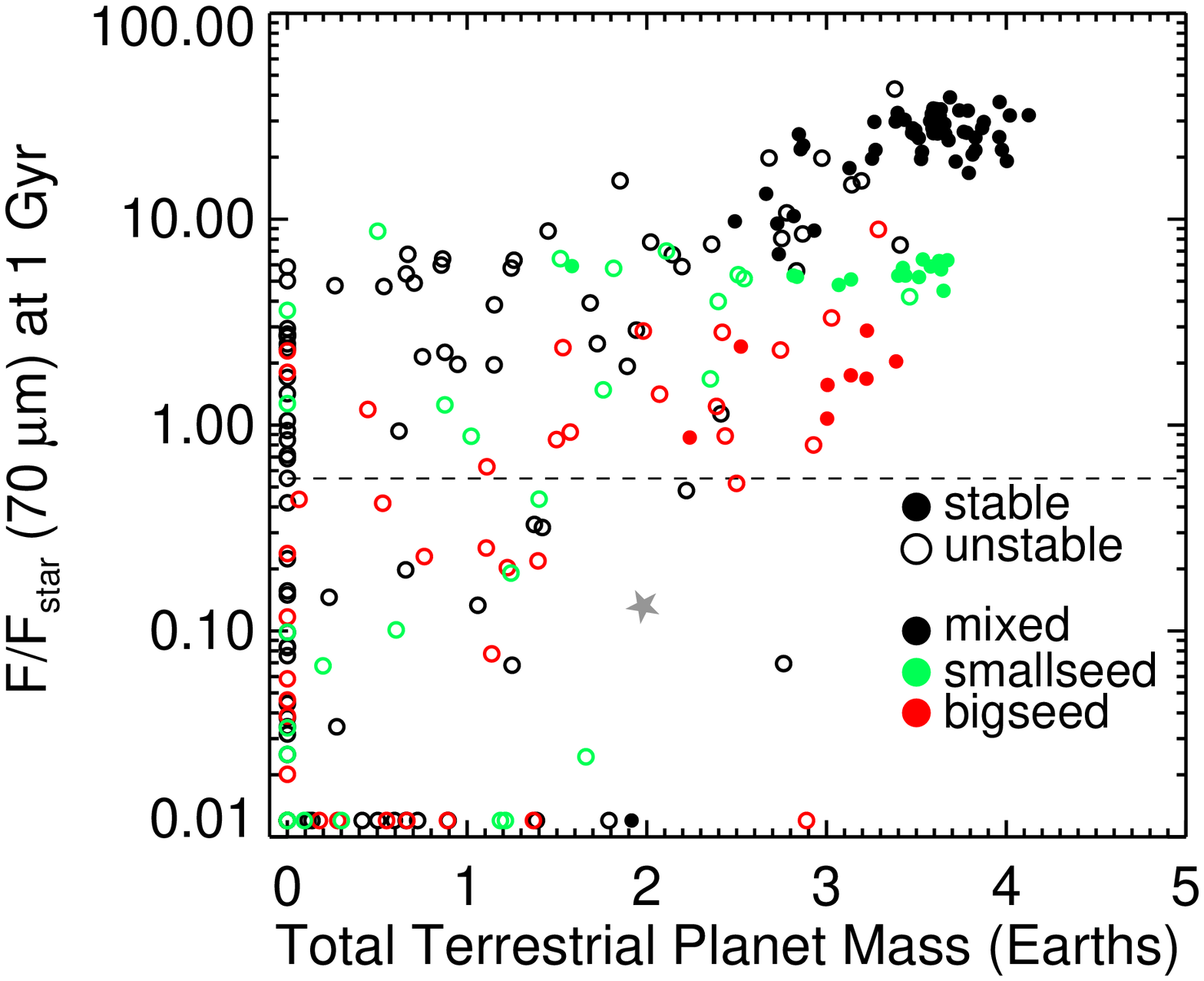}
\\
\includegraphics[width=0.48\textwidth]{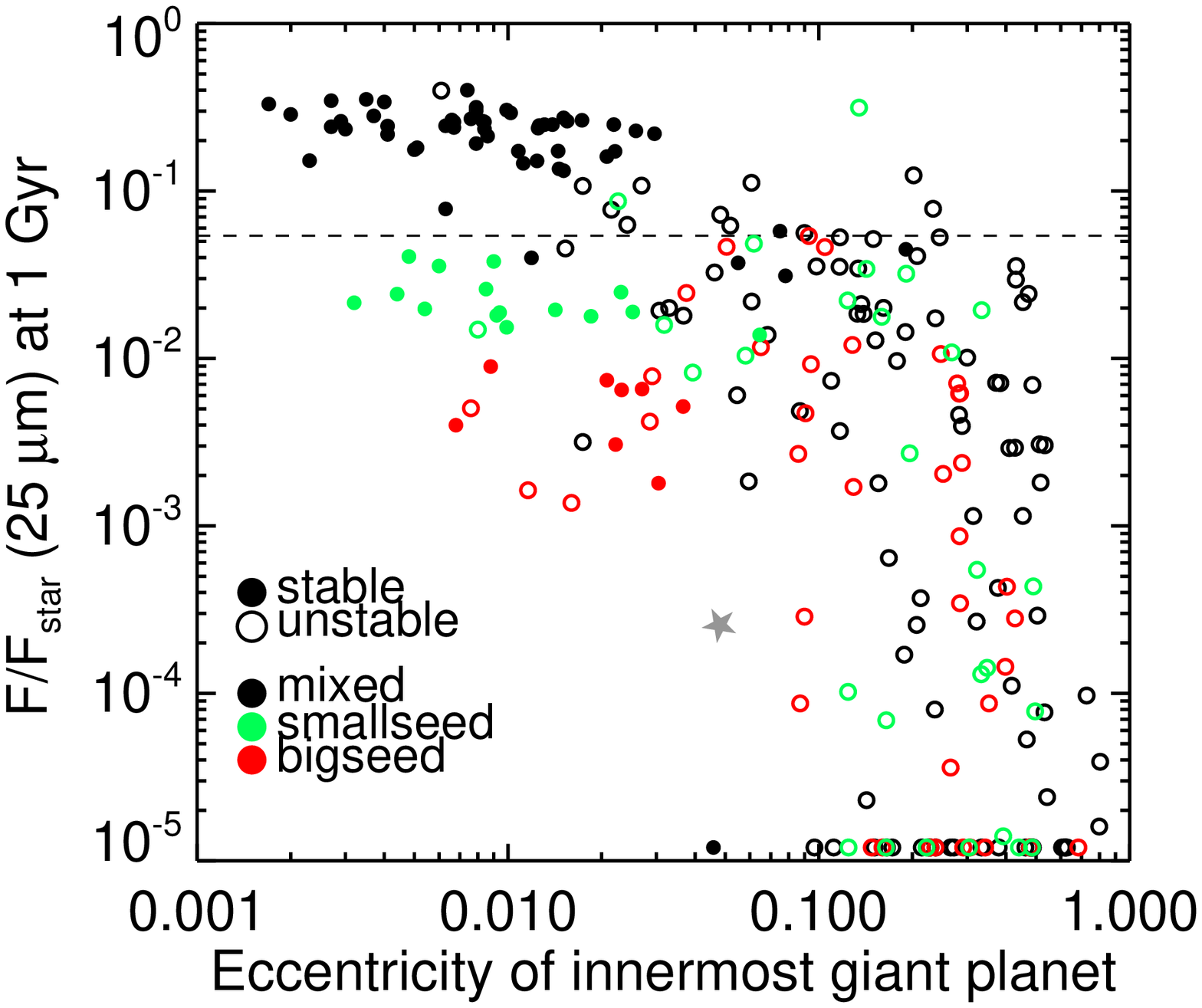}
\hfill
\includegraphics[width=0.48\textwidth]{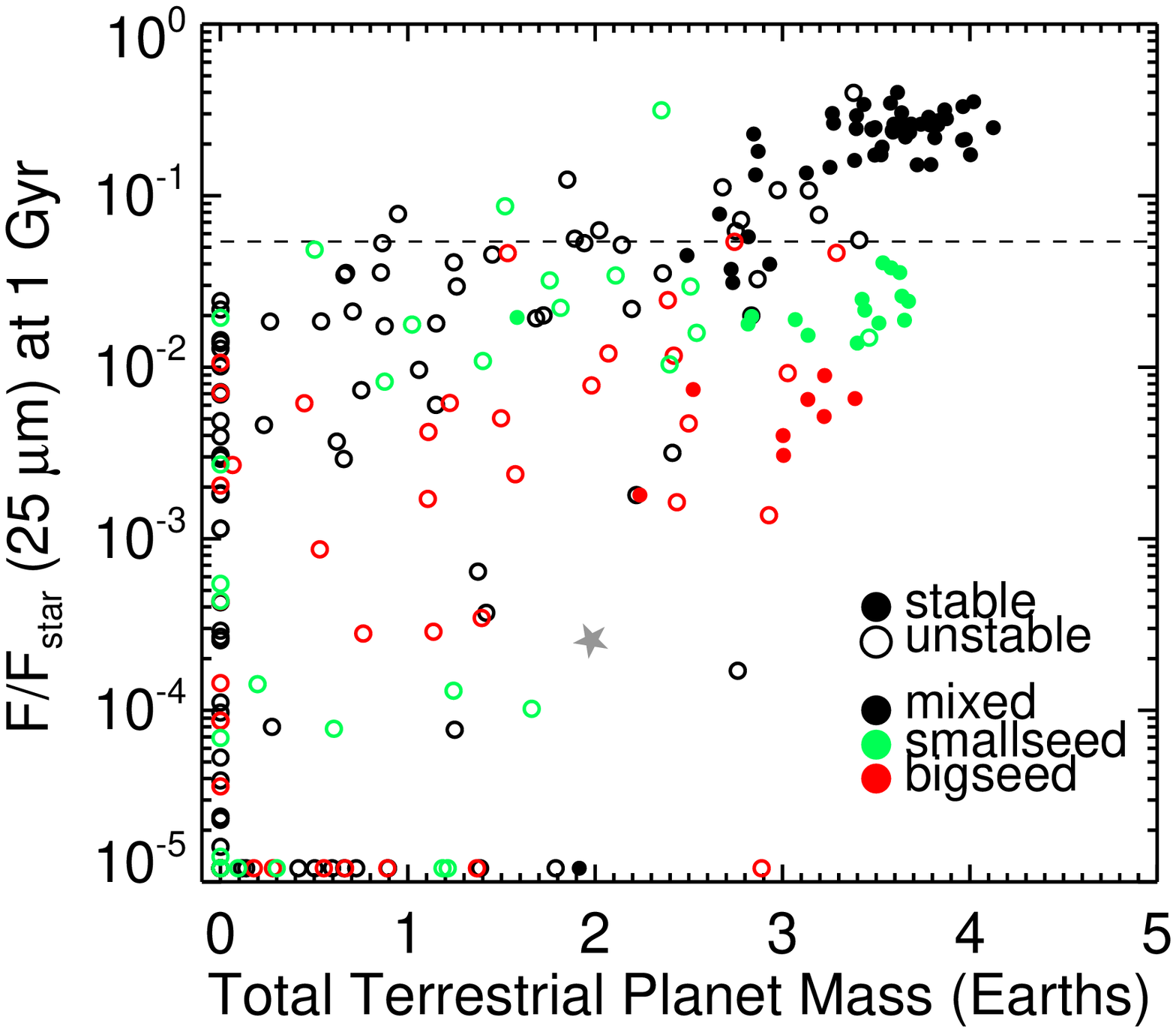}

\caption{Correlations with the dust flux after 1 Gyr of dynamical and collisional evolution for the {\tt seeds} simulations, as compared with the {\tt mixed} simulations.  The top panels show $F/F_{star}$ at $70 \micron$ vs. the eccentricity of the innermost surviving giant planet (left) and the total mass in surviving terrestrial planets (right).  The bottom panels show the same comparisons but at $25 \micron$.  The {\tt mixed} simulations are in black, the {\tt smallseed} simulations in green, and the {\tt bigseed} simulations in red.  Filled circles represent stable simulations and open circles unstable simulations.  }
\label{fig:ff_seed}
\end{figure*}

\begin{figure*}
\includegraphics[width=0.48\textwidth]{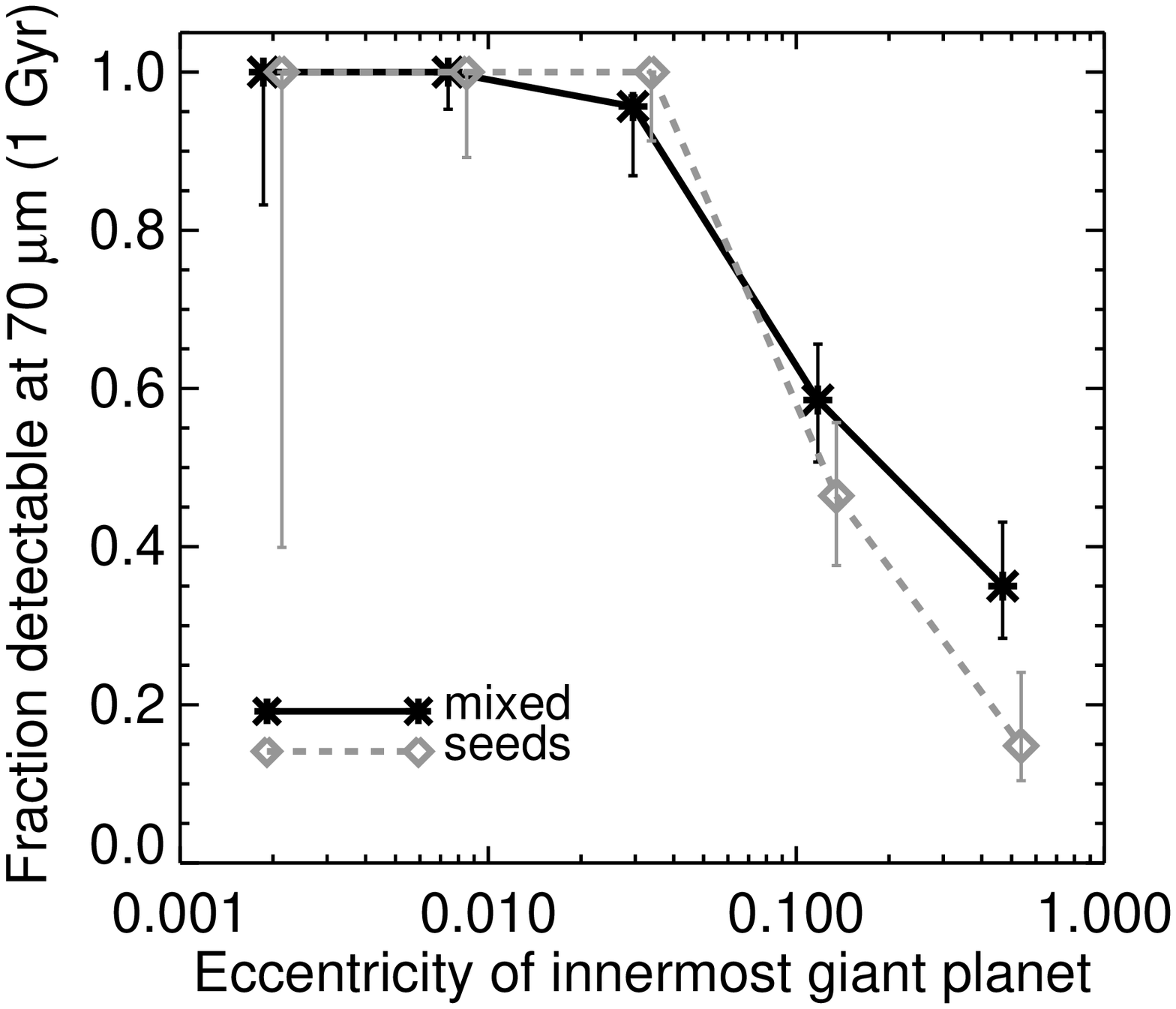}
\hfill
\includegraphics[width=0.48\textwidth]{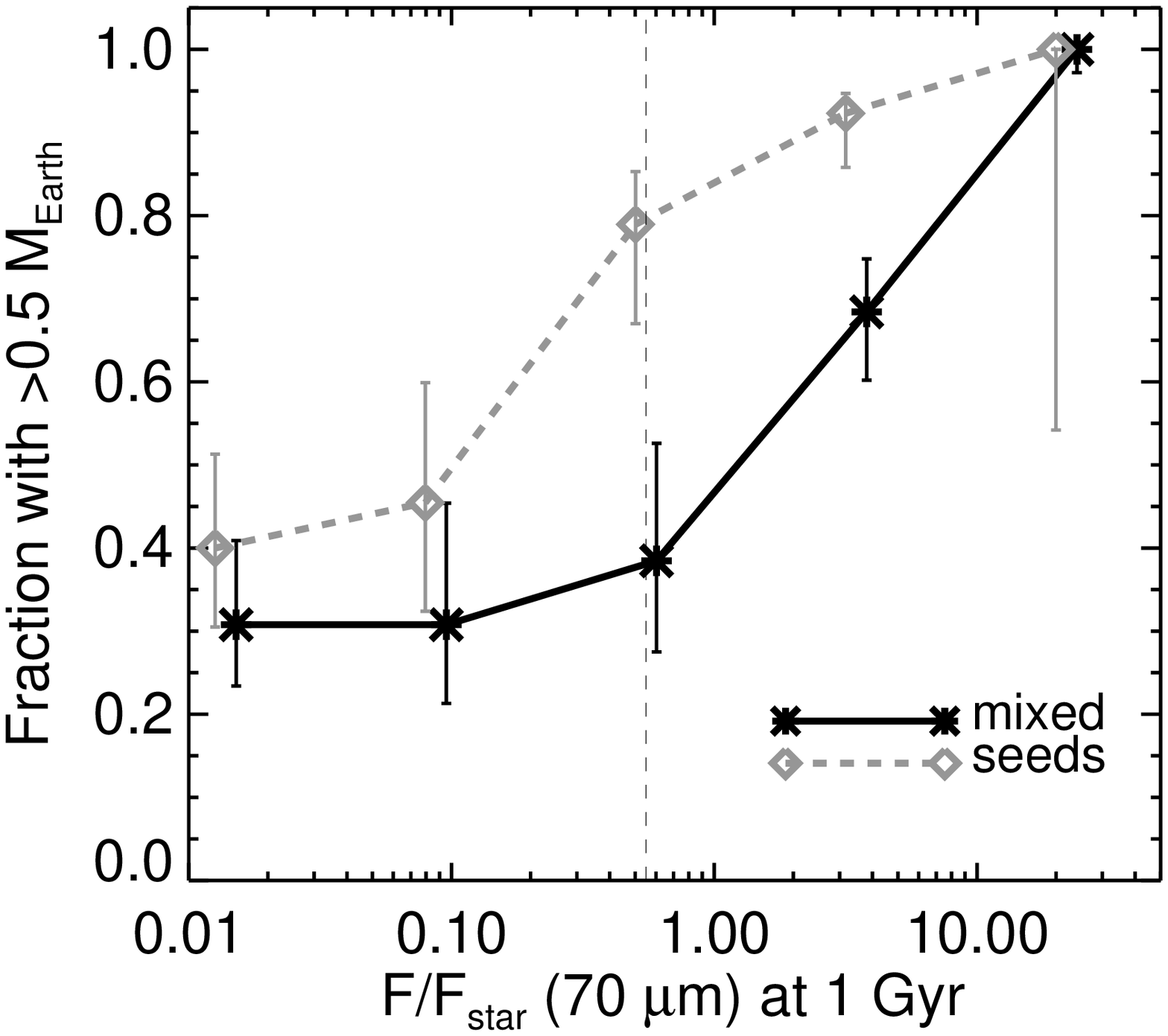}
\caption{{\bf Left:} The fraction of systems that would be detectable with {\it Spitzer} (with $F/F_{star} (70 \micron) \geq 0.55$ after 1 Gyr of collisional and dynamical evolution) as a function of the eccentricity of the innermost giant planet $e_g$ for the {\tt mixed} and combined {\tt seeds} simulations.  The error bars are based on binomial statistics ~\citep[see][]{burgasser03}.  This essentially represents a horizontal slice through the top left panel of Fig~\ref{fig:ff_seed}. {\bf Right:} The fraction of systems with $0.5 \mearth$ or more in surviving terrestrial planets as a function of $F/F_{star} (70 \micron) \geq 0.55$ (1 Gyr) for the {\tt mixed} and {\tt seeds} simulations.  Systems with $F/F_{star} < 10^{-2}$ are included in the bin at $F/F_{star} \approx 10^{-2}$.  The Spitzer detection limit is shown as the dashed line.  This represents a vertical slice through the bottom right panel of Fig.~\ref{fig:ff_seed}. }
\label{fig:hist-seeds}
\end{figure*}

The shortened lifetime of warm dust is reflected in its changing spectral energy distribution
(Fig.~\ref{fig:aeit_seed}).  Compared with simulations without seeds, there is a much faster decrease in flux at
$\lambda \lesssim 100 \micron$ at early times as the region just exterior to the outermost giant planet is cleared
much more efficiently and to a larger radial separation.  At longer wavelengths the flux also decreases more rapidly
due to depletion by scattering from icy embryos, although at long wavelengths this is counteracted in part by the
expansion of the dust disk to larger radii and therefore increased surface area (though lower temperature).  There
are spikes in the flux (seen at $25 \micron$) when objects enter the inner planetary system to encounter and be
ejected by the innermost giant planet. These spikes are due to the fact that each particle represents a
distribution of smaller bodies and, with a higher numerical resolution, these spikes would be 
less pronounced. There is a large-scale decrease in flux after the instability and all
dust disappears when the last planetesimal is ejected after $\sim$ 100 Myr.  

The giant and terrestrial planet evolution differed only slightly between the {\tt mixed} and {\tt seeds} simulations. (Recall that the giant planets in the {\tt seeds} simulations are identical to the {\tt mixed} simulations in terms of their total mass and mass distribution.)  However, the giant planet instabilities in the {\tt seeds} simulations appeared slightly weaker.  The mean innermost giant planet eccentricity was 0.21 for the {\tt seeds} simulations compared with 0.27 for the {\tt mixed} simulations.\footnote{A somewhat higher fraction of the {\tt bigseed} simulations were unstable compared with the {\tt mixed} and {\tt smallseed} simulations.  This difference is only moderately statistically significant and might be due in part to a small glitch in our initial conditions for the {\tt seeds} simulations: the inner edge of the outer planetesimal disk was always 4 Hill radii exterior to the outermost giant planet but the icy embryos were always initially between 10-20 AU.  Thus, in many cases the innermost one to two seeds were in immediate dynamical contact with the outermost giant planet.  This preferentially occurred when the outermost planet was very massive (and hence on a more distant initial orbit) and this glitch does not appear to have contaminated our results.  In fact, the median instability time was {\em later} for the {\tt bigseed} simulations than the {\tt mixed} simulations, which is the opposite of what would be expected if the instabilities were systematically driven by seed-giant planet interactions.}  The cause of this difference appears to be late encounters between a giant planet and a seed embryo, after the end of giant planet-giant planet scattering.  To test this, we sub-divide the 39 unstable {\tt bigseed} simulations into the 22 simulations in which a giant planet underwent at least one planet-seed scattering event after the completion of planet-planet encounters and the 17 simulations that did not (i.e., the last planet-planet scattering event in these 17 simulations came after the last planet-seed scattering event).  The simulations with late planet-seed encounters had systematically shorter instabilities (as measured by the time between the first and last {\em planet-planet} scattering events) and lower final giant planet eccentricities.  We think that what happens is that in these systems a planet-seed scattering event can lead to a small readjustment in one giant planet's orbit that separates it sufficiently from other giant planets to stabilize the system.  However, we note that these samples are relatively small and we cannot rule out that the weaker {\tt seeds} instabilities are a product of small number statistics.

Given the weaker giant planet instabilities in the {\tt seeds} simulations, terrestrial planet formation was correspondingly more efficient: only about 1/4 of the unstable {\tt seeds} simulations destroyed all of their terrestrial material compared with more than 40\% for the unstable {\tt mixed} simulations.  However, the differences between the planetary evolution in the {\tt seeds} and {\tt mixed} simulations are small compared with those between some of the other sets of simulations such as the {\tt lowmass} and {\tt equal} runs.

Despite their influence on the outer planetesimal disk, seeds underwent little accretion.  Among all fifty {\tt bigseed} simulations, no seed accreted more than five planetesimals, and there was just a single seed-seed collision and 5 giant planet-seed collisions.  There was slightly more accretion among the seeds in the {\tt smallseed} simulations, which had a comparable rate of planetesimal-seed impacts but a higher rate of giant impacts, with 4 seed-seed collisions and 12 giant planet-seed collisions among the fifty simulations, although we note that all but 3 of the giant planet-seed collisions occurred very early and were probably caused by the seed being placed on an orbit that was initially very close to a giant planet.

Figure~\ref{fig:ff_seed} shows the correlations between the dust-to-stellar flux ratio $F/F_{star} (70 \micron)$ after 1 Gyr, and either the innermost giant planet eccentricity or the total terrestrial planet mass.  We plot results for the {\tt smallseed}, {\tt bigseed}, and {\tt mixed} simulations.  The most important difference between the {\tt seeds} and {\tt mixed} simulations is that the {\tt seeds} simulations produce less dust at late times, especially warm dust that is observable at wavelengths shorter than about $100 \micron$.  Among just the unstable subset of simulations, 53 of 96 {\tt mixed} simulations ($55.2\%^{+4.9\%}_{-5.1\%}$) had dust fluxes that would be detectable with {\em Spitzer} after 1 Gyr of evolution, i.e., with $F/F_{star} (70 \micron)  \ge 0.55$~\citep{trilling08}.  In comparison, 14 of 29 ($48.3\%^{+9\%}_{-8.7\%}$) unstable {\tt smallseed} and 16 of 39 ($= 41\%^{+8.1\%}_{-7.2\%}$) {\tt bigseed} simulations were detectable at $70 \micron$ after 1 Gyr.  Given the statistical error bars, the decreased detection rate compared with the {\tt mixed} simulations is only a $1\sigma$ result for the {\tt smallseed} simulations and $2\sigma$ for the {\tt bigseed} simulations.  

The effect of the seeds is more apparent when considering the stable simulations.  The stable systems remain detectable at 98\% or higher rate for each of the {\tt mixed}, {\tt smallseed} and {\tt bigseed} sets of simulations.  However, the actual dust brightness decreases dramatically for simulations with seeds.  The median $F/F_{star} (70\micron)$ at 1 Gyr was 26.2 for the stable {\tt mixed} simulations, 5.3 for {\tt smallseed}, and 1.7 for {\tt bigseed}.  Given the scatter, the difference between the {\tt mixed} and {\tt seeds} simulations is significant at the $3 \sigma$ level and the difference between the {\tt bigseed} and {\tt smallseed} is significant at $5 \sigma$.  

At $25 \micron$ the differences are even more striking.  In unstable systems, 12/96 ($12.5 \%^{+4.1\%}_{-2.6\%}$) {\tt mixed} simulations, 2/29 ($6.9 \%^{+7.8\%}_{-2.1\%}$) {\tt smallseed} simulations, and 0/39 ($0^{+4.4\%}$) {\tt bigseed} simulations were above the {\it Spitzer} detection threshold of $F/F_{star} (25 \micron) \ge 0.054$ after 1 Gyr~\citep{trilling08}.  This constitutes a $1-2\sigma$ difference.  Among the stable systems, 51/56 ($91.1\%^{+2.4\%}_{-5.3\%}$) {\tt mixed} simulations were detectable at $25 \micron$ after 1 Gyr but not a single stable {\tt bigseed} or {\tt smallseed} was detectable ($0^{+7.3\%}$ for the combined {\tt seeds} simulations).  

With {\it Spitzer}'s detection limits, debris disks at $70 \micron$ vastly outnumber those at $24 \micron$.  Around stars older than 300 Myr the frequency of $24 \micron$ dust excesses was estimated at $2.8\%^{+2.4\%}_{-0.8\%}$ by~\cite{carpenter09} and at $1.9\%\pm 1.2\%$ by~\cite{gaspar09}. However, these estimates are based on just a handful of detections (from more than 100 targets).  Among the $24\micron$ detections are several systems such as $\eta$ Corvi~\citep{wyatt05} and HD 69830~\citep{beichman05,bryden06} that appear to contain dust at $\sim$ 1 AU; this dust has been interpreted as either being transient~\citep{wyatt07a} or due to a very peculiar outcome of planet formation~\citep{wyatt10}.  Thus, the frequency of systems with $24\micron$ excess due to collisional equilibrium processes is significantly smaller than the quoted values.  The frequency of dust excesses at $70 \micron$ is $16.4\%^{+2.8\%}_{-2.9\%}$~\citep{trilling08}.  This is at least 5-8 times higher than at $24 \micron$, and the removal of systems with potentially transient dust can only cause this ratio to increase.

The {\tt mixed} simulations produce an overabundance of $25 \micron$ dust excesses.  After 1 Gyr of evolution, the frequency of detectable dust at $25\micron$ was $91\%_{-5.3\%}^{+2.4\%}$ and $12.5\%^{+4.1\%}_{-2.6\%}$ ($1-\sigma$ error bars) for stable and unstable systems, respectively.  At $70 \micron$ the frequency of detectable dust was $98.2\%^{+0.5\%}_{-3.8\%}$ for stable and $55.2\%_{-5.1\%}^{+4.9\%}$ for unstable systems.  Thus, the ratio of the fraction of systems that were detectable after 1 Gyr at $70\micron$ to $25\micron$ is 1.08 for the stable {\tt mixed} systems and 4.4 for unstable {\tt mixed} systems.  The higher ratio for the unstable systems is a simple consequence of the instability preferentially clearing out the inner portion of outer planetesimal disks and leaving behind the colder part of the disk that does not emit much flux at $25\micron$.  Nonetheless, no combination of these ratios for the {\tt mixed} simulations can match the observed ratio of 5-10 after 1 Gyr of evolution.\footnote{We note that the unstable {\tt mixed} systems with detectable $25\micron$ flux are all quite close to the detection limit (Fig.~\ref{fig:ff_seed}), and the fraction of unstable systems that is detectable at $25\micron$ drops drastically to 2/96 = $2.1\%_{-0.6\%}^{+2.6\%}$ after 3 Gyr of evolution, and the ratio between the detectable frequency at $70 \micron$ to $25\micron$ increases to 25.5, well within the range allowed by observations.  However, what is lacking in the {\tt mixed} simulations is the ability to account for stable systems without $25\micron$ flux, as the fraction of stable systems with detectable $25 \micron$ flux remains very high, at $83.9\%^{+3.7\%}_{-6.0\%}$.}

The {\tt seeds} simulations may explain the dearth of $25 \micron$ dust excesses because there is a natural suppression of the $25 \micron$ flux for both stable and unstable systems.  From the 44 stable and unstable {\tt smallseed} simulations, only 2 were detectable at $25\micron$.  None of the {\tt bigseed} simulations were detectable at $25 \micron$.  The reason for the lack of $25 \micron$ flux is that the viscous-like spreading out of the outer planetesimal disk acts to both deplete the inner part of the disk by inward scattering and to push the outer part of the disk to ever colder temperatures.  The net effect is the near complete frustration of the $25 \micron$ flux.  However, the detection rates at $70 \micron$ are higher than 50\% for both the {\tt smallseed} and {\tt bigseed} simulations, including both the stable and unstable systems.  Thus, the existence of $\sim$ Earth-mass seeds in outer planetesimal disks may provide a natural explanation for the very low frequency of $25 \micron$ excesses compared with $70 \micron$ excesses.  

As seen in Fig.~\ref{fig:ff_seed}, the anti-correlation between giant planet eccentricity $e_g$ and debris disks still holds for the {\tt seeds} simulations.  For eccentricities larger than 0.1, the dust fluxes show a rapid decrease for the sets of simulations with and without seeds because the dynamics of unstable giant planets dominates the survival of planetesimals.  However, for smaller eccentricities there is a clear segregation: the {\tt bigseed} systems have the smallest dust fluxes, the {\tt mixed} systems have the largest, and the {\tt smallseed} are in the middle.  In this realm the stirring up of outer planetesimal disks dominates the dust flux, and as we've seen before the {\tt seeds} simulations create a lower-mass and colder planetesimal disk than the {\tt mixed} systems, leading to lower dust fluxes in proportion to the seeds' mass (not number).  In fact, many of the unstable {\tt bigseed} systems with modestly eccentric giant planets ($e_g \sim 0.1$) have dust fluxes as high as the stable systems.  However, despite their differences, Figure~\ref{fig:hist-seeds} shows that the fraction of systems that is detectable at $70 \micron$ as a function of $e_g$ is very similar for the {\tt seeds} and {\tt mixed} systems.  

The debris disk-terrestrial planet correlation still holds for the {\tt seeds} simulations.  As with the $e_g$, the correlation between the dust flux at $70 \micron$ after 1 Gyr and the total surviving terrestrial planet mass is less evident because stable {\tt seeds} systems that efficiently form terrestrial planets produce far less dust than their {\tt mixed} counterparts.  Again, for the stable systems there is a clear segregation between the sets of simulations, with the largest seed mass {\tt bigseed} having the smallest flux.  However, Fig.~\ref{fig:hist-seeds} shows that the fraction of systems that form at least $0.5 \mearth$ in terrestrial planets increases for both the {\tt seeds} and {\tt mixed} simulations. However, the {\tt seeds} curve is $1-2 \sigma$ higher than the {\tt mixed} curve close to the detection limit.  This reflects the fact that the {\tt seeds} simulations deplete their outer planetesimal disks far more than the {\tt mixed} simulations: only very calm systems preserve enough planetesimals to produce dust.  The systems that are observed at a given dust flux therefore represent more stable systems for the {\tt seeds} simulations than the {\tt mixed} ones.  Thus, the {\tt seeds} simulations predict an even stronger correlation between stars with observed debris disks and yet-to-be-discovered terrestrial planets.

\subsection{The width of the outer planetesimal disk (the {\tt widedisk} simulations)}
\begin{figure}
\includegraphics[width=0.45\textwidth]{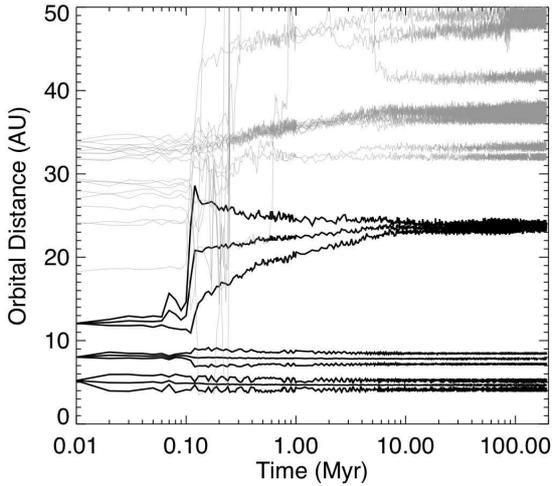}
\caption{Orbital evolution of a {\tt widedisk} simulation that underwent a Nice model-like instability.  The evolution of the semimajor axes, perihelion and aphelion distances of the three giant planets are shown in black, and the semimajor axes of the 19 surviving outer planetesimals in grey (note that one additional planetesimal survived with $a$ = 59 AU).  Two terrestrial planets formed in this system, at 0.71 AU ($1.4\mearth$)and 1.76 AU ($0.13 \mearth$): accretion was perturbed by the perturbations moderately eccentric inner giant planets at 4.7 AU ($0.78 M_J$, $e$ oscillates between 0.03 and 0.17) and 7.8 AU ($2.4 M_J$, $e$ oscillates between 0.06 and 0.1).   }
\label{fig:at_widedisk}
\end{figure}

\begin{figure*}
\includegraphics[width=0.48\textwidth]{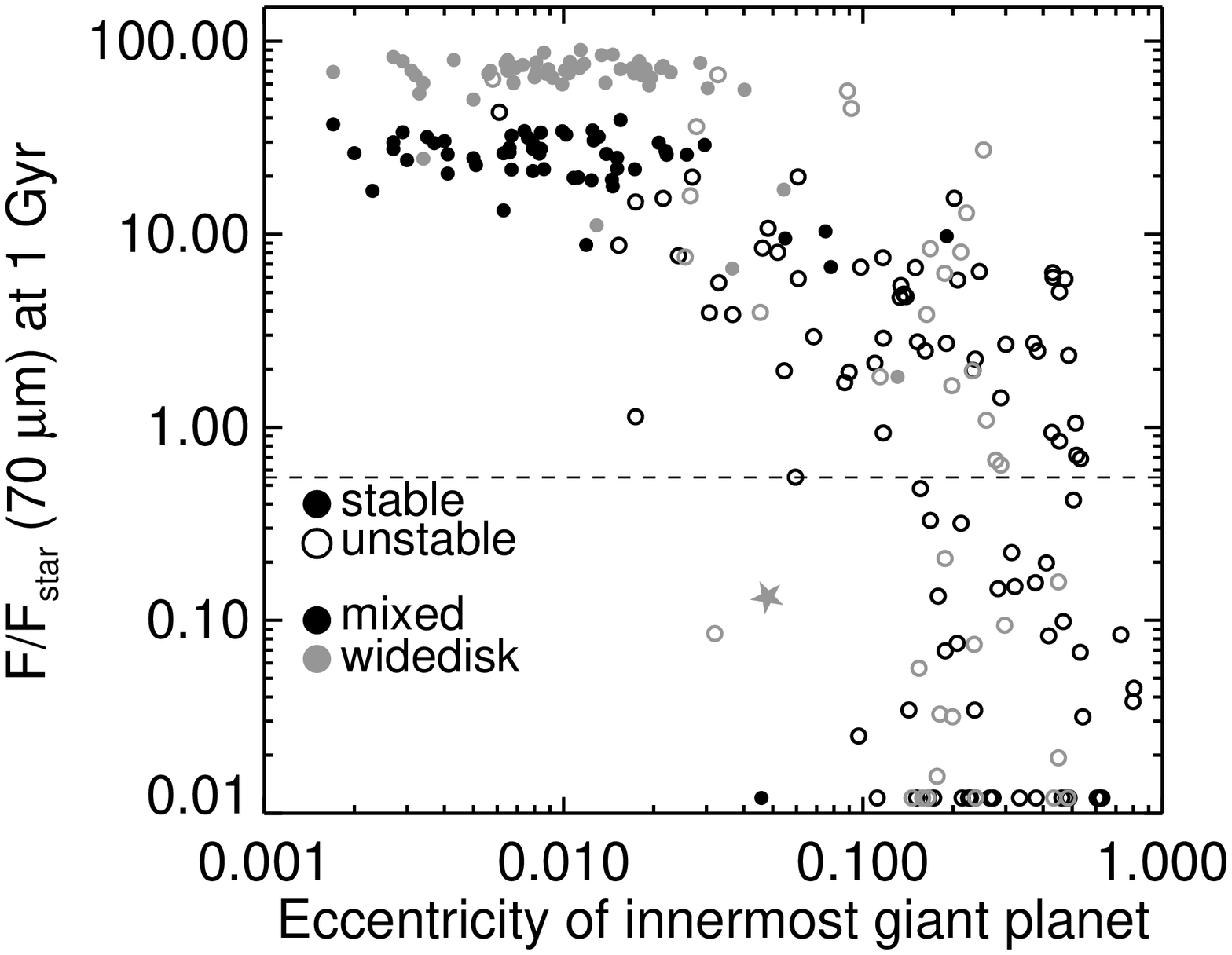}
\hfill
\includegraphics[width=0.48\textwidth]{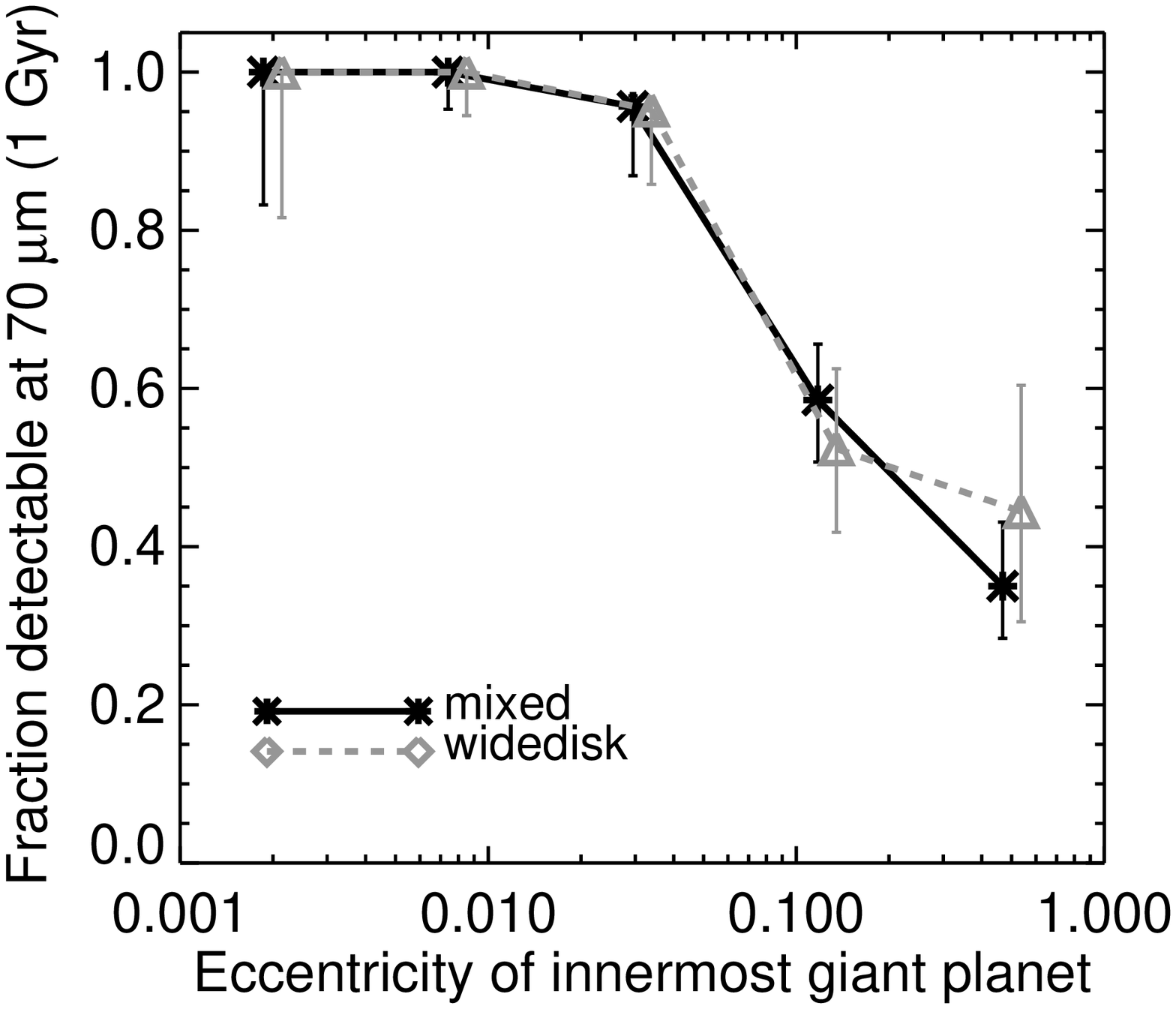}
\caption{{\bf Left:} The dust-to-stellar flux ratio at 1 Gyr at $70\micron$ vs, the innermost giant planet's eccentricity for the {\tt widedisk} (grey) and {\tt mixed} (black) simulations.  Filled circles represent stable simulations and open circles unstable ones.  The Solar System is shown with the grey star. {\bf Right:} Histogram of the fraction of systems that are detectable at $70\micron$ as a function of the innermost giant planet's eccentricity; this is essentially a horizontal slice through the left panel.  }
\label{fig:eg_widedisk}
\end{figure*}

The {\tt widedisk} simulations allow us to test the effects of a higher-mass, wider outer planetesimal disk.  Compared with the {\tt mixed} simulations, the outer planetesimal disk in the {\tt widedisk} simulations was twice as wide, 20 AU rather than 10 AU, and contained twice the total mass in planetesimals, $100 \mearth$ instead of $50 \mearth$.  The inner 10 AU of the outer planetesimal disk is therefore the same for the {\tt widedisk} and {\tt mixed} simulations (although the numerical resolution is halved in the {\tt widedisk} runs) but the {\tt widedisk} systems contain an additional 10 AU of planetesimals.  In addition, the outer boundary of each simulation -- the limit beyond which particles are considered to be ejected -- was 1000 AU rather than 100 AU.  

The {\tt widedisk} simulations behaved similarly to the {\tt mixed} simulations in most respects.  However, the more massive outer planetesimal disk did cause a few notable differences between the {\tt widedisk} and {\tt mixed} simulations.  Given the much larger angular momentum reservoir contained in the outer planetesimal disk, $\sim$Saturn-mass giant planets on eccentric orbits can be captured in the outer system by planetesimal scattering.  Figure~\ref{fig:at_widedisk} shows the evolution of one such system, in which the two inner planets ($M_{inner} = 0.78 M_J, M_{outer} = 2.4 M_J$) started just interior to the 2:1 mean motion resonance.  The giant planets' clearing out of the inner portion of the outer planetesimal disk drove the two inner giant planets across the 2:1 MMR after $\sim 0.1$ Myr.  This caused a perturbative increase in their eccentricities and a corresponding increase in the eccentricity of the outer, Saturn-mass planet.  The outer planet's semimajor axis increased quickly and over the next 10 Myr its eccentricity slowly decreased by  secular friction with the outer disk of planetesimals.  Despite the perturbative evolution of the system, 19 planetesimals totaling $1.9 \mearth$ survived in the outer planetesimal disk (their orbital evolution is shown in grey in Fig.~\ref{fig:at_widedisk}).  Most of these started the simulation in the outer parts of the planetesimal disk (one at 18 AU).  As the outer giant planet migrated outward, it shepherded many of these planetesimals in its 3:2, 2:1, and even its 3:1 mean motion resonances, located at 30.8, 37.2 and 48.9 AU at the end of the simulation, respectively.  This is analogous to the shepherding of Kuiper belt objects such as Pluto during Neptune's planetesimal-driven migration~\citep{levison08}.  The surviving planetesimal disk in the simulation from Fig.~\ref{fig:at_widedisk} is massive enough to remain detectable at $70 \micron$ for 3 Gyr but is cold enough not to be detectable at shorter wavelengths.  Planetesimal-driven migration of a massive planet therefore represents another mechanism -- in addition to the presence of seeds in outer planetesimal disks -- to deplete outer planetesimal disks and to push them outward to colder temperatures.  However, this mechanism operated efficiently in only a small fraction of {\tt widedisk} simulations.  

\begin{figure*}
\includegraphics[width=0.48\textwidth]{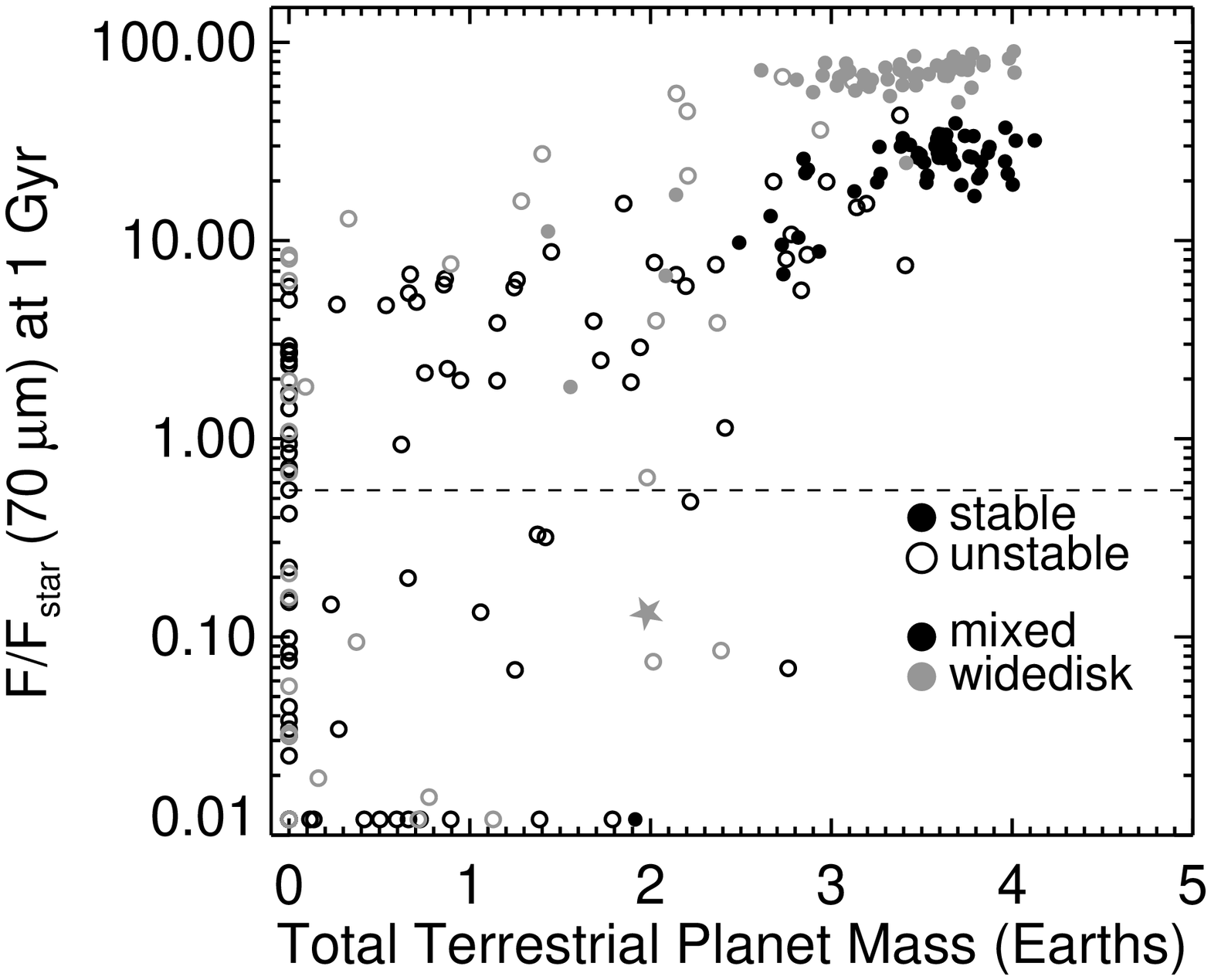}
\hfill
\includegraphics[width=0.48\textwidth]{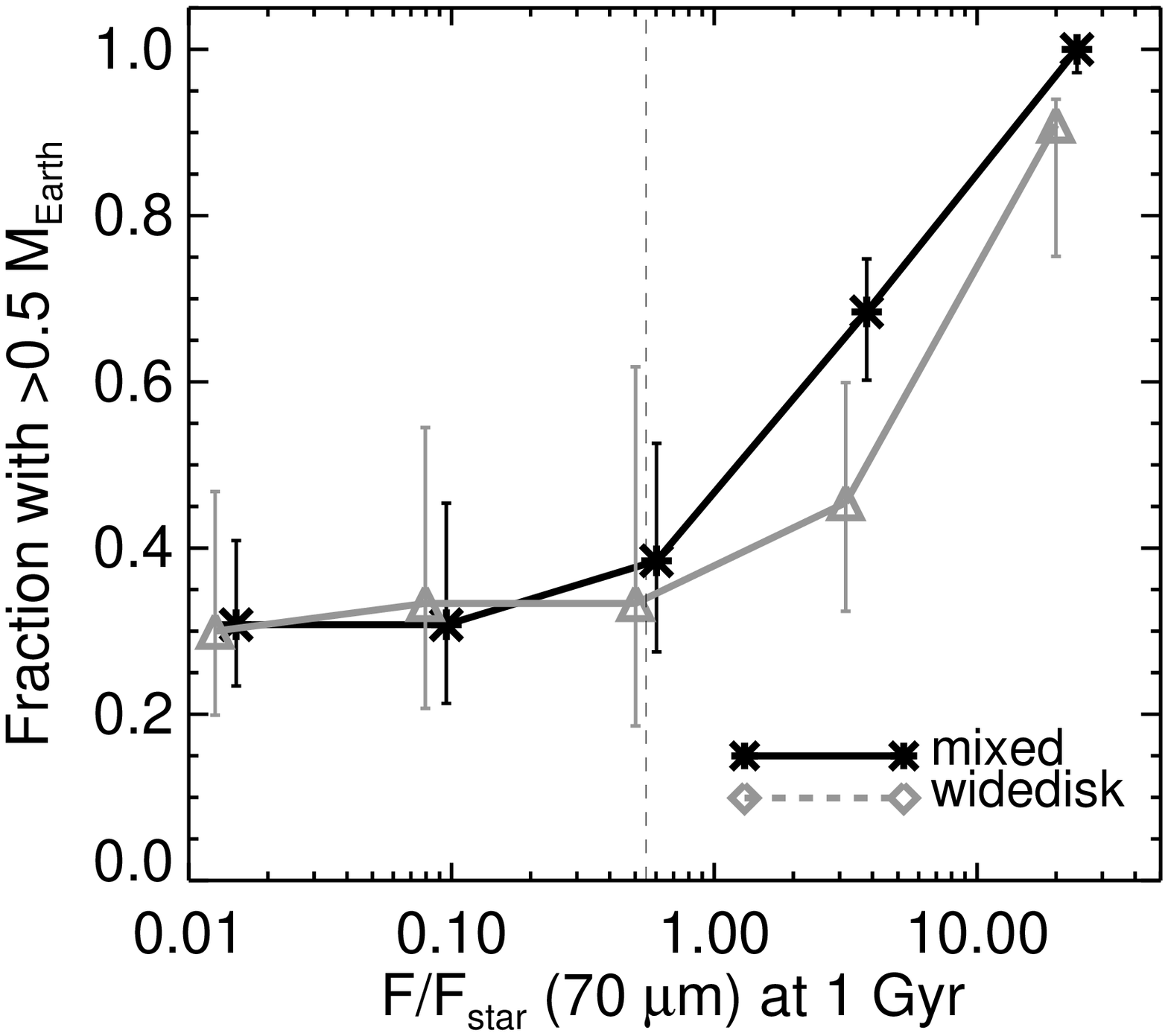}
\caption{{\bf Left:} The dust-to-stellar flux ratio at 1 Gyr at $70\micron$ vs, the total terrestrial planet mass for the {\tt widedisk} (grey) and {\tt mixed} (black) simulations.  Filled circles represent stable simulations and open circles unstable ones.  The Solar System is shown with the grey star. {\bf Right:} Histogram of the fraction of systems that contain $0.5 \mearth$ or more in terrestrial planets as a function of $F/F_{star} (70\micron)$ after 1 Gyr; this is essentially a vertical slice through the left panel.  }
\label{fig:mt_widedisk}
\end{figure*}

Despite having identical giant planet initial conditions, a significantly smaller fraction of {\tt widedisk} simulations went unstable compared with the {\tt mixed} simulations ($40.7\%_{-4.9\%}^{+5.2\%}$ for {\tt widedisk} vs. $63.2\%_{-4.1\%}^{+3.6\%}$ for {\tt mixed}).  This is because of the larger angular momentum reservoir in the outer planetesimal disks.  For cases when the instability starts in the outer portion of the planetary system, as the outer planet's orbit becomes eccentric the disk's ability to transfer orbital angular momentum (as well as energy) to damp the outer planet's eccentricity and also to increase its semimajor axis increases for a more massive planetesimal disk.  Thus, if all systems of giant planets formed on orbits that would be unstable in the absence of planetesimal disks, their long-term orbital evolution should vary with the outer disk mass.  The distribution of outer disk masses may therefore play a critical role in shaping the dynamics of inner planetary systems.  

Figure~\ref{fig:eg_widedisk} compares the debris disk - giant planet eccentricity anti-correlation for the {\tt
widedisk} and {\tt mixed} simulations.  Both distributions follow the same shape as the other sets of simulations: at $e_g \lesssim 0.05$ the vast majority of systems were dynamically stable and therefore have high fluxes, and for $e_g
\gtrsim 0.05$ unstable systems dominate and the probability of having a significant dust flux decreases strongly
with increasing $e_g$.  For stable systems, the $70 \micron$ fluxes are generally 2-3 times larger for the {\tt
widedisk} systems because their outer planetesimal disks are more massive and the dust flux is dominated by the
outer regions of the disk where the collisional evolution is slow.  However, the ratio of the median
$F/F_{star}(70\micron)$ after 1 Gyr for the {\tt widedisk} to the {\tt mixed} simulations was 2.7, but at
$25\micron$ the value was only 1.6.  This shows that in the inner regions of the outer planetesimal disk, where
collisional evolution is faster, the higher-mass {\tt widedisk} disks approach the dust production level of the
lower-mass {\tt mixed} disks.

The distribution of fluxes for the unstable systems is the same for the {\tt widedisk} and {\tt mixed} systems, meaning that the rate of survival of planetesimal disks is dominated by the giant
planet dynamics rather than the initial conditions, even the total mass and radial extent. For these unstable
systems, the distributions of instability times for the {\tt mixed} and {\tt widedisk} simulations were
indistinguishable. The distribution of the fraction of systems that is detectable with {\it Spitzer} is almost
identical for the two sets of simulations: in both cases there is a plateau at $\sim 100\%$ for $e_g \lesssim 0.05$
and a sharp decrease toward higher eccentricities.  

Figure~\ref{fig:mt_widedisk} compares the debris disk - terrestrial planet correlation for the {\tt widedisk} and {\tt mixed} simulations.  Again, the two distributions have the same shape but the dust fluxes for the stable systems (i.e., for those with the highest surviving terrestrial planet mass) are 2-3 times higher for the {\tt widedisk} systems.  The distributions are almost indistinguishable for systems with less than about $2\mearth$ in surviving terrestrial planets.  For both sets of simulations the fraction of systems with $0.5 \mearth$ or more in terrestrial planets increases monotonically with $F/F_{star}(70\micron)$ at 1 Gyr.  However, above the detection limit a slightly higher fraction of {\tt mixed} systems contain terrestrial planets compared with {\tt widedisk}.  This is the opposite of the effect that we saw in section 4.1 for the {\tt seeds} systems.  The {\tt widedisk} systems form very bright dust disks and require somewhat stronger giant planet perturbations to decrease the flux below the detection limit.  Thus, systems with bright debris disks at $70\micron$ are {\it less} sensitive to the presence of terrestrial planets than the {\tt mixed} systems.  Note that this difference comes from the slightly higher fraction of {\tt widedisk} systems with eccentric giant planets that produce $70 \micron$ excesses (see Fig.~\ref{fig:eg_widedisk})

\section{Effect of the gas disk during instabilities (the {\tt gas} simulations)}

In the {\tt gas} set of simulations we added additional forces to the Mercury integrator~\citep{chambers99} that
acted on planetesimal and embryo particles to mimic the effects of the dissipating gaseous protoplanetary disk from
which the planets formed.  In these simulations we included two effects: 1) aerodynamic gas drag due to the headwind
felt by bodies orbiting at the Keplerian speed while gas orbits slower due to pressure support.  Aerodynamic drag
acts most strongly on small objects, i.e. planetesimals, and leads to a rapid decay in eccentricity and inclination
as well as a slower decay of the semimajor axis; and 2) tidal damping (also called ``type 1 damping'') due to
gravitational interactions between objects and the disk.  Type 1 damping increases for more massive bodies because
it is caused by waves excited in the disk, meaning that this was an important source of dissipation for embryos but
not for planetesimals.  Aerodynamic drag was
calculated using standard models~\citep{adachi76} assuming planetesimals to be spheres with radii of 10~km.  Type~1 damping was included based on linear calculations for
planets embedded within isothermal disks~\citep{tanaka04}.  We included additional terms for large eccentricities and
inclinations that were derived by~\cite{cresswell08}.  We included type 1 damping but not type 1 migration both to maintain a clearer comparison with the simulations without gas forces and because eccentricity damping from the disk is roughly 2 orders of magnitude faster than radial migration~\citep{tanaka04,cresswell08}.

We assumed the presence of an underlying gas disk that corresponds to roughly half the minimum-mass solar nebula~\citep{weidenschilling77,hayashi81}, with surface density profile $\Sigma (r ) = 875 \, (r / 1 {\rm AU})^{-3/2} \, \, g \, cm^{-2}$ and vertical density distribution $\rho(z) = exp(z^2/z_0^2)$, where $z_0 (r ) = 0.0472 \, (r / 1 {\rm AU})^{5/4} \, {\rm AU}$~\citep[see also][]{thommes03,raymond06,mandell07}.  We note that the distribution of solid mass in our initial conditions is distributed according to a different density profile, $\Sigma \propto r^{-1}$, in both the terrestrial and outer planetesimal zones.  We used a steeper radial surface density profile for the gas in order to increase the gas density in the inner disk to maximize the effect of gas drag on the survival of terrestrial bodies.  However, we note that our initial conditions for the inner and outer disks -- which were chosen as represent approximate guesses for the Solar System's primordial disk -- do not even follow the same global surface density profile because there is far too little mass in the terrestrial zone.  The solution to this problem may lie with variations in the efficiency of planetesimal formation at different orbital radii within protoplanetary disks~\citep[e.g.][]{chambers10}.  

To model the final stage in the lifetime of the gaseous disk, the disk's surface density was dissipated linearly and uniformly in 500,000 years.  This is slightly longer than most estimates of the final dissipation phase~\citep{simon95,wolk96,chiang07,currie09} and should thus maximize the importance of the gas disk phase.  This situation is roughly consistent with models for dynamical instabilities among planets in the presence of gas disks~\citep{chatterjee08,moeckel08,matsumura10,marzari10,moeckel12}, which predict that instabilities should preferentially occur late in the disk phase.  The power-law gas density profile probably overestimates the amount of gas interior to the giant planets~\citep{crida07}, so these simulations should provide an upper limit to the effects of gas on terrestrial bodies.  The initial conditions for the {\tt gas} simulations were drawn directly from the {\tt mixed} set of simulations with one important change: the middle giant planet's eccentricity was increased to make its orbit cross the orbit of the innermost planet.  This was to ensure that the system would be immediately unstable so that we could test the effects of the relatively short-lived gaseous disk.

\begin{figure}
\includegraphics[width=0.45\textwidth]{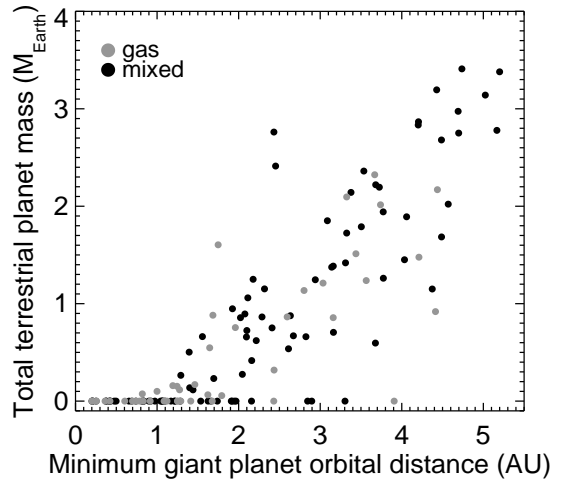}
\caption{Total mass in surviving terrestrial planets as a function of the minimum perihelion distance of any giant planet during the simulation for the unstable {\tt mixed} (black dots) and the {\tt gas} (grey dots) simulations. }
\label{fig:gasevol}
\end{figure}

The goal of the {\tt gas} simulations is to test the effect of damping from the gas disk on the dynamics and survival of rocky and icy bodies in the inner and outer planetary system.  To accomplish this, we want the giant planet instabilities to be the same as for the {\tt mixed} set, to isolate the effects of the disk. Thus, we neglected type~1 and type~2 radial migration of giant planets (and embryos) in the {\tt gas} simulations, although they would certainly occur in a realistic minimum-mass disk~\citep[e.g.,][]{lin86,ward97}.  Indeed, in a more self-consistent setting the giant planets would likely be trapped in resonance at early times and the instability would be triggered by either eccentricity excitation~\citep{marzari10,libert11} or the dispersal of the gas disk~\citep{moeckel08,chatterjee08,moeckel12}.  

The giant planets behaved similarly in the {\tt gas} and the unstable {\tt mixed} simulations (note that in this section we compare with only the unstable portion of the {\tt mixed} simulations because all of the {\tt gas} simulations were unstable).  The median eccentricity of the surviving giant planets in the {\tt gas} simulations is slightly higher -- 0.26 vs. 0.22.  The reason for the stronger instabilities in the {\tt gas} simulations is that they were initially placed on strongly unstable, crossing orbits and thus were unable to undergo weaker instabilities that can occur when they develop more slowly (and recall that no damping was felt by the giant planets in the {\tt gas} simulations).  The {\tt gas} giants provided a comparable fit to the exoplanet distribution ($p$ value from K-S test of 0.67 for {\tt gas}, 0.49 for {\tt mixed}).  A notable difference between the distributions is a population of low-eccentricity ($e \lesssim 0.05$) giant planets that is significantly more abundant in the unstable {\tt mixed} than the {\tt gas} simulations.  This population was generated by weakly unstable systems and these systems are the most efficient at both forming terrestrial planets and producing debris disks.  Although the instabilities in the {\tt gas} simulations occurred very early by design, the distributions of the duration of instabilities were virtually identical for the {\tt gas} and unstable {\tt mixed} simulations, extending from $10^4$ to $10^7$ years with a median of slightly more than 300,000 years.  This is due in part to the fact that we have not included appropriate drag forces acting on the giant planets as these are difficult to estimate without hydrodynamical simulations~\citep[see e.g.,][]{moeckel08,marzari10}.  

\begin{figure}
\includegraphics[width=0.45\textwidth]{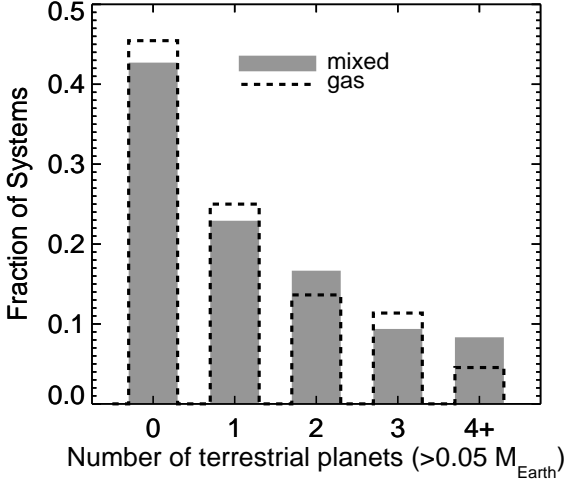}
\caption{Distribution of the number of surviving terrestrial planets per system for the unstable {\tt mixed} (grey) and {\tt gas} (dashed line) simulations.}
\label{fig:gas_hist_nterr}
\end{figure}

The effect of the giant planets on terrestrial planet formation was similar for the {\tt gas} and {\tt mixed} simulations.  Figure~\ref{fig:gasevol} shows that the sculpting of the terrestrial zone by the giant planets -- as measured by the minimum giant planet perihelion distance during the simulation -- is the same for the unstable {\tt mixed} and the {\tt gas} simulations.  The only slight difference comes from two {\tt gas} simulations in which a giant planet came closer than 1 AU to the star (in one case for a prolonged period of almost 1 Myr) but that succeeded in forming a terrestrial planet.  In both of these cases the surviving planet was roughly a Mars mass (although in one case the planet accreted another embryo) and underwent large-scale oscillations in eccentricity and inclination.  The survival of these planets may be due in part to gas drag from the disk, although they may simply represent a tail of the {\tt mixed} distribution.  

Figure~\ref{fig:gas_hist_nterr} shows that the distributions of the number of surviving terrestrial planets in the {\tt gas} and the unstable {\tt mixed} systems are very close~\citep[$1\sigma$ is 5-8\% for most bins as calculated using binomial statistics; see][]{burgasser03}.  The {\tt gas} and unstable {\tt mixed} simulations each formed a mean of 1.2 planets planets per unstable system, although there is a slight tendency for more zero- and one-planet systems in the {\tt gas} simulations.  We attribute this to the fact that in the {\tt gas} simulations the giant planets started on orbits that were already strongly unstable and the instability involved the innermost giant planet with the strongest influence on the terrestrial planet zone.  Thus, outward-directed instabilities and weakly unstable systems were less frequent in the {\tt gas} simulations.

Although the number of surviving planets was similar, the surviving terrestrial planets in the unstable {\tt mixed} simulations were significantly more massive than in the {\tt gas} simulations.  The median terrestrial planet mass was $0.43\mearth$ for {\tt gas} and $0.73\mearth$ for {\tt mixed} (counting only simulations that were integrated for $>$100 Myr and planets $>0.1\mearth$).  In addition, 28 of the 90 unstable {\tt mixed} terrestrial planets were more than $1\mearth$ ($31.1\%^{+5.2\%}_{-4.4\%}$) compared with 4 of 33 ($12.1\%^{+7.9\%}_{-3.5\%}$) for the {\tt gas} terrestrial planets.  The larger masses of the unstable {\tt mixed} simulations come from the contribution from weakly unstable systems, i.e. those with minimum giant planet perihelion distances larger than about 4 AU in Fig.~\ref{fig:gasevol}.  For systems where a giant planet entered within 4 AU of the star, the two sets had the same median terrestrial planet mass.  

\begin{figure}
\includegraphics[width=0.45\textwidth]{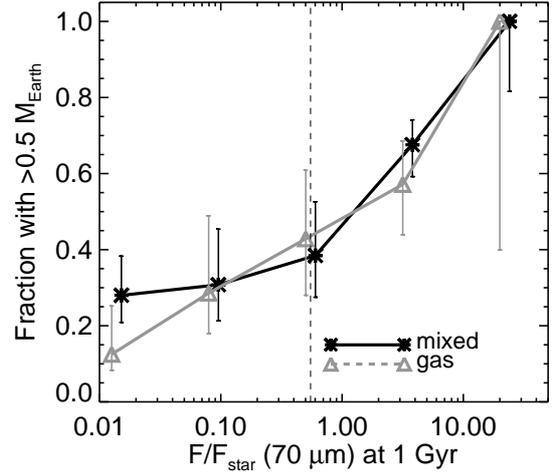}
\caption{The fraction of systems that contain $\ge 0.5 \mearth$ in surviving terrestrial planets as a function of $F/F_{star} (70\micron)$ after 1 Gyr for the unstable {\tt mixed} (black) and {\tt gas} (grey) sets of simulations.  Error bars are $1\sigma$ values calculated with binomial statistics.  }
\label{fig:gas_hist_mtff}
\end{figure}

A small fraction of unstable systems produced asteroid belts without terrestrial planets.  In these systems a number of rocky planetesimals were the only survivors in the inner planetary system, as all terrestrial embryos had been destroyed.  This occurred in 14 of 299 ($4.7 \pm 1.2 \%$) of unstable simulations across all the sets of simulations (excluding the {\tt lowmass} simulations).  The {\tt gas} simulations had a slightly higher rate of production of asteroid belt-only systems ($3/45 = 6.7\%_{-2.0\%}^{+5.7\%}$), presumably because in a few cases gas drag was able to stabilize the orbits of planetesimals that were marginally unstable.  These asteroid belts are low-mass, containing only 1-20 asteroid particles (each $5\times 10^{-3} \mearth$), albeit typically on excited orbits with moderate eccentricities ($e \sim 0.2-0.3$) and inclinations ($i \sim 20-30^\circ$).  Given their rapid collisional evolution, these belts probably become quickly dominate by a few relatively large objects and are probably not detectable with current instruments.  

The surviving terrestrial planets in the {\tt gas} and the unstable {\tt mixed} simulations had similar median eccentricities $e$ and inclinations $i$ (median values of $e\approx0.1$ and $i \approx 5-6^\circ$).  However, the {\tt gas} simulations tend to have significantly higher oscillation amplitudes in $e$ and $i$.  Although the median oscillation amplitudes are relatively close (median peak-to-peak $e_{osc} = 0.13$ and $i_{osc} = 7^\circ$ for {\tt gas} vs. 0.10  and $5^\circ$ for {\tt mixed}), planets in the {\tt gas} simulations are shifted to higher values.  Again, this difference is simply due to the lack of weakly unstable systems in the {\tt gas} simulations; when a cut in the minimum giant planet perihelion of $<$4 AU is applied the oscillation amplitudes are a match.  

The anti-correlation between the giant planet eccentricity and the dust flux at $70 \micron$ is very similar between the {\tt gas} and unstable {\tt mixed} simulations.  The positive correlation between the mass in surviving terrestrial planets and dust flux is also preserved in the {\tt gas} simulations.  Figure~\ref{fig:gas_hist_mtff} shows the fraction of systems that formed at least $0.5\mearth$ in terrestrial planets as a function of $F/F_{star}(70\micron)$ at 1 Gyr.  The two distributions are almost identical.  The only slight divergence is at small $F/F_{star}$ values, where the {\tt mixed} simulations are about $1\sigma$ higher than the {\tt gas} simulations.  This is explained by the fact that the {\tt gas} simulations are inward-directed by design, because the instability is triggered by a close encounter between the inner two giant planets (simply because our initial conditions put the middle giant planet on an orbit that crosses the inner one's).  In contrast, the {\tt mixed} instabilities include both inward- and outward-directed instabilities, i.e., instabilities that can be triggered in, and largely confined to, either the inner or outer parts of the system.  Inward-directed instabilities that perturb the outer planetesimal disk are necessarily very strong, somewhat stronger than equivalent outward-directed instabilities that perturb the outer disk.  This only appears to be important (and then only at $1\sigma$) for the lowest values of $F/F_{star}$, where the disk is most strongly depleted (note that the lowest bin includes systems with $F/F_{star} < 0.01$).  Thus, the presence of disk gas at the time of giant planet instabilities does not appear to have a significant effect on the debris disk-terrestrial planet correlation.  

We conclude that there are no strong systematic differences between the {\tt gas} and the unstable {\tt mixed} sets of simulations.

\section{Implications of dynamically active giant planets}

We now explore the implications of our results for expected correlations between extra-solar terrestrial planets, giant planets and debris disks.  We emphasize that our conclusions are of course determined in part by our chosen initial conditions, which are poorly-constrained observationally.  Several aspects of the initial conditions have considerable uncertainties: 1) the mass and mass distribution in the terrestrial planet zone, 2) the mass, mass distribution and extent of the outer planetesimal disk, 3) the number, masses and initial spacing of the giant planets, and 4) the relative masses and spacing of these three components.  We have chosen what we consider to be reasonable values of the different components, but these values certainly vary from system to system and cover a far wider range than the subset included here. In addition, some systems with qualitatively different properties probably exist, such as disks with very widely-spaced giant planets or in which planetesimals only form in narrow regions.  We discuss the impact of these assumptions on our results in section 6.4 below.  

With these limitations in mind, we now perform a simple experiment based on the results of our simulations.  The goal of the experiment is to use the giant planets to match the observed exoplanet mass and eccentricity distributions and to then test different correlations within that framework.  We construct two samples of systems that provide an adequate match to observations using simple, non-fine-tuned mixes of different sets of simulations.  We then explore the implications of those samples.  


\subsection{Observational constraints}

Any sample we construct is constrained by observations of giant exoplanets, debris disks, and correlations between those two.  We now summarize these key characteristics. 

First, we are constrained by the distribution of known giant exoplanets, particularly those beyond 0.2 AU that have presumably not been affected by tidal interactions with their host stars.  The mass distribution can be fit with a simple power law: $dN/dM \sim M^{-1.1}$\citep{butler06,udry07b}. This is the mass distribution of the entire sample. It is likely similar to the ensemble-averaged mass distribution of planets prior to scattering (with some modification due to
mass-dependent planetary ejections and collisions with the star) but, as discussed below, it need not be the mass
distribution prior to scattering in any individual system (i.e. on a system-by-system level, planet masses may be
correlated). The frequency of giant planets is very low ($<1\%$) close-in, but
increases sharply at 0.5-1 AU and appears to remain at a high level out to at least 3 AU~\citep{cumming08,mayor11}.  The median giant exoplanet eccentricity is $\sim 0.25$ and the distribution extends to above
0.9~\citep{butler06,udry07b}. The eccentricity distribution is independent of orbital distance~\citep[for planets not affected by tides][]{ford08}.  In addition, observations show that more massive giant planets ($M_p \gtrsim M_J$) have higher eccentricities than lower-mass giants~\citep[$M_p \lesssim M_J$;][]{jones06,ribas07,ford08,wright09}.   

\begin{figure*}
\hskip .5in \includegraphics[width=0.9\textwidth]{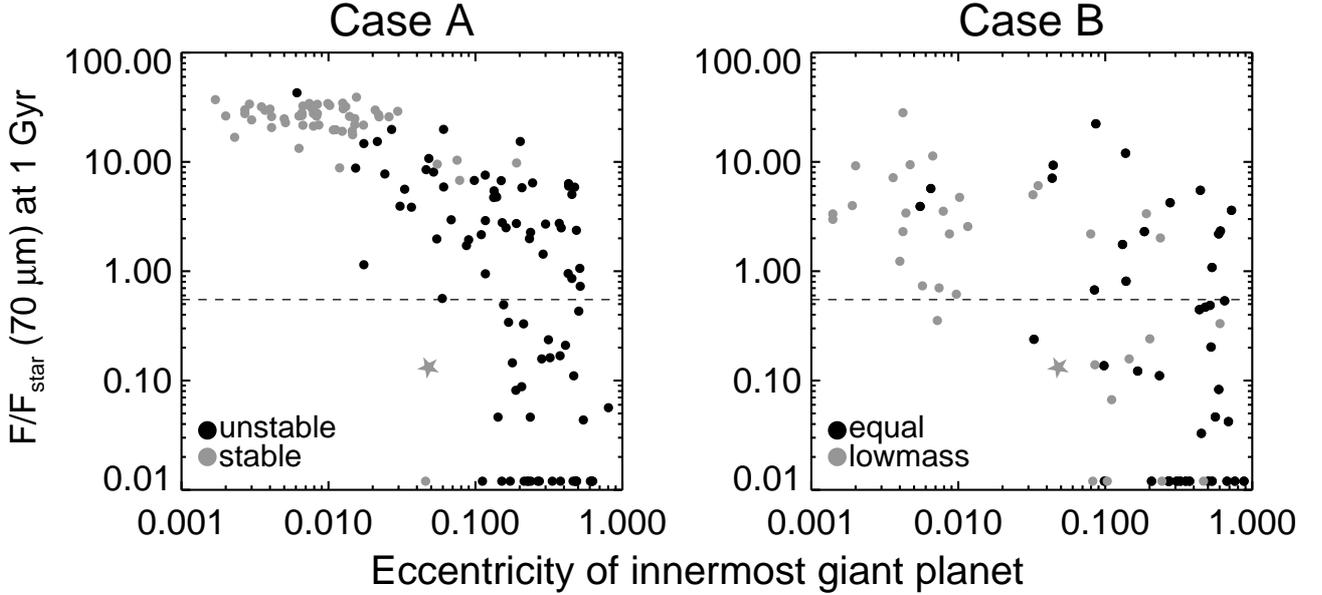}
\caption{The dust-to-stellar flux ratio at $70 \micron$ after 1 Gyr of evolution as a function of the eccentricity of the innermost giant planet $e_{g,in}$ for cases A and B.  Systems in which the innermost giant planet is exterior to 8 AU have been excluded from this plot.  }
\label{fig:eg-ff70_cases}
\end{figure*}

Second, we are constrained by debris disk statistics, and we focus on observations at $70 \micron$ made primarily with {\it Spitzer}.  Observations show that $16.4\% ^{+2.8}_{-2.9}$\% of solar-type stars have detectable dust emission at $70 \micron$~\citep{trilling08}.  There is no observed variation in this fraction with age, although the upper envelope of actual fluxes decreases for stars older than 1 Gyr or so~\citep{hillenbrand08,carpenter09}.  

Finally, we are constrained by any connection that might exist between the presence of giant exoplanets and debris disks.  Debris disks have been detected around more than 20 stars with known exoplanets.  However, there is currently no correlation between the presence of planets and debris disks: the incidence of debris disks is about 15\% for both stars with and without planets~\citep[see Table 1 in ][]{kospal09,moromartin07,bryden09}.  We also note that the strong observed correlation between the fraction of stars with currently-known exoplanets (hot Jupiters in particular) and the stellar metallicity~\citep{gonzalez97,santos01,fischer05} is not apparent in the sample of known debris disks~\citep{beichman06,greaves06,bryden06,kospal09}.

\subsection{Consistency with known giant planet properties}

The masses of individual giant planets, as well as the mass ratio between planets in a given system, are the most important factor governing the outcome of planet-planet scattering~\citep{raymond10}.  Equal-mass giant planets provide the strongest instabilities for a given mass, and the most eccentric surviving planets~\citep{ford03,raymond10}.  Scattered equal-mass planets are also more widely-spaced than planets with mass ratios of a few~\citep{raymond09b,raymond10}.  And among equal-mass unstable systems, more massive planets yield larger eccentricities but smaller inclinations~\citep{raymond10}.  The dynamics of scattering is only weakly dependent on the planet masses; Neptune-mass planets at a few AU require far more close encounters to eject one another than Jupiter-mass planets but their final orbital distributions are similar.  The number of giant planets also plays a role; in general, more giant planets lead to more scattering events and higher final eccentricities~\citep{juric08}.  

We construct two mixtures of our simulations to reproduce the observations:
\begin{itemize}
\item {\bf Case A} is based on the {\tt mixed} simulations.  The eccentricity distribution of surviving giant planets in the simulations (considering just the innermost planet as it provides the closest match to radial velocity observations) provides a quantitative match to the observed distribution with a probability value $p$ of 0.49 calculated from a K-S test.  The best match is found by including only unstable systems, but the $p$ value is still an acceptable 5-25\% if the giant planet sample includes a 5-10\% contribution from stable systems, with a higher $p$ for smaller contributions of stable systems (see Fig.~19 in Paper~1).  Case A includes a 10\% contribution from stable systems.  Note that, since case A is built on the {\tt mixed} simulations and that several other sets of simulations share the same giant planet characteristics ({\tt smallseed}, {\tt bigseed}, {\tt widedisk}, and {\tt gas}), variations on Case A can be constructed by substituting a different set of simulations for the {\tt mixed} set.   
\item {\bf Case B} is constructed from a combination of the {\tt equal} and {\tt lowmass} simulations.  The simplest scenario to explain the observation that higher-mass planets have higher eccentricities is for massive planets to form in systems with multiple, roughly equal-mass planets~\citep[see section 5 of ][]{raymond10}.  At low planet masses the eccentricities are larger than observed, so to balance the sample a contribution of systems with lower-mass ($M \leq M_{Sat}-M_{Jup}$) planets with significant mass ratios is needed -- these are represented with our {\tt lowmass} set. The exact number of low-mass systems needed is poorly constrained.  In practice, we divide the {\tt lowmass} in two based on the mass of the innermost surviving giant planet (dividing at $50 \mearth$); case B includes an equal number of systems at $M \leq M_{Jup}$ from the {\tt equal} and unstable {\tt lowmass} simulations.   
\end{itemize}

\begin{figure*}
\includegraphics[width=0.48\textwidth]{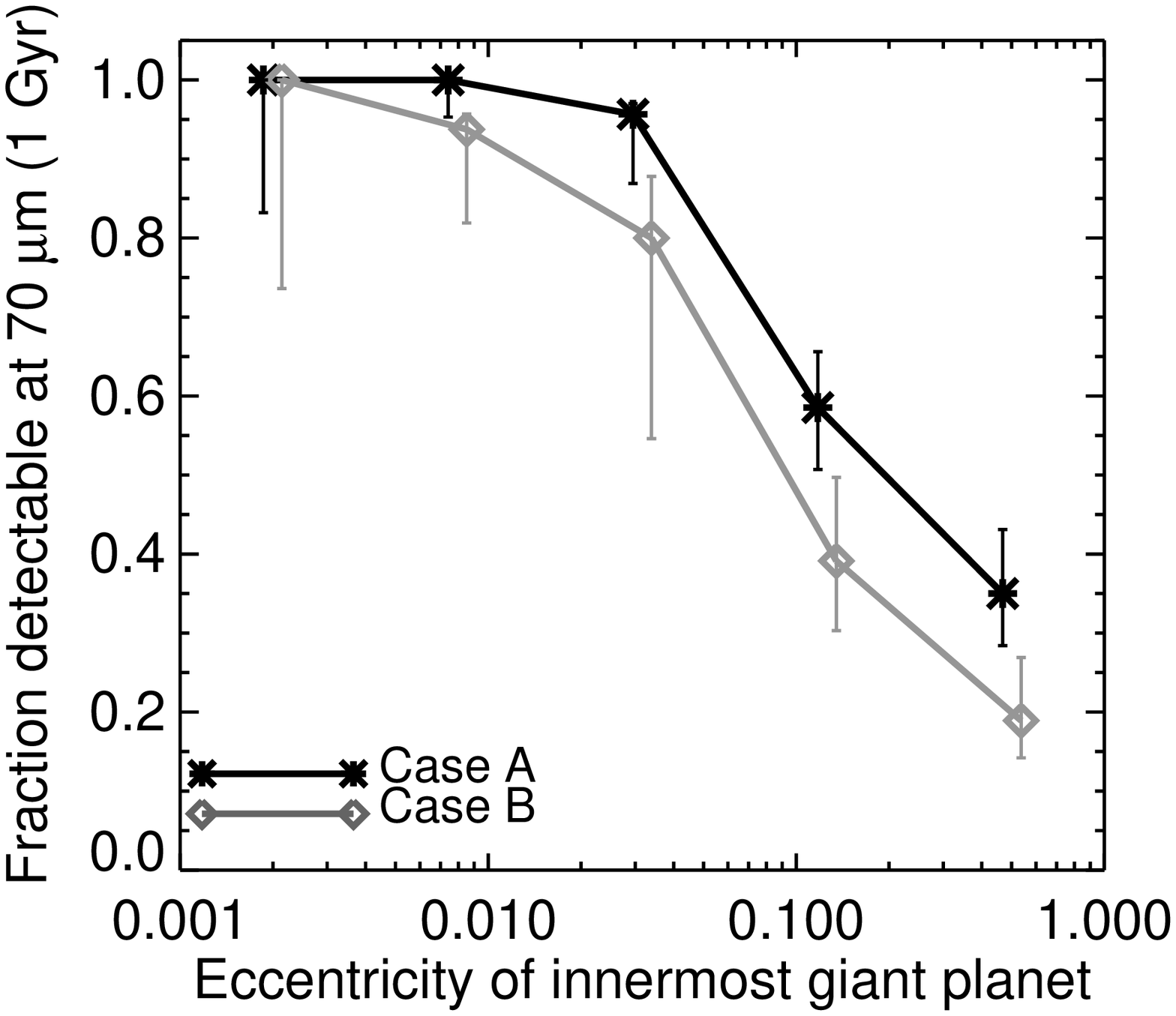}
\includegraphics[width=0.48\textwidth]{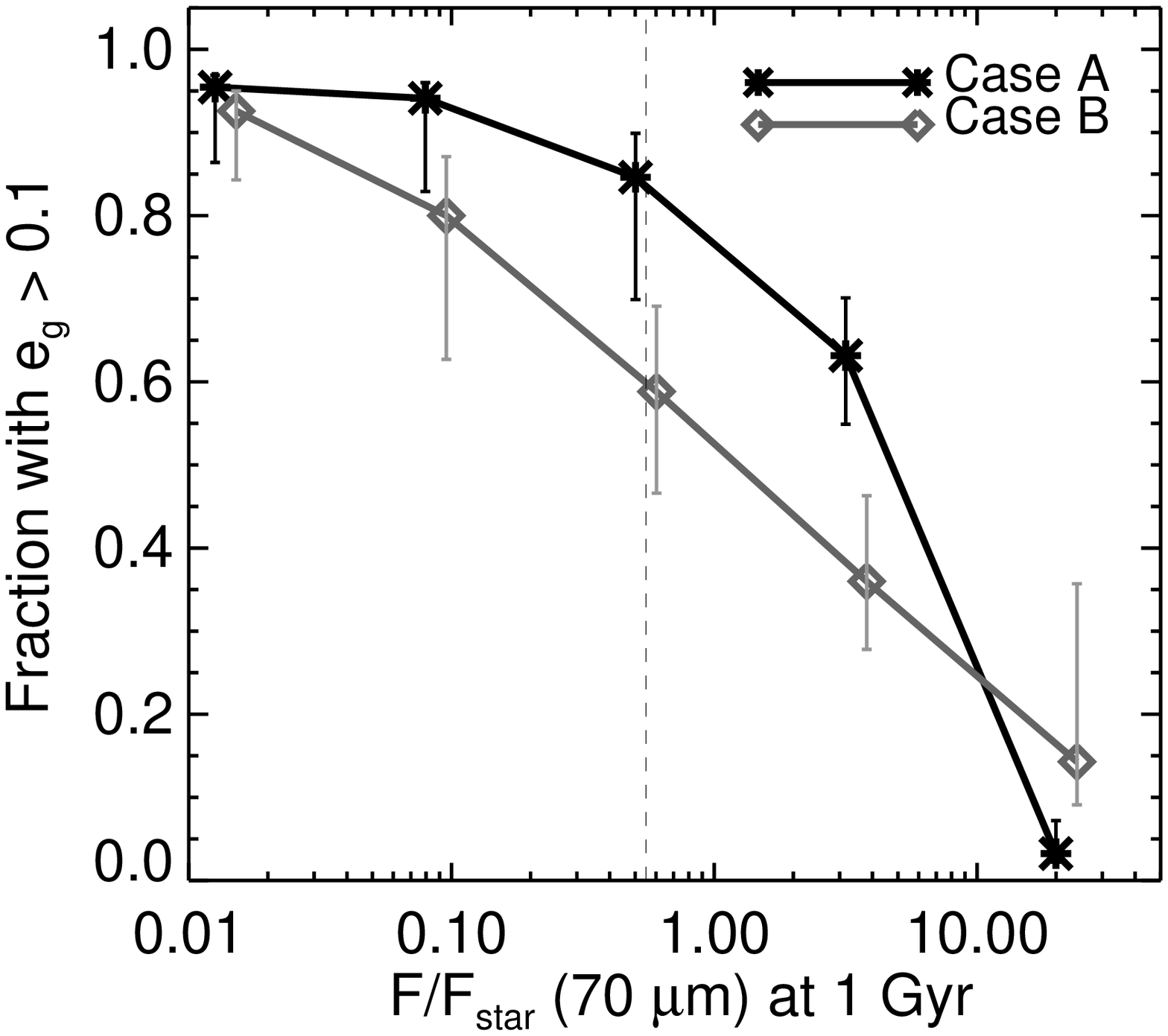}
\caption{{\bf Left:} The fraction of systems that would be detectable with {\it Spitzer} (i.e., with $F/F_{star} (70 \micron) \geq 0.55$ after 1 Gyr of collisional and dynamical evolution) as a function of the eccentricity of the innermost giant planet $e_g$, for cases A (black) and B (grey).  The error bars are based on binomial statistics ~\citep[see][]{burgasser03}.  This essentially represents a horizontal slice through Fig~\ref{fig:eg-ff70_cases}.  Note that cases A and B include {\em all} relevant simulations; there is no weighting of stable/unstable or equal/lowmass simulations.  {\bf Right:} The fraction of systems with $e_g > 0.1$ as a function of $F/F_{star} (70 \micron) \geq 0.55$ (1 Gyr), again for cases A and B.  Systems with $F/F_{star} < 10^{-2}$ are included in the bin at $F/F_{star} \approx 10^{-2}$.  The Spitzer detection limit is shown as the dashed line.  This represents a vertical slice through Fig.~\ref{fig:eg-ff70_cases}. }
\label{fig:hist_egff_cases}
\end{figure*}

Cases A and B each provide a marginally acceptable match to the observed exoplanet eccentricity distribution.  Case A matches the giant exoplanet distribution with $p = 0.08$ from K-S tests, and if we allow a 5-10\% increase in the number of planets with $e=0$~\citep[as suggested by][]{zakamska10} Case B also matches the distribution, with $p \approx 0.1$.  Given the uncertainties in orbital fitting of exoplanet eccentricities~\citep{shen08,zakamska10} we do not attempt to fine-tune our samples to better fit the observations.  Case A naturally matches the observed mass distribution (except for a bias due to the mass dependence of ejected planets) and as the outcomes for the {\tt equal} simulations were mostly mass-independent, weighting of different outcomes with the {\tt equal} contribution is not necessary, and case B can be also considered to provide a match to the mass distribution~\citep[see section 5 in][ for more details]{raymond10}. However, the two cases have different implications for the nature of planetary systems.  In case A, all planetary systems experience the same qualitative evolution because nearly all of them become dynamically unstable.  In case B, the evolution of systems is divided according to the planetary masses: high-mass planetary systems undergo extremely violent instabilities, but the evolution of lower-mass systems is much calmer and many such systems are dynamically stable.  Current observations favor case B over case A because it reproduces the higher eccentricities of more massive planets~\citep{ribas07,wright09} while for case A higher-mass planets have {\em lower} eccentricities than low-mass planets~\citep{raymond10}.  

\subsection{Debris disk correlations in cases A and B}
The expected trends -- an anti-correlation between giant planet eccentricity and debris disks and a positive correlation between terrestrial planets and debris disks -- are clearly seen in both the case A and B systems.  

Figure~\ref{fig:eg-ff70_cases} shows the dust flux at $70 \micron$ after 1 Gyr vs. the innermost giant planet's eccentricity for all the simulations in each case without the weighting described above (e.g., Case B still only contains the unstable {\tt lowmass} simulations with inner planet masses greater than $50 \mearth$ but there is no weighting between the {\tt equal} and {\tt lowmass} components).  The anti-correlation is clear in the case A simulations but the large intrinsic scatter in the {\tt lowmass} component of case B makes the trend less evident for case B.  The scatter is caused by the fact that the case B simulations are susceptible to planetesimal-driven radial migration that allows for a wide range in the depletion of the outer planetesimal disk depending on the outer planet's orbital history.  

The correlation between $e_g$ and $F/F_{star} (70 \micron)$ is clearer when the data are binned.  Figure~\ref{fig:hist_egff_cases} (left panel) shows that cases A and B are very similar in terms of the fraction of systems that is detectable at $70 \micron$ as a function of $e_g$: both cases show the expected clear anti-correlation between debris disks and eccentric giant planets including a rapid decrease in the detectable fraction for $e_g > 0.03-0.1$.  When considering the fraction of systems with $e_g > 0.1$ as a function of $F/F_{star} (70 \micron)$ (right panel of Fig.~\ref{fig:hist_egff_cases}) there is an offset of roughly $1-\sigma$ between the two cases that is again due to the larger inherent scatter in the case B simulations, in particular the high-mass, unstable component of the {\tt lowmass} simulations.   

\begin{figure*}
\hskip .5in \includegraphics[width=0.9\textwidth]{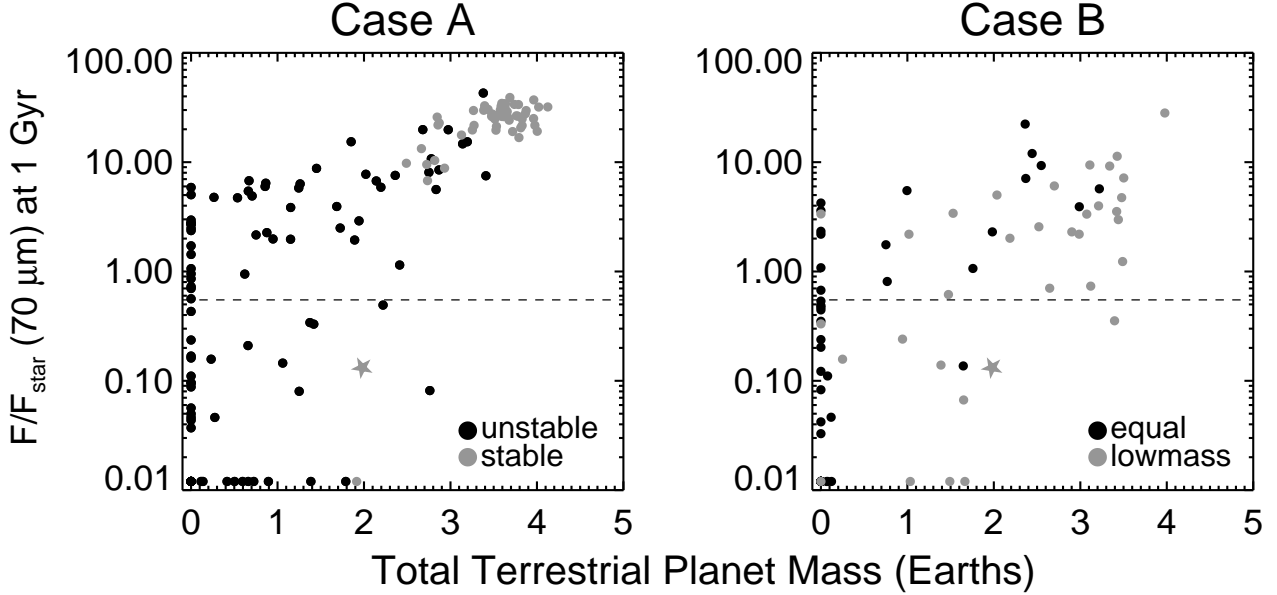}
\caption{The dust-to-stellar flux ratio at $70 \micron$ after 1 Gyr of evolution as a function of the total mass in surviving terrestrial planets for cases A and B.  Again, the Solar System is represented by the grey star.  }
\label{fig:mterr-ff70_cases}
\end{figure*}

We now turn to the debris disk - terrestrial planet correlation shown in Paper 1.  This correlation arises naturally from the destructive role of dynamically active giant planets for both terrestrial planet formation and the survival of outer planetesimal disks.  Indeed, the giant planets' role in terrestrial planet formation is almost purely antagonistic: giant planets may quench or stifle terrestrial planet formation but have not been shown to help it along in any significant way~\citep[see Paper~1 and also][]{chambers02,levison03,raymond04,raymond06a,raymond06,raymond09c}, except in some circumstances to promote runaway growth of planetesimals~\citep{kortenkamp01}.  Nonetheless, if we assume that giant planets generally form outside the snow line, moderately high eccentricities are needed before the impact on terrestrial planet formation becomes deleterious. 

Figure~\ref{fig:mterr-ff70_cases} shows the terrestrial planet-debris disk correlation for cases A and B.  As before, the correlation between $F/F_{star} (70 \micron)$ after 1 Gyr and the total mass in surviving terrestrial planets is clearer for case A.  Again, this comes from the {\tt low mass} simulations' much larger scatter in $F/F_{star}$ and somewhat lower typical values for $F/F_{star}$.

\begin{figure*}
\includegraphics[width=0.48\textwidth]{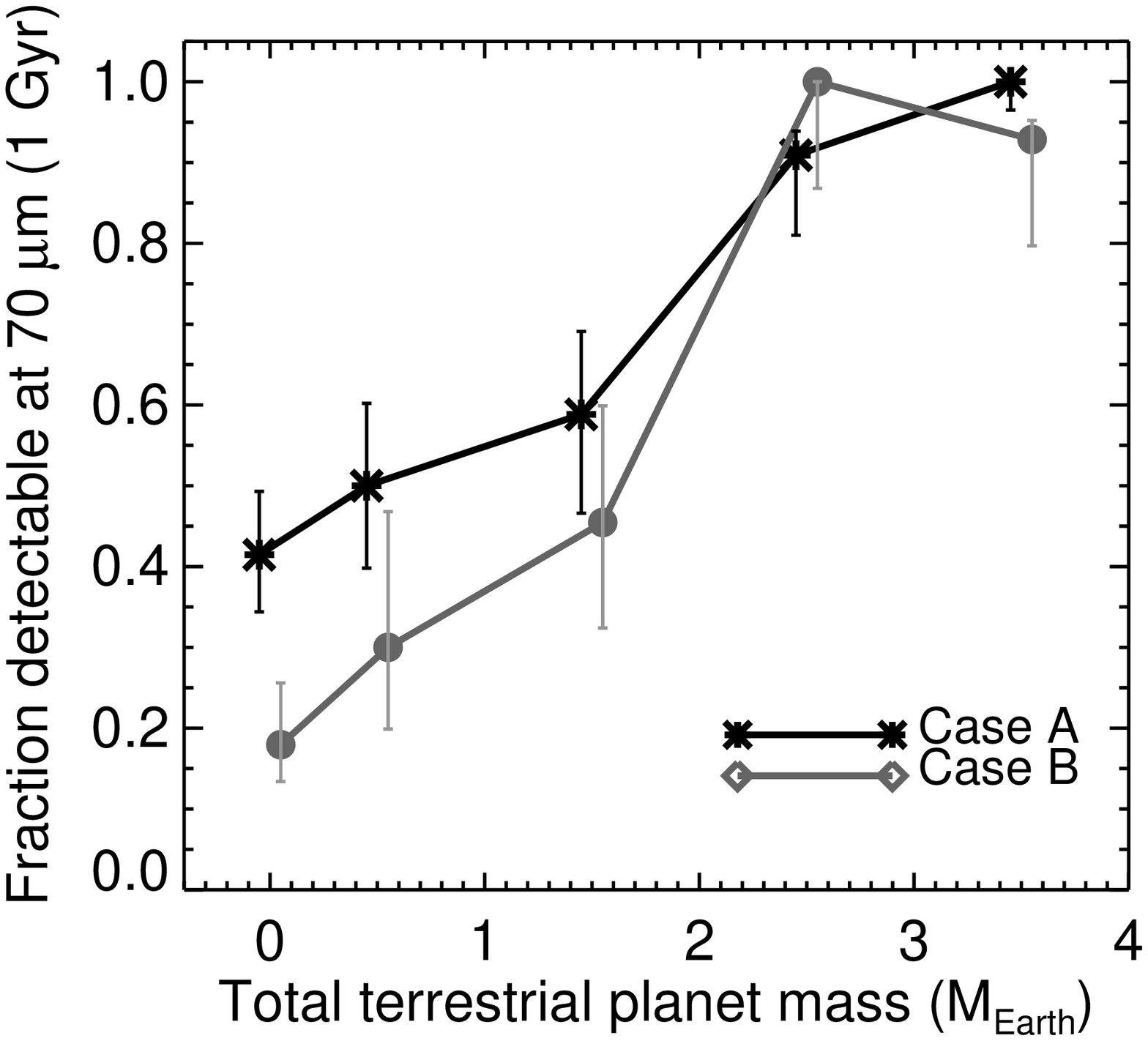}
\includegraphics[width=0.48\textwidth]{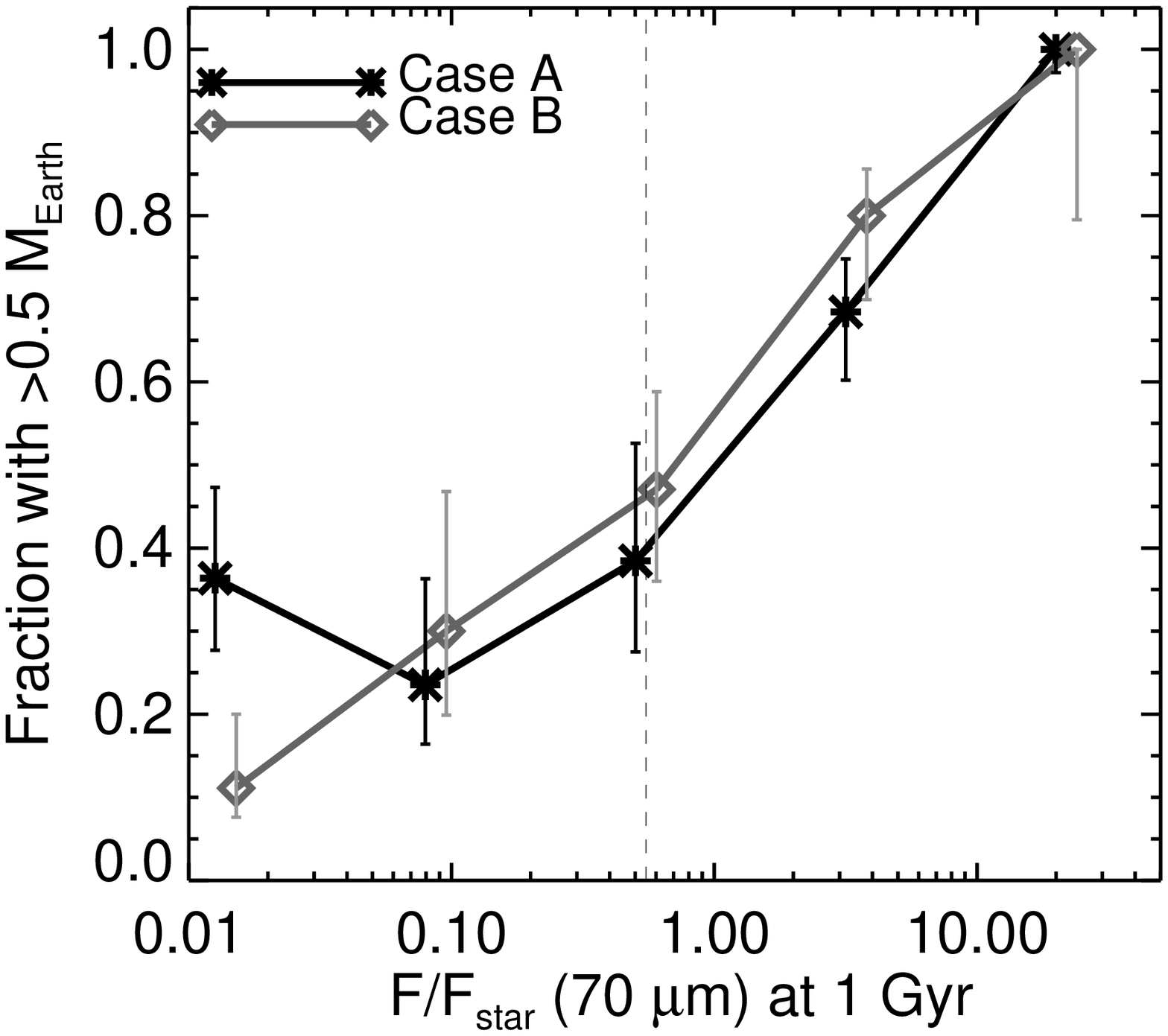}
\caption{{\bf Left:} The fraction of systems that would be detectable with {\it Spitzer} ($F/F_{star} (70 \micron) \geq 0.55$ after 1 Gyr) as a function of the total mass in surviving terrestrial planets for cases A (black) and B (grey).  This represents a horizontal slice through Fig~\ref{fig:mterr-ff70_cases}.  {\bf Right:} The fraction of systems with $0.5 \mearth$ or more in terrestrial planets as a function of $F/F_{star} (70 \micron) \geq 0.55$ (1 Gyr) for cases A and B.  Systems with $F/F_{star} < 10^{-2}$ are included in the bin at $F/F_{star} \approx 10^{-2}$.  The Spitzer detection limit is shown as the dashed line.  This represents a vertical slice through Fig.~\ref{fig:hist_mtff_cases}. }
\label{fig:hist_mtff_cases}
\end{figure*}

When the data are binned, the terrestrial planet-debris disk correlation is clearer for both cases.  Figure~\ref{fig:hist_mtff_cases} (left panel) shows the fraction of systems with $F/F_{star} (70 \micron)$ above the {\it Spitzer} detection limit as a function of the total terrestrial planet mass (i.e., a horizontal slice through Fig.~\ref{fig:mterr-ff70_cases}).  The correlations are similar for cases A and B, with only small differences for small terrestrial planet masses, for which a higher fraction of systems remains detectable for case A.  We attribute this difference to the more dynamic evolution of case B systems, which deplete or destroy their outer planetesimal disks at a much higher rate than case A systems.  Cases A and B follow nearly identical trends in terms of the fraction of systems with at least 0.5 $\mearth$ in terrestrial planets as a function of $F/F_{star} (70 \micron)$ (i.e., a vertical slice through Fig.~\ref{fig:mterr-ff70_cases}).  The only modest discrepancy that at very low fluxes ($F/F_{star} (70\micron) \lesssim 0.05$) there are more case A systems with terrestrial planets.  Once again, we attribute this to the more dynamic evolution of case B systems: the typical case B system that destroys its outer planetesimal disk is more likely to also destroy its inner rocky material than a corresponding case A system because in this regime case B is dominated by the very violently unstable {\tt equal} systems.

To summarize, we conclude from Figs~\ref{fig:eg-ff70_cases}-\ref{fig:mterr-ff70_cases} that the debris disk - eccentric giant planet anti-correlation and the debris disk - terrestrial planet correlation are clear in both cases A and B.  

\subsection{Discussion}

Could our two predicted correlations be artifacts of our initial conditions?  Is there any reasonable scenario that could remove these correlations? 

The debris disk - eccentric giant planet anti-correlation exists because giant planet instabilities tend to clear out outer planetesimal disks, mainly by dynamical ejection.  For eccentric giant planets not to be anti-correlated with debris disks, something fundamental about our proposed scenario must change.  
To start with, other mechanisms have been proposed to explain the large eccentricities of the observed exoplanets~\citep[for an exhaustive list, see Section 1 of][]{ford08}.  However, to date the planet-planet scattering model is the only mechanism that has been shown to be physically viable and to fully reproduce the currently-observed characteristics of the giant exoplanet population~\citep{ford08,chatterjee08,juric08,raymond10}.  

The initial conditions of our simulations certainly play a role in producing the correlations we find; for example, if there exist wide radial gaps between the giant planets and outer planetesimal disks then the giant planets' influence on debris disks would be weaker.  The giant planets would appear to be basically irrelevant for the existence of debris disks.  However, even a distant giant planet has a destructive influence on a planetesimal disk by increasing eccentricities sufficiently that inter-particle collisions become destructive~\citep{mustill09}.  A simple fit to the observed distribution of debris disks using a self-stirred model like the one presented in section 2 shows that outer planetesimal disks are typically located at 15-120 AU~\citep{kennedy10}.  Thus, it may be possible for debris disks and giant planets at a few AU or less to appear uncorrelated~\citep[as is the case in the current sample][]{bryden09,kospal09} but the anti-correlation between eccentric giants and debris disks should be clear for giant planets beyond roughly 5-10 AU.  In addition, there should be a minimal orbital radius (and a maximum temperature) for an outer planetesimal disk in a given system, depending on the giant planet architecture.  The statistics of debris disks in known giant planet systems are currently too poor to provide constraints (see discussion in section 5.2 of Paper 1).  

The debris disk - terrestrial planet correlation exists for three reasons.  First, terrestrial planet formation is less efficient in the presence of eccentric giant planets.  Second, the destruction rate of outer planetesimal disks by ejection increases under the influence of eccentric giant planets.  Third, we have assumed that any system that can form terrestrial planets can also produce a debris disk.  

The first reason is well-established from many dynamical studies~\citep{chambers02,levison03,raymond04,raymond06a,raymond06,raymond09c,morishima10}.  As discussed above, the dynamics behind the second reason is sound and the only reasonable way to negate or temper the eccentric giant planet - debris disk anti-correlation is for there to exist a large radial gap between giant planets and outer planetesimal disks.  The third reason is hard to constrain.  As a blanket statement this assumption is almost certainly incorrect, as the observed frequency of debris disks of $\sim$16\%~\citep{trilling08,carpenter09} is smaller than the currently-estimated frequency of close-in super Earths of 20-50\%~\citep{howard10,howard11,mayor11}.  Thus, there are probably many systems that can form terrestrial planets but without a massive outer planetesimal disk.  The reason for the lower frequency of outer disks is unclear; it could be related to external perturbations from passing stars during the embedded cluster phase~\citep{malmberg07,malmberg11}.  Or perhaps outer planetesimal disks simply are not a common initial condition for planet formation, although that assertion appears to be at odds with the high frequency of disks around young stars~\citep[e.g][]{hillenbrand08}.  One can imagine that there could also exist systems with outer planetesimal disks but very little mass in the inner disk, for example if a giant planet migrated through and depleted the inner disk~\citep[although][shows that this depletion is much weaker than one would naively expect]{raymond06b,mandell07}.  Indeed, in our own Solar System Jupiter's inward-then-outward migration may have removed most of the mass from the asteroid belt~\citep{walsh11}.  However, there is no evidence in the exoplanet distribution for systematic depletion of inner disks by giant planet migration.  Indeed, the radial distribution of giant exoplanets increases sharply beyond about 1 AU~\citep{mayor11} such that terrestrial planets closer than 0.5-0.7 AU to their stars are probably only weakly affected by giant planets, at least those with low to moderate eccentricities.  In contrast,  inner disks certainly have a wide mass range and high-mass disks may form super Earths or Neptune-like planets rather than Earth-like planets~\citep{ikoma01,raymond08a}.  

We think that the most likely interpretation of current observations and theory is roughly as follows.  Protoplanetary disks start with a variety of masses and mass distributions, and are subsequently divided into different regions by any giant planets that form.  Formation models suggest that giant planets may preferentially form at a few to ten AU~\citep{kokubo02,thommes08,levison10}, essentially dividing their disks into the distinct regions we have assumed: the inner terrestrial zone, the giant planet zone and the outer planetesimal disk.  In some systems these zones are not cleanly separated, for example if the giant planets migrate very far outward or inward or if a relatively close stellar encounter disrupts the outer planetesimal disk.  Indeed, observations suggest that there is probably an abundance of systems with inner terrestrial planet-forming disks but without the outer planetesimals to produce debris disks.  However, we see no evidence or clear theory to contradict our assumption that all or at least most stars with debris disks also have inner disks of protoplanets.  Thus, we think that debris disks can indeed act as signposts for systems that should have formed terrestrial planets.  In contrast, in systems without debris disks eccentric giant planets may act as signposts of terrestrial planet {\em destruction}.  

To summarize, we see no compelling argument against the assumptions we have made and we think that our two proposed correlations are reasonable.  

\section{Conclusions}
In Paper 1~\citep{raymond11} we showed that old solar-type stars with bright cold dust correlate strongly with dynamically calm environments that are conducive to efficient terrestrial accretion.  The fact that both the inner and outer parts of planetary systems are sculpted by what lies in between -- the giant planets -- yields a natural connection between terrestrial planet and debris disks.  We predicted two observational correlations: an anti-correlation between eccentric giant planets and debris disks, and a positive correlation between terrestrial planets and debris disks.  We also showed that the Solar System appears to be a somewhat unusual case in terms of having a rich system of terrestrial planets but a severely depleted Kuiper belt with little cold dust, which is probably a result of the outward-directed, relatively weak dynamical instability that caused the late heavy bombardment~\citep{morby10}.  

In this paper we used six new sets of simulations to test the effect of several system parameters on the results from Paper~1.  None of the parameters qualitatively changed the eccentric giant planet-debris disk anti-correlation or the debris disk-terrestrial planet correlation but their effects were diverse and interesting: 
\begin{itemize}
\item Low-mass giant planets undergo planetesimal-driven migration that radially spreads the giant planets (the {\tt lowmass} simulations).  Systems with low-mass giant planet are very efficient at forming terrestrial planets and also at producing abundant observable dust simply because their weak giant planet perturbations do not stifle these processes.  Systems with low-mass outer planets often produce isolated belts of planetesimals orbiting between the giant planets.  The probability of a system containing such a belt increases strongly with decreasing outer giant planet mass and increasing separation between the giant planets.  
\item In contrast, systems with equal-mass giant planets undergo the strongest instabilities. In more than 2/3 of the {\tt equal} systems, all terrestrial material was destroyed, and all planetesimal disks were ejected in a similar fraction of cases (though not with a one to one correspondence).  Given the tendency for more massive exoplanets to have more eccentric orbits~\citep{wright09}, the {\tt equal} systems may be representative of the high-mass ($M_p \gtrsim M_J$) exoplanet systems.  
\item The presence of $0.5-2 \mearth$ objects in outer planetesimal disks cause the disks to radially spread (the {\tt seeds} simulations).  The inward-spreading part of the disk is removed by interactions with giant planets (or, in systems without giant planets, presumably by accelerated collisional evolution) whereas the outer part of the disk spreads to larger orbital distances.  Such cold disks are detectable at long but not short wavelengths.  The presence of seeds may therefore explain the very low frequency of stars with $25 \micron$ excess compared with the frequency of $70 \micron$ excesses~\citep{bryden06,trilling08,carpenter09}.  
\item The {\tt widedisk} simulations showed that a more massive and extended outer planetesimal disk stabilizes a significant fraction of systems because planet-planetesimal interactions allow higher-mass unstable giant planet systems to radially spread out and damp their eccentricities.  Thus, if all systems of giant planet systems form in similar, near-unstable configurations, the outer planetesimal disk mass may be a key factor in the fraction of systems that end up being unstable.  More massive outer disks produce larger quantities of dust, in particular at long wavelengths that are dominated by the outermost part of the planetesimal disks where the collisional timescales are the longest.  
\item The presence of disk gas for the first 0.5 Myr of the system's evolution does not have a strong effect on the outcome (the {\tt gas} simulations).  The debris disk-terrestrial planet correlation and the eccentric giant planet - debris disk anti-correlation are not affected.  
\end{itemize}

We constructed two samples of simulations called cases A and B that matched the observed exoplanet mass and eccentricity distributions.  Case A was built on the {\tt mixed} simulations and case B on a combination of the {\tt equal} and the high-mass unstable component of the {\tt lowmass} simulations.  These cases attempt to bracket the likely initial conditions for planet-planet scattering in the known systems~\citep[e.g.,][]{chatterjee08,juric08,raymond10}.  
Each of these cases clearly show the eccentric giant planet - debris disk anti-correlation and the debris disk-terrestrial planet correlation. 

The only plausible alternative to the eccentric giant planet - debris disk anti-correlation is for outer planetesimal disks to be located far away from the giant planets and thus to be extremely cold (see section 6).  Given that the known exoplanets are dominated by relatively close-in planets (typically within a few AU) and that the planetesimals responsible for the known debris disks are at 15-120 AU~\citep{kennedy10}, it is not surprising that the expected eccentric giant planet - debris disk anti-correlation has not yet been detected~\citep[][see also section 5.2 of Paper 1]{bryden09}.  However, we predict that it will be seen in upcoming datasets that include more distant planets.  

The only plausible alternative to the debris disk-terrestrial planet correlation is for star with debris disks to be systematically depleted in their inner (terrestrial) regions {\em and} for outer planetesimal disks to be systematically cold enough to be uncorrelated with eccentric giant planets.  Although it is easy to imagine scenarios in which this may occur -- for example, a system in which an inward-migrating giant planet depletes the inner disk -- there is no observational evidence or self-consistent theoretical model that suggests that such scenarios should be common or correlated with debris disks.  Thus, we predict that upcoming datasets will find a strong correlation between debris disks and the presence of terrestrial planets.

The frequency of debris disks around Sun-like stars older than 1 Gyr is $\sim$16\%~\citep{trilling08}, which is lower than the observed frequency of close-in super Earth planets~\citep{howard10,howard11,mayor11}.  Thus, there is certainly a large population of systems that can form terrestrial planets but without the outer planetesimal disks needed to produce debris disks.  In these systems, direct radial velocity observations or constraints on the giant planet architecture may offer the best insight for constraining the existence of terrestrial planets.  But stars with debris disks most likely also had abundant material for building terrestrial planets.  Thus, debris disks can indeed act as signposts of (past) terrestrial planet formation.

To sum up, our main result is a prediction that debris disks should be anti-correlated with systems containing eccentric giant planets and correlated with the presence of terrestrial planets.  Solar-type stars with bright cold debris disks and no giant planets are excellent candidates to search for Earth-like planets.  In contrast, systems without debris disks and with eccentric giant planets are probably not good candidates for terrestrial planets. Upcoming observations will test our predictions.

\begin{acknowledgements}
We thank the referee, Hal Levison, for a careful and thorough report that helped us improve the paper. 
Simulations were run at Weber State University and at Purdue University (supported in part by the NSF through
TeraGrid resources).  S.N.R. acknowledges the CNRS's PNP and EPOV programs, the Conseil Regional d'Aquitaine, and NASA Astrobiology
Institute's Virtual Planetary Laboratory lead team. P.J.A. acknowledges funding from NASA's Origins of Solar Systems
program (NNX09AB90G), NASA's Astrophysics Theory program (NNX11AE12G), and the NSF's Division
of Astronomical Sciences (0807471).  F. S. acknowledges support from the European Research Council (ERC Grant 209622: E$_3$ARTHs)

This paper is dedicated to S.N.R.'s son Zachary Max Raymond, whose birth on June 14, 2011 caused a long but absolutely worthwhile delay of the publication of this paper.
\end{acknowledgements}

\bibliographystyle{aa}

\begin{thebibliography}{127}
\expandafter\ifx\csname natexlab\endcsname\relax\def\natexlab#1{#1}\fi

\bibitem[{{Adachi} {et~al.}(1976){Adachi}, {Hayashi}, \& {Nakazawa}}]{adachi76}
{Adachi}, I., {Hayashi}, C., \& {Nakazawa}, K. 1976, Progress of Theoretical
  Physics, 56, 1756

\bibitem[{{Armitage}(2011)}]{armitage11}
{Armitage}, P.~J. 2011, \araa, 49, 195

\bibitem[{{Beichman} {et~al.}(2005){Beichman}, {Bryden}, {Gautier},
  {Stapelfeldt}, {Werner}, {Misselt}, {Rieke}, {Stansberry}, \&
  {Trilling}}]{beichman05}
{Beichman}, C.~A., {Bryden}, G., {Gautier}, T.~N., {et~al.} 2005, \apj, 626,
  1061

\bibitem[{{Beichman} {et~al.}(2006){Beichman}, {Bryden}, {Stapelfeldt},
  {Gautier}, {Grogan}, {Shao}, {Velusamy}, {Lawler}, {Blaylock}, {Rieke},
  {Lunine}, {Fischer}, {Marcy}, {Greaves}, {Wyatt}, {Holland}, \&
  {Dent}}]{beichman06}
{Beichman}, C.~A., {Bryden}, G., {Stapelfeldt}, K.~R., {et~al.} 2006, \apj,
  652, 1674

\bibitem[{{Booth} {et~al.}(2009){Booth}, {Wyatt}, {Morbidelli},
  {Moro-Mart{\'{\i}}n}, \& {Levison}}]{booth09}
{Booth}, M., {Wyatt}, M.~C., {Morbidelli}, A., {Moro-Mart{\'{\i}}n}, A., \&
  {Levison}, H.~F. 2009, \mnras, 399, 385

\bibitem[{{Bryden} {et~al.}(2009){Bryden}, {Beichman}, {Carpenter}, {Rieke},
  {Stapelfeldt}, {Werner}, {Tanner}, {Lawler}, {Wyatt}, {Trilling}, {Su},
  {Blaylock}, \& {Stansberry}}]{bryden09}
{Bryden}, G., {Beichman}, C.~A., {Carpenter}, J.~M., {et~al.} 2009, \apj, 705,
  1226

\bibitem[{{Bryden} {et~al.}(2006){Bryden}, {Beichman}, {Trilling}, {Rieke},
  {Holmes}, {Lawler}, {Stapelfeldt}, {Werner}, {Gautier}, {Blaylock}, {Gordon},
  {Stansberry}, \& {Su}}]{bryden06}
{Bryden}, G., {Beichman}, C.~A., {Trilling}, D.~E., {et~al.} 2006, \apj, 636,
  1098

\bibitem[{{Burgasser} {et~al.}(2003){Burgasser}, {Kirkpatrick}, {Reid},
  {Brown}, {Miskey}, \& {Gizis}}]{burgasser03}
{Burgasser}, A.~J., {Kirkpatrick}, J.~D., {Reid}, I.~N., {et~al.} 2003, \apj,
  586, 512

\bibitem[{{Butler} {et~al.}(2006){Butler}, {Wright}, {Marcy}, {Fischer},
  {Vogt}, {Tinney}, {Jones}, {Carter}, {Johnson}, {McCarthy}, \&
  {Penny}}]{butler06}
{Butler}, R.~P., {Wright}, J.~T., {Marcy}, G.~W., {et~al.} 2006, \apj, 646, 505

\bibitem[{{Carpenter} {et~al.}(2009){Carpenter}, {Bouwman}, {Mamajek}, {Meyer},
  {Hillenbrand}, {Backman}, {Henning}, {Hines}, {Hollenbach}, {Kim},
  {Moro-Martin}, {Pascucci}, {Silverstone}, {Stauffer}, \&
  {Wolf}}]{carpenter09}
{Carpenter}, J.~M., {Bouwman}, J., {Mamajek}, E.~E., {et~al.} 2009, \apjs, 181,
  197

\bibitem[{{Chambers}(1999)}]{chambers99}
{Chambers}, J.~E. 1999, \mnras, 304, 793

\bibitem[{{Chambers}(2010)}]{chambers10}
{Chambers}, J.~E. 2010, \icarus, 208, 505

\bibitem[{{Chambers} \& {Cassen}(2002)}]{chambers02}
{Chambers}, J.~E. \& {Cassen}, P. 2002, Meteoritics and Planetary Science, 37,
  1523

\bibitem[{{Chambers} {et~al.}(1996){Chambers}, {Wetherill}, \&
  {Boss}}]{chambers96}
{Chambers}, J.~E., {Wetherill}, G.~W., \& {Boss}, A.~P. 1996, Icarus, 119, 261

\bibitem[{{Chatterjee} {et~al.}(2008){Chatterjee}, {Ford}, {Matsumura}, \&
  {Rasio}}]{chatterjee08}
{Chatterjee}, S., {Ford}, E.~B., {Matsumura}, S., \& {Rasio}, F.~A. 2008, \apj,
  686, 580

\bibitem[{{Chiang} \& {Murray-Clay}(2007)}]{chiang07}
{Chiang}, E. \& {Murray-Clay}, R. 2007, Nature Physics, 3, 604

\bibitem[{{Chiang} \& {Youdin}(2010)}]{chiang10}
{Chiang}, E. \& {Youdin}, A.~N. 2010, Annual Review of Earth and Planetary
  Sciences, 38, 493

\bibitem[{{Cresswell} \& {Nelson}(2008)}]{cresswell08}
{Cresswell}, P. \& {Nelson}, R.~P. 2008, \aap, 482, 677

\bibitem[{{Crida} \& {Morbidelli}(2007)}]{crida07}
{Crida}, A. \& {Morbidelli}, A. 2007, \mnras, 377, 1324

\bibitem[{{Cumming} {et~al.}(2008){Cumming}, {Butler}, {Marcy}, {Vogt},
  {Wright}, \& {Fischer}}]{cumming08}
{Cumming}, A., {Butler}, R.~P., {Marcy}, G.~W., {et~al.} 2008, \pasp, 120, 531

\bibitem[{{Currie} {et~al.}(2009){Currie}, {Lada}, {Plavchan}, {Robitaille},
  {Irwin}, \& {Kenyon}}]{currie09}
{Currie}, T., {Lada}, C.~J., {Plavchan}, P., {et~al.} 2009, \apj, 698, 1

\bibitem[{{Dauphas} \& {Pourmand}(2011)}]{dauphas11}
{Dauphas}, N. \& {Pourmand}, A. 2011, \nat, 473, 489

\bibitem[{{Dohnanyi}(1969)}]{dohnanyi69}
{Dohnanyi}, J.~S. 1969, \jgr, 74, 2531

\bibitem[{{Dominik} \& {Decin}(2003)}]{dominik03}
{Dominik}, C. \& {Decin}, G. 2003, \apj, 598, 626

\bibitem[{{Duncan} \& {Levison}(1997)}]{duncan97}
{Duncan}, M.~J. \& {Levison}, H.~F. 1997, Science, 276, 1670

\bibitem[{{Fernandez} \& {Ip}(1984)}]{fernandez84}
{Fernandez}, J.~A. \& {Ip}, W. 1984, Icarus, 58, 109

\bibitem[{{Fischer} \& {Valenti}(2005)}]{fischer05}
{Fischer}, D.~A. \& {Valenti}, J. 2005, \apj, 622, 1102

\bibitem[{{Ford} \& {Rasio}(2008)}]{ford08}
{Ford}, E.~B. \& {Rasio}, F.~A. 2008, \apj, 686, 621

\bibitem[{{Ford} {et~al.}(2003){Ford}, {Rasio}, \& {Yu}}]{ford03}
{Ford}, E.~B., {Rasio}, F.~A., \& {Yu}, K. 2003, in Astronomical Society of the
  Pacific Conference Series, Vol. 294, Scientific Frontiers in Research on
  Extrasolar Planets, ed. {D.~Deming \& S.~Seager}, 181--188

\bibitem[{{G{\'a}sp{\'a}r} {et~al.}(2009){G{\'a}sp{\'a}r}, {Rieke}, {Su},
  {Balog}, {Trilling}, {Muzzerole}, {Apai}, \& {Kelly}}]{gaspar09}
{G{\'a}sp{\'a}r}, A., {Rieke}, G.~H., {Su}, K.~Y.~L., {et~al.} 2009, \apj, 697,
  1578

\bibitem[{{Gladman}(1993)}]{gladman93}
{Gladman}, B. 1993, Icarus, 106, 247

\bibitem[{{Gomes} {et~al.}(2004){Gomes}, {Morbidelli}, \& {Levison}}]{gomes04}
{Gomes}, R.~S., {Morbidelli}, A., \& {Levison}, H.~F. 2004, \icarus, 170, 492

\bibitem[{{Gonzalez}(1997)}]{gonzalez97}
{Gonzalez}, G. 1997, \mnras, 285, 403

\bibitem[{{Gould} {et~al.}(2010){Gould}, {Dong}, {Gaudi}, {Udalski}, {Bond},
  {Greenhill}, {Street}, {Dominik}, {Sumi}, {Szyma{\'n}ski}, {Han}, {Allen},
  {Bolt}, {Bos}, {Christie}, {DePoy}, {Drummond}, {Eastman}, {Gal-Yam},
  {Higgins}, {Janczak}, {Kaspi}, {Koz{\l}owski}, {Lee}, {Mallia}, {Maury},
  {Maoz}, {McCormick}, {Monard}, {Moorhouse}, {Morgan}, {Natusch}, {Ofek},
  {Park}, {Pogge}, {Polishook}, {Santallo}, {Shporer}, {Spector}, {Thornley},
  {Yee}, {{$\mu$}FUN Collaboration}, {Kubiak}, {Pietrzy{\'n}ski},
  {Soszy{\'n}ski}, {Szewczyk}, {Wyrzykowski}, {Ulaczyk}, {Poleski}, {OGLE
  Collaboration}, {Abe}, {Bennett}, {Botzler}, {Douchin}, {Freeman}, {Fukui},
  {Furusawa}, {Hearnshaw}, {Hosaka}, {Itow}, {Kamiya}, {Kilmartin}, {Korpela},
  {Lin}, {Ling}, {Makita}, {Masuda}, {Matsubara}, {Miyake}, {Muraki}, {Nagaya},
  {Nishimoto}, {Ohnishi}, {Okumura}, {Perrott}, {Philpott}, {Rattenbury},
  {Saito}, {Sako}, {Sullivan}, {Sweatman}, {Tristram}, {von Seggern}, {Yock},
  {MOA Collaboration}, {Albrow}, {Batista}, {Beaulieu}, {Brillant}, {Caldwell},
  {Calitz}, {Cassan}, {Cole}, {Cook}, {Coutures}, {Dieters}, {Dominis Prester},
  {Donatowicz}, {Fouqu{\'e}}, {Hill}, {Hoffman}, {Jablonski}, {Kane}, {Kains},
  {Kubas}, {Marquette}, {Martin}, {Martioli}, {Meintjes}, {Menzies},
  {Pedretti}, {Pollard}, {Sahu}, {Vinter}, {Wambsganss}, {Watson}, {Williams},
  {Zub}, {PLANET Collaboration}, {Allan}, {Bode}, {Bramich}, {Burgdorf},
  {Clay}, {Fraser}, {Hawkins}, {Horne}, {Kerins}, {Lister}, {Mottram},
  {Saunders}, {Snodgrass}, {Steele}, {Tsapras}, {RoboNet Collaboration},
  {J{\o}rgensen}, {Anguita}, {Bozza}, {Calchi Novati}, {Harps{\o}e}, {Hinse},
  {Hundertmark}, {Kj{\ae}rgaard}, {Liebig}, {Mancini}, {Masi}, {Mathiasen},
  {Rahvar}, {Ricci}, {Scarpetta}, {Southworth}, {Surdej}, {Th{\"o}ne}, \&
  {MiNDSTEp Consortium}}]{gould10}
{Gould}, A., {Dong}, S., {Gaudi}, B.~S., {et~al.} 2010, \apj, 720, 1073

\bibitem[{{Greaves} {et~al.}(2006){Greaves}, {Fischer}, \& {Wyatt}}]{greaves06}
{Greaves}, J.~S., {Fischer}, D.~A., \& {Wyatt}, M.~C. 2006, \mnras, 366, 283

\bibitem[{{Habing} {et~al.}(2001){Habing}, {Dominik}, {Jourdain de Muizon},
  {Laureijs}, {Kessler}, {Leech}, {Metcalfe}, {Salama}, {Siebenmorgen},
  {Trams}, \& {Bouchet}}]{habing01}
{Habing}, H.~J., {Dominik}, C., {Jourdain de Muizon}, M., {et~al.} 2001, \aap,
  365, 545

\bibitem[{{Hahn} \& {Malhotra}(1999)}]{hahn99}
{Hahn}, J.~M. \& {Malhotra}, R. 1999, \aj, 117, 3041

\bibitem[{{Hansen}(2009)}]{hansen09}
{Hansen}, B.~M.~S. 2009, \apj, 703, 1131

\bibitem[{{Hayashi}(1981)}]{hayashi81}
{Hayashi}, C. 1981, Progress of Theoretical Physics Supplement, 70, 35

\bibitem[{{Hillenbrand} {et~al.}(2008){Hillenbrand}, {Carpenter}, {Kim},
  {Meyer}, {Backman}, {Moro-Mart{\'{\i}}n}, {Hollenbach}, {Hines}, {Pascucci},
  \& {Bouwman}}]{hillenbrand08}
{Hillenbrand}, L.~A., {Carpenter}, J.~M., {Kim}, J.~S., {et~al.} 2008, \apj,
  677, 630

\bibitem[{{Holman} \& {Wisdom}(1993)}]{holman93}
{Holman}, M.~J. \& {Wisdom}, J. 1993, \aj, 105, 1987

\bibitem[{{Howard} {et~al.}(2011){Howard}, {Marcy}, {Bryson}, {Jenkins},
  {Rowe}, {Batalha}, {Borucki}, {Koch}, {Dunham}, {Gautier}, {Van Cleve},
  {Cochran}, {Latham}, {Lissauer}, {Torres}, {Brown}, {Gilliland}, {Buchhave},
  {Caldwell}, {Christensen-Dalsgaard}, {Ciardi}, {Fressin}, {Haas}, {Howell},
  {Kjeldsen}, {Seager}, {Rogers}, {Sasselov}, {Steffen}, {Basri},
  {Charbonneau}, {Christiansen}, {Clarke}, {Dupree}, {Fabrycky}, {Fischer},
  {Ford}, {Fortney}, {Tarter}, {Girouard}, {Holman}, {Johnson}, {Klaus},
  {Machalek}, {Moorhead}, {Morehead}, {Ragozzine}, {Tenenbaum}, {Twicken},
  {Quinn}, {Isaacson}, {Shporer}, {Lucas}, {Walkowicz}, {Welsh}, {Boss},
  {Devore}, {Gould}, {Smith}, {Morris}, {Prsa}, \& {Morton}}]{howard11}
{Howard}, A.~W., {Marcy}, G.~W., {Bryson}, S.~T., {et~al.} 2011, ArXiv e-prints

\bibitem[{{Howard} {et~al.}(2010){Howard}, {Marcy}, {Johnson}, {Fischer},
  {Wright}, {Isaacson}, {Valenti}, {Anderson}, {Lin}, \& {Ida}}]{howard10}
{Howard}, A.~W., {Marcy}, G.~W., {Johnson}, J.~A., {et~al.} 2010, Science, 330,
  653

\bibitem[{{Ikoma} {et~al.}(2001){Ikoma}, {Emori}, \& {Nakazawa}}]{ikoma01}
{Ikoma}, M., {Emori}, H., \& {Nakazawa}, K. 2001, \apj, 553, 999

\bibitem[{{Jones} {et~al.}(2006){Jones}, {Butler}, {Tinney}, {Marcy}, {Carter},
  {Penny}, {McCarthy}, \& {Bailey}}]{jones06}
{Jones}, H.~R.~A., {Butler}, R.~P., {Tinney}, C.~G., {et~al.} 2006, \mnras,
  369, 249

\bibitem[{{Juri{\'c}} \& {Tremaine}(2008)}]{juric08}
{Juri{\'c}}, M. \& {Tremaine}, S. 2008, \apj, 686, 603

\bibitem[{{Kains} {et~al.}(2011){Kains}, {Wyatt}, \& {Greaves}}]{kains11}
{Kains}, N., {Wyatt}, M.~C., \& {Greaves}, J.~S. 2011, \mnras, 414, 2486

\bibitem[{{Kennedy} \& {Wyatt}(2010)}]{kennedy10}
{Kennedy}, G.~M. \& {Wyatt}, M.~C. 2010, \mnras, 405, 1253

\bibitem[{{Kenyon} \& {Bromley}(2008)}]{kenyon08}
{Kenyon}, S.~J. \& {Bromley}, B.~C. 2008, \apjs, 179, 451

\bibitem[{{Kenyon} \& {Bromley}(2010)}]{kenyon10}
{Kenyon}, S.~J. \& {Bromley}, B.~C. 2010, \apjs, 188, 242

\bibitem[{{Kirsh} {et~al.}(2009){Kirsh}, {Duncan}, {Brasser}, \&
  {Levison}}]{kirsh09}
{Kirsh}, D.~R., {Duncan}, M., {Brasser}, R., \& {Levison}, H.~F. 2009, \icarus,
  199, 197

\bibitem[{{Kokubo} \& {Ida}(2002)}]{kokubo02}
{Kokubo}, E. \& {Ida}, S. 2002, \apj, 581, 666

\bibitem[{{Kortenkamp} {et~al.}(2001){Kortenkamp}, {Wetherill}, \&
  {Inaba}}]{kortenkamp01}
{Kortenkamp}, S.~J., {Wetherill}, G.~W., \& {Inaba}, S. 2001, Science, 293,
  1127

\bibitem[{{K{\'o}sp{\'a}l} {et~al.}(2009){K{\'o}sp{\'a}l}, {Ardila},
  {Mo{\'o}r}, \& {{\'A}brah{\'a}m}}]{kospal09}
{K{\'o}sp{\'a}l}, {\'A}., {Ardila}, D.~R., {Mo{\'o}r}, A., \&
  {{\'A}brah{\'a}m}, P. 2009, \apjl, 700, L73

\bibitem[{{Krivov}(2010)}]{krivov10}
{Krivov}, A.~V. 2010, Research in Astronomy and Astrophysics, 10, 383

\bibitem[{{Krivov} {et~al.}(2006){Krivov}, {L{\"o}hne}, \& {Srem{\v
  c}evi{\'c}}}]{krivov06}
{Krivov}, A.~V., {L{\"o}hne}, T., \& {Srem{\v c}evi{\'c}}, M. 2006, \aap, 455,
  509

\bibitem[{{Krivov} {et~al.}(2005){Krivov}, {Srem{\v c}evi{\'c}}, \&
  {Spahn}}]{krivov05}
{Krivov}, A.~V., {Srem{\v c}evi{\'c}}, M., \& {Spahn}, F. 2005, \icarus, 174,
  105

\bibitem[{{Levison} \& {Agnor}(2003)}]{levison03}
{Levison}, H.~F. \& {Agnor}, C. 2003, \aj, 125, 2692

\bibitem[{{Levison} {et~al.}(2011){Levison}, {Morbidelli}, {Tsiganis},
  {Nesvorn{\'y}}, \& {Gomes}}]{levison11}
{Levison}, H.~F., {Morbidelli}, A., {Tsiganis}, K., {Nesvorn{\'y}}, D., \&
  {Gomes}, R. 2011, \aj, 142, 152

\bibitem[{{Levison} {et~al.}(2008){Levison}, {Morbidelli}, {Vanlaerhoven},
  {Gomes}, \& {Tsiganis}}]{levison08}
{Levison}, H.~F., {Morbidelli}, A., {Vanlaerhoven}, C., {Gomes}, R., \&
  {Tsiganis}, K. 2008, \icarus, 196, 258

\bibitem[{{Levison} \& {Stewart}(2001)}]{levison01}
{Levison}, H.~F. \& {Stewart}, G.~R. 2001, \icarus, 153, 224

\bibitem[{{Levison} {et~al.}(2010){Levison}, {Thommes}, \&
  {Duncan}}]{levison10}
{Levison}, H.~F., {Thommes}, E., \& {Duncan}, M.~J. 2010, \aj, 139, 1297

\bibitem[{{Libert} \& {Tsiganis}(2011)}]{libert11}
{Libert}, A.-S. \& {Tsiganis}, K. 2011, \mnras, 412, 2353

\bibitem[{{Lin} \& {Papaloizou}(1986)}]{lin86}
{Lin}, D.~N.~C. \& {Papaloizou}, J. 1986, \apj, 309, 846

\bibitem[{{Lissauer}(1993)}]{lissauer93}
{Lissauer}, J.~J. 1993, \araa, 31, 129

\bibitem[{{Lisse} {et~al.}(2008){Lisse}, {Chen}, {Wyatt}, \&
  {Morlok}}]{lisse08}
{Lisse}, C.~M., {Chen}, C.~H., {Wyatt}, M.~C., \& {Morlok}, A. 2008, \apj, 673,
  1106

\bibitem[{{L{\"o}hne} {et~al.}(2008){L{\"o}hne}, {Krivov}, \&
  {Rodmann}}]{lohne08}
{L{\"o}hne}, T., {Krivov}, A.~V., \& {Rodmann}, J. 2008, \apj, 673, 1123

\bibitem[{{Luu} {et~al.}(1997){Luu}, {Marsden}, {Jewitt}, {Trujillo},
  {Hergenrother}, {Chen}, \& {Offutt}}]{luu97}
{Luu}, J., {Marsden}, B.~G., {Jewitt}, D., {et~al.} 1997, \nat, 387, 573

\bibitem[{{Malhotra}(1993)}]{malhotra93}
{Malhotra}, R. 1993, \nat, 365, 819

\bibitem[{{Malmberg} {et~al.}(2011){Malmberg}, {Davies}, \&
  {Heggie}}]{malmberg11}
{Malmberg}, D., {Davies}, M.~B., \& {Heggie}, D.~C. 2011, \mnras, 411, 859

\bibitem[{{Malmberg} {et~al.}(2007){Malmberg}, {de Angeli}, {Davies}, {Church},
  {Mackey}, \& {Wilkinson}}]{malmberg07}
{Malmberg}, D., {de Angeli}, F., {Davies}, M.~B., {et~al.} 2007, \mnras, 378,
  1207

\bibitem[{{Mandell} {et~al.}(2007){Mandell}, {Raymond}, \&
  {Sigurdsson}}]{mandell07}
{Mandell}, A.~M., {Raymond}, S.~N., \& {Sigurdsson}, S. 2007, \apj, 660, 823

\bibitem[{{Marchal} \& {Bozis}(1982)}]{marchal82}
{Marchal}, C. \& {Bozis}, G. 1982, Celestial Mechanics, 26, 311

\bibitem[{{Marzari} {et~al.}(2010){Marzari}, {Baruteau}, \&
  {Scholl}}]{marzari10}
{Marzari}, F., {Baruteau}, C., \& {Scholl}, H. 2010, \aap, 514, L4+

\bibitem[{{Masset} \& {Snellgrove}(2001)}]{masset01}
{Masset}, F. \& {Snellgrove}, M. 2001, \mnras, 320, L55

\bibitem[{{Matsumura} {et~al.}(2010){Matsumura}, {Thommes}, {Chatterjee}, \&
  {Rasio}}]{matsumura10}
{Matsumura}, S., {Thommes}, E.~W., {Chatterjee}, S., \& {Rasio}, F.~A. 2010,
  \apj, 714, 194

\bibitem[{{Mayor} {et~al.}(2011){Mayor}, {Marmier}, {Lovis}, {Udry},
  {S{\'e}gransan}, {Pepe}, {Benz}, {Bertaux}, {Bouchy}, {Dumusque}, {Lo Curto},
  {Mordasini}, {Queloz}, \& {Santos}}]{mayor11}
{Mayor}, M., {Marmier}, M., {Lovis}, C., {et~al.} 2011, ArXiv e-prints

\bibitem[{{Moeckel} \& {Armitage}(2012)}]{moeckel12}
{Moeckel}, N. \& {Armitage}, P.~J. 2012, \mnras, 419, 366

\bibitem[{{Moeckel} {et~al.}(2008){Moeckel}, {Raymond}, \&
  {Armitage}}]{moeckel08}
{Moeckel}, N., {Raymond}, S.~N., \& {Armitage}, P.~J. 2008, \apj, 688, 1361

\bibitem[{{Mo{\'o}r} {et~al.}(2006){Mo{\'o}r}, {{\'A}brah{\'a}m}, {Derekas},
  {Kiss}, {Kiss}, {Apai}, {Grady}, \& {Henning}}]{moor06}
{Mo{\'o}r}, A., {{\'A}brah{\'a}m}, P., {Derekas}, A., {et~al.} 2006, \apj, 644,
  525

\bibitem[{{Morbidelli} {et~al.}(2010){Morbidelli}, {Brasser}, {Gomes},
  {Levison}, \& {Tsiganis}}]{morby10}
{Morbidelli}, A., {Brasser}, R., {Gomes}, R., {Levison}, H.~F., \& {Tsiganis},
  K. 2010, \aj, 140, 1391

\bibitem[{{Morbidelli} {et~al.}(2007){Morbidelli}, {Tsiganis}, {Crida},
  {Levison}, \& {Gomes}}]{morbidelli07}
{Morbidelli}, A., {Tsiganis}, K., {Crida}, A., {Levison}, H.~F., \& {Gomes}, R.
  2007, \aj, 134, 1790

\bibitem[{{Morishima} {et~al.}(2010){Morishima}, {Stadel}, \&
  {Moore}}]{morishima10}
{Morishima}, R., {Stadel}, J., \& {Moore}, B. 2010, Icarus, 207, 517

\bibitem[{{Moro-Mart{\'{\i}}n} {et~al.}(2007){Moro-Mart{\'{\i}}n}, {Carpenter},
  {Meyer}, {Hillenbrand}, {Malhotra}, {Hollenbach}, {Najita}, {Henning}, {Kim},
  {Bouwman}, {Silverstone}, {Hines}, {Wolf}, {Pascucci}, {Mamajek}, \&
  {Lunine}}]{moromartin07}
{Moro-Mart{\'{\i}}n}, A., {Carpenter}, J.~M., {Meyer}, M.~R., {et~al.} 2007,
  \apj, 658, 1312

\bibitem[{{Mustill} \& {Wyatt}(2009)}]{mustill09}
{Mustill}, A.~J. \& {Wyatt}, M.~C. 2009, \mnras, 399, 1403

\bibitem[{{Nagasawa} {et~al.}(2005){Nagasawa}, {Lin}, \&
  {Thommes}}]{nagasawa05}
{Nagasawa}, M., {Lin}, D.~N.~C., \& {Thommes}, E. 2005, \apj, 635, 578

\bibitem[{{Nesvorn{\'y}} {et~al.}(2010){Nesvorn{\'y}}, {Jenniskens}, {Levison},
  {Bottke}, {Vokrouhlick{\'y}}, \& {Gounelle}}]{nesvorny10}
{Nesvorn{\'y}}, D., {Jenniskens}, P., {Levison}, H.~F., {et~al.} 2010, \apj,
  713, 816

\bibitem[{{Pierens} \& {Raymond}(2011)}]{pierens11}
{Pierens}, A. \& {Raymond}, S.~N. 2011, \aap, 533, A131

\bibitem[{{Raymond}(2006)}]{raymond06a}
{Raymond}, S.~N. 2006, \apjl, 643, L131

\bibitem[{{Raymond} {et~al.}(2009{\natexlab{a}}){Raymond}, {Armitage}, \&
  {Gorelick}}]{raymond09b}
{Raymond}, S.~N., {Armitage}, P.~J., \& {Gorelick}, N. 2009{\natexlab{a}},
  \apjl, 699, L88

\bibitem[{{Raymond} {et~al.}(2010){Raymond}, {Armitage}, \&
  {Gorelick}}]{raymond10}
{Raymond}, S.~N., {Armitage}, P.~J., \& {Gorelick}, N. 2010, \apj, 711, 772

\bibitem[{{Raymond} {et~al.}(2011){Raymond}, {Armitage}, {Moro-Mart{\'{\i}}n},
  {Booth}, {Wyatt}, {Armstrong}, {Mandell}, {Selsis}, \& {West}}]{raymond11}
{Raymond}, S.~N., {Armitage}, P.~J., {Moro-Mart{\'{\i}}n}, A., {et~al.} 2011,
  \aap, 530, A62

\bibitem[{{Raymond} {et~al.}(2008{\natexlab{a}}){Raymond}, {Barnes},
  {Armitage}, \& {Gorelick}}]{raymond08b}
{Raymond}, S.~N., {Barnes}, R., {Armitage}, P.~J., \& {Gorelick}, N.
  2008{\natexlab{a}}, \apjl, 687, L107

\bibitem[{{Raymond} {et~al.}(2008{\natexlab{b}}){Raymond}, {Barnes}, \&
  {Mandell}}]{raymond08a}
{Raymond}, S.~N., {Barnes}, R., \& {Mandell}, A.~M. 2008{\natexlab{b}}, \mnras,
  384, 663

\bibitem[{{Raymond} {et~al.}(2009{\natexlab{b}}){Raymond}, {Barnes}, {Veras},
  {Armitage}, {Gorelick}, \& {Greenberg}}]{raymond09a}
{Raymond}, S.~N., {Barnes}, R., {Veras}, D., {et~al.} 2009{\natexlab{b}},
  \apjl, 696, L98

\bibitem[{{Raymond} {et~al.}(2006{\natexlab{a}}){Raymond}, {Mandell}, \&
  {Sigurdsson}}]{raymond06}
{Raymond}, S.~N., {Mandell}, A.~M., \& {Sigurdsson}, S. 2006{\natexlab{a}},
  Science, 313, 1413

\bibitem[{{Raymond} {et~al.}(2009{\natexlab{c}}){Raymond}, {O'Brien},
  {Morbidelli}, \& {Kaib}}]{raymond09c}
{Raymond}, S.~N., {O'Brien}, D.~P., {Morbidelli}, A., \& {Kaib}, N.~A.
  2009{\natexlab{c}}, Icarus, 203, 644

\bibitem[{{Raymond} {et~al.}(2004){Raymond}, {Quinn}, \& {Lunine}}]{raymond04}
{Raymond}, S.~N., {Quinn}, T., \& {Lunine}, J.~I. 2004, Icarus, 168, 1

\bibitem[{{Raymond} {et~al.}(2006{\natexlab{b}}){Raymond}, {Quinn}, \&
  {Lunine}}]{raymond06b}
{Raymond}, S.~N., {Quinn}, T., \& {Lunine}, J.~I. 2006{\natexlab{b}}, \icarus,
  183, 265

\bibitem[{{Raymond} {et~al.}(2007){Raymond}, {Quinn}, \& {Lunine}}]{raymond07a}
{Raymond}, S.~N., {Quinn}, T., \& {Lunine}, J.~I. 2007, Astrobiology, 7, 66

\bibitem[{{Ribas} \& {Miralda-Escud{\'e}}(2007)}]{ribas07}
{Ribas}, I. \& {Miralda-Escud{\'e}}, J. 2007, \aap, 464, 779

\bibitem[{{Santos} {et~al.}(2001){Santos}, {Israelian}, \& {Mayor}}]{santos01}
{Santos}, N.~C., {Israelian}, G., \& {Mayor}, M. 2001, \aap, 373, 1019

\bibitem[{{Schlaufman}(2010)}]{schlaufman10}
{Schlaufman}, K.~C. 2010, \apj, 719, 602

\bibitem[{{Shen} \& {Turner}(2008)}]{shen08}
{Shen}, Y. \& {Turner}, E.~L. 2008, \apj, 685, 553

\bibitem[{{Simon} \& {Prato}(1995)}]{simon95}
{Simon}, M. \& {Prato}, L. 1995, \apj, 450, 824

\bibitem[{{Su} {et~al.}(2009){Su}, {Rieke}, {Stapelfeldt}, {Malhotra},
  {Bryden}, {Smith}, {Misselt}, {Moro-Martin}, \& {Williams}}]{su09}
{Su}, K.~Y.~L., {Rieke}, G.~H., {Stapelfeldt}, K.~R., {et~al.} 2009, \apj, 705,
  314

\bibitem[{{Tanaka} \& {Ward}(2004)}]{tanaka04}
{Tanaka}, H. \& {Ward}, W.~R. 2004, \apj, 602, 388

\bibitem[{{Thommes} {et~al.}(1999){Thommes}, {Duncan}, \&
  {Levison}}]{thommes99}
{Thommes}, E.~W., {Duncan}, M.~J., \& {Levison}, H.~F. 1999, \nat, 402, 635

\bibitem[{{Thommes} {et~al.}(2003){Thommes}, {Duncan}, \&
  {Levison}}]{thommes03}
{Thommes}, E.~W., {Duncan}, M.~J., \& {Levison}, H.~F. 2003, Icarus, 161, 431

\bibitem[{{Thommes} {et~al.}(2008){Thommes}, {Matsumura}, \&
  {Rasio}}]{thommes08}
{Thommes}, E.~W., {Matsumura}, S., \& {Rasio}, F.~A. 2008, Science, 321, 814

\bibitem[{{Triaud} {et~al.}(2010){Triaud}, {Collier Cameron}, {Queloz},
  {Anderson}, {Gillon}, {Hebb}, {Hellier}, {Loeillet}, {Maxted}, {Mayor},
  {Pepe}, {Pollacco}, {S{\'e}gransan}, {Smalley}, {Udry}, {West}, \&
  {Wheatley}}]{triaud10}
{Triaud}, A.~H.~M.~J., {Collier Cameron}, A., {Queloz}, D., {et~al.} 2010,
  \aap, 524, A25+

\bibitem[{{Trilling} {et~al.}(2008){Trilling}, {Bryden}, {Beichman}, {Rieke},
  {Su}, {Stansberry}, {Blaylock}, {Stapelfeldt}, {Beeman}, \&
  {Haller}}]{trilling08}
{Trilling}, D.~E., {Bryden}, G., {Beichman}, C.~A., {et~al.} 2008, \apj, 674,
  1086

\bibitem[{{Tsiganis} {et~al.}(2005){Tsiganis}, {Gomes}, {Morbidelli}, \&
  {Levison}}]{tsiganis05}
{Tsiganis}, K., {Gomes}, R., {Morbidelli}, A., \& {Levison}, H.~F. 2005, \nat,
  435, 459

\bibitem[{{Udry} \& {Santos}(2007)}]{udry07b}
{Udry}, S. \& {Santos}, N.~C. 2007, \araa, 45, 397

\bibitem[{{Walsh} {et~al.}(2011){Walsh}, {Morbidelli}, {Raymond}, {O'Brien}, \&
  {Mandell}}]{walsh11}
{Walsh}, K.~J., {Morbidelli}, A., {Raymond}, S.~N., {O'Brien}, D.~P., \&
  {Mandell}, A.~M. 2011, \nat, 475, 206

\bibitem[{{Ward}(1997)}]{ward97}
{Ward}, W.~R. 1997, \icarus, 126, 261

\bibitem[{{Weidenschilling}(1977)}]{weidenschilling77}
{Weidenschilling}, S.~J. 1977, \apss, 51, 153

\bibitem[{{Winn} {et~al.}(2010){Winn}, {Fabrycky}, {Albrecht}, \&
  {Johnson}}]{winn10}
{Winn}, J.~N., {Fabrycky}, D., {Albrecht}, S., \& {Johnson}, J.~A. 2010, \apjl,
  718, L145

\bibitem[{{Wolk} \& {Walter}(1996)}]{wolk96}
{Wolk}, S.~J. \& {Walter}, F.~M. 1996, \aj, 111, 2066

\bibitem[{{Wright} {et~al.}(2009){Wright}, {Upadhyay}, {Marcy}, {Fischer},
  {Ford}, \& {Johnson}}]{wright09}
{Wright}, J.~T., {Upadhyay}, S., {Marcy}, G.~W., {et~al.} 2009, \apj, 693, 1084

\bibitem[{{Wyatt}(2008)}]{wyatt08}
{Wyatt}, M.~C. 2008, \araa, 46, 339

\bibitem[{{Wyatt} {et~al.}(2010){Wyatt}, {Booth}, {Payne}, \&
  {Churcher}}]{wyatt10}
{Wyatt}, M.~C., {Booth}, M., {Payne}, M.~J., \& {Churcher}, L.~J. 2010, \mnras,
  402, 657

\bibitem[{{Wyatt} {et~al.}(1999){Wyatt}, {Dermott}, {Telesco}, {Fisher},
  {Grogan}, {Holmes}, \& {Pi{\~n}a}}]{wyatt99}
{Wyatt}, M.~C., {Dermott}, S.~F., {Telesco}, C.~M., {et~al.} 1999, \apj, 527,
  918

\bibitem[{{Wyatt} {et~al.}(2005){Wyatt}, {Greaves}, {Dent}, \&
  {Coulson}}]{wyatt05}
{Wyatt}, M.~C., {Greaves}, J.~S., {Dent}, W.~R.~F., \& {Coulson}, I.~M. 2005,
  \apj, 620, 492

\bibitem[{{Wyatt} {et~al.}(2007{\natexlab{a}}){Wyatt}, {Smith}, {Greaves},
  {Beichman}, {Bryden}, \& {Lisse}}]{wyatt07a}
{Wyatt}, M.~C., {Smith}, R., {Greaves}, J.~S., {et~al.} 2007{\natexlab{a}},
  \apj, 658, 569

\bibitem[{{Wyatt} {et~al.}(2007{\natexlab{b}}){Wyatt}, {Smith}, {Su}, {Rieke},
  {Greaves}, {Beichman}, \& {Bryden}}]{wyatt07b}
{Wyatt}, M.~C., {Smith}, R., {Su}, K.~Y.~L., {et~al.} 2007{\natexlab{b}}, \apj,
  663, 365

\bibitem[{{Zakamska} {et~al.}(2010){Zakamska}, {Pan}, \& {Ford}}]{zakamska10}
{Zakamska}, N.~L., {Pan}, M., \& {Ford}, E.~B. 2010, \mnras, 1566

\end{thebibliography}

\end{document}